# Gravitational fingerprints of a black hole solution in Palatini Kalb-Ramond gravity

Carlos F. S. Pereira,[1,*] A. A. Araújo Filho,[2,3,4,†] Marcos V. de S. Silva,[5,‡] and N. Heidari[6,4,7,§]

[1] *Departamento de Física, Universidade Federal do Espírito Santo, Avenida Fernando Ferrari, 514, Goiabeiras, 29060-900, Vitória, Espiríto Santo, Brazil.*

[2] *Departamento de Física, Universidade Federal da Paraíba, Caixa Postal 5008, 58051–970, João Pessoa, Paraíba, Brazil.*

[3] *Departamento de Física, Universidade Federal de Campina Grande Caixa Postal 10071, 58429-900 Campina Grande, Paraíba, Brazil.*

[4] *Center for Theoretical Physics, Khazar University, 41 Mehseti Street, Baku, AZ-1096, Azerbaijan.*

[5] *Department of Theoretical Physics, Atomic and Optics, Campus Miguel Delibes, University of Valladolid UVA, Paseo Belén, 7, 47011 - Valladolid, Spain*

[6] *Departamento de Física, Universidade Federal de Campina Grande, Caixa Postal 10071, 58429-900 Campina Grande, Paraíba, Brazil.*

[7] *School of Physics, Damghan University, Damghan, 3671641167, Iran.*

We investigate particle motion, optical observables, and linear wave propagation around a static black hole in Palatini Kalb Ramond gravity with a vanishing cosmological constant. The analysis distinguishes the conserved mass $M$ from the orbital mass $\bar{M} = M/\rho$ and isolates the radial deformation $\sigma^2 = 1 - q/2$, where $q = \ell_2/(1 - \ell_1)$. At fixed $\bar{M}$, the photon sphere, critical impact parameter, and innermost stable circular orbit retain their Schwarzschild values. Radial frequency ratios also preserve their Schwarzschild profiles for the same areal release and particle radii, while proper distances, infall times, and accumulated orbital angles depend on $\sigma$. For radial fall, the tidal eigenvalues acquire a factor $\sigma^{-2}$, and the instability exponent of the photon orbit scales as $\sigma^{-1}$. In addition, we examine light deflection in the weak and strong regimes with an explicit account of the conical asymptotics. The shadow comparison with Sgr A* and M87* exhibits a coupling degeneracy when the orbital mass and areal distance are set. Scalar, electromagnetic, and massless Dirac perturbations probe the physical metric, whereas a homogeneous axial gravitational sector with $\delta B_{\mu\nu} = 0$ admits an exact Schwarzschild spectral map. In this sector, both oscillation and attenuation rates scale as $\sigma^{-1}$ at fixed $\bar{M}$. Third order WKB estimates agree with frequencies extracted from time evolution to within 0.7% for the reported fundamental modes.

* carlosfisica32@gmail.com
† dilto@fisica.ufc.br
‡ marcos.sousa@uva.es
§ heidari.n@gmail.com

## CONTENTS

## I. INTRODUCTION

Black hole observations constrain gravity through quantities that depend on different parts of spacetime geometry. Orbital motion fixes a mass scale, the shadow boundary selects critical null trajectories, and wave relaxation probes propagation near an effective potential barrier. The Event Horizon Telescope observations of M87* and Sgr A* bring these distinctions into contact with astronomical measurements [1–4]. The geometry of photon surfaces and analytical studies of black hole shadows establish the relation between null trapping and the critical image boundary [5, 6]. The observed bright ring also depends on the emitting material and on the trajectories that connect it to the observer [7]. Shadow tests of gravity therefore require an interpretation of the image together with its mass and distance calibration [8].

Lorentz symmetry offers a direct connection between gravitational phenomenology and field theories with tensor backgrounds. Spontaneous violation arises when a dynamical field acquires a vacuum value that selects preferred local directions. This possibility was investigated in string theory and subsequently incorporated into the gravitational sector of the Standard Model Extension [9, 10]. Among the fields that realize such a background, the antisymmetric tensor $B_{\mu\nu}$ has a natural origin in the Kalb Ramond description of string interactions [11]. Its coupling to curvature permits a vacuum tensor configuration to influence the gravitational field. The general structure of Lorentz violation with an antisymmetric tensor, including its excitations and gravitational couplings, was developed in Ref. [12].

Black hole solutions with a Kalb Ramond background have since provided explicit geometries in which these effects can be studied. Early work constructed a modified spherical solution supported by the tensor vacuum [13], followed by further static configurations with spontaneous Lorentz violation [14, 15] and electrically charged solutions [16]. Recent analyses have revisited the relation between the curvature couplings, conserved charges, and black hole thermodynamics [17]. The optical and dynamical consequences have also been examined through geodesics, deflection, scattering, and quasinormal modes [18]. Extensions involving nonlinear electrodynamics and noncommutative geometry have broadened this investigation to other Kalb Ramond backgrounds [19, 20]. Their predictions depend on the curvature couplings, the choice of connection, and the matter action.

The choice of geometric variables is essential in this comparison. In metric affine gravity, the metric and affine connection are independent, allowing torsion and nonmetricity to enter the geometric description [21, 22]. Nonmetricity, defined here by $Q_{\alpha\mu\nu} = -\nabla^{(\Gamma)}_{\alpha} g_{\mu\nu}$, measures the failure of the independent connection to preserve the metric under parallel transport. In the Palatini formulation considered below, the connection is torsion free. Nonminimal curvature couplings can nevertheless make it differ from the Levi Civita connection of the physical metric, as occurs in broader Palatini extensions of general relativity [23]. A related development appears in symmetric teleparallel gravity, where a flat and torsion free connection attributes the gravitational description to nonmetricity [24]. Spherical solutions and Solar System tests in $f(Q)$ gravity illustrate the observable role of the connection under those geometric restrictions [25, 26]. The present theory instead retains the curvature of the independent connection and relates it to an auxiliary metric.

Spontaneous Lorentz violation has already been explored within this broader geometry through metric affine bumblebee gravity [27]. Exact vacuum solutions, including stationary axisymmetric configurations, established concrete backgrounds for studying its gravitational consequences [28, 29]. More recently, She, Niu, and Yang obtained static black holes in Palatini Kalb Ramond gravity with two nonzero curvature couplings and determined their conserved mass through the Iyer Wald construction [30, 31]. Rotating configurations have also been investigated in the Palatini Kalb Ramond setting [32]. The static family considered here admits a vanishing cosmological constant and separates the mass entering the gravitational potential from the conserved charge. This distinction affects both the comparison between solutions and the interpretation of astronomical mass estimates.

The coupling of matter also needs to be specified when the connection is independent. Minimal coupling in a geometry with nonmetricity requires care at the level of the matter action [33]. We adopt neutral particles and electromagnetic radiation coupled minimally to the physical metric. Their trajectories follow its Levi Civita connection, while their measured frequencies are evaluated with the physical metric and the observer four velocity. The auxiliary metric enters the gravitational equations and the axial gravitational sector studied below.

Gravitational lensing supplies several ways to examine the resulting geometry. In the strong deflection regime, photons approach an unstable circular orbit and can generate relativistic images. The Schwarzschild lens analysis and the analytical expansion around the critical impact parameter established the basis for calculating their positions and magnifications [34–36]. Subsequent developments refined this expansion for general static spherical geometries with flat asymptotics and organized its applications to black hole lensing [37, 38]. In the weak deflection regime, the Gauss Bonnet approach relates the bending of light to the curvature and boundary terms of the optical geometry [39]. Its extension to rotating optical geometries and the treatment of sources and observers at finite distance enlarged the range of applications [40–43].

These geometric definitions become especially relevant when the spatial metric retains a radial deformation at large distance. Studies of lensing by global monopoles, cosmic strings, and backgrounds with Lorentz violation have demonstrated the role of the asymptotic geometry in the angular calculation [44–47]. For the Palatini Kalb Ramond solution, the total azimuthal sweep must be interpreted with the conical reference geometry kept explicit. The local emission and reception angles, the accumulated coordinate azimuth, and the angular excess relative to a chosen background enter distinct stages of the lensing calculation. Keeping track of them is necessary when relating an orbital integral to an observed image.

Particle motion and tidal measurements provide complementary information. Radial frequency shifts depend on both the gravitational field and the relative motion of emitter and receiver [48, 49]. Geodesic deviation instead probes the curvature measured along the falling trajectory. Analyses in charged and more general spherical black hole geometries have shown how radial stretching and transverse compression respond to changes in the metric [50, 51]. These quantities are useful here because a deformation of the radial metric can affect proper distances, elapsed times, and tidal strengths even when selected circular orbit quantities retain their Schwarzschild form.

Wave propagation supplies a further comparison through the oscillation frequencies and attenuation rates of black hole perturbations [52, 53]. The connection between unstable null orbits and quasinormal modes explains the leading behavior of many test fields at large angular momentum [54]. Its application to gravitational perturbations depends on the underlying field equations, and modified theories can violate the usual correspondence [55, 56]. For the antisymmetric field itself, recent analyses of particular metric theories with nonminimal couplings have also identified strong coupling and unstable modes [57]. Those results concern different dynamical settings and reinforce the need to state the perturbation sector under consideration. Here we distinguish scalar, electromagnetic, and Dirac test fields from a homogeneous axial gravitational sector with the covariant Kalb Ramond field held fixed. We compare WKB frequencies with characteristic time evolution and spectral extraction [58–61], restricting the stability statements to these specified sectors.

The question addressed in this work is which signatures of the Palatini Kalb Ramond background survive once the mass entering orbital dynamics has been calibrated. We consider the static solution with $\Lambda = 0$, write its orbital mass as $\bar{M} = M/\rho$, and use $\sigma^2 = 1 - q/2$, with $q = \ell_2/(1-\ell_1)$, to describe the residual radial deformation. We examine null and timelike trajectories, radial frequency shifts, tidal forces, shadows, lensing, and linear wave propagation within this common normalization. The shadow comparison with Sgr A* and M87* is interpreted using the mass and distance information that actually enters the angular observable. The parameter values used to display the exact geometry illustrate its mathematical dependence; their interpretation as viable universal couplings remains subject to the Solar System bounds obtained for this solution [30].

## II. GENERAL RELATIONS

As a starting point, we consider a static and spherically symmetric line element, whose structure is characterized by metric functions that depend exclusively on the radial coordinate. This class of geometries is widely employed to describe static gravitational configurations, which allows us to perform the investigation of various geometric and physical properties of spacetime. In this context, the line element can be written as:

$$\mathrm{d}s^2 = -F(r)\mathrm{d}t^2 + G(r)\mathrm{d}r^2 + r^2\left(\mathrm{d}\theta^2 + \sin^2\theta\,\mathrm{d}\varphi^2\right), \tag{1}$$

where $G(r) = \sigma^2/F(r)$ and $f(r)$ is a radial function that depends only on the radial coordinate. The parameter $\sigma$ connects the effects of Lorentz violation and will be defined below. In this sense, we define a smooth curve in this generic spacetime, Eq. (1), which has a length $S$ given by

$$S = \int \sqrt{\left(g_{\mu\nu}\frac{\mathrm{d}x^{\mu}}{\mathrm{d}\lambda}\frac{\mathrm{d}x^{\nu}}{\mathrm{d}\lambda}\right)}\mathrm{d}\lambda, \tag{2}$$

where $\lambda$ is an affine parameter that can represent the observer's proper time. Taking $S$ as the affine parameter itself, we can show that the curve that minimizes Eq. (2) is the same one that minimizes

$$\int \left(g_{\mu\nu}\frac{\mathrm{d}x^{\mu}}{\mathrm{d}\lambda}\frac{\mathrm{d}x^{\nu}}{\mathrm{d}\lambda}\right)\mathrm{d}\lambda = \int \mathcal{L}\,\mathrm{d}\lambda. \tag{3}$$

Now, considering the analysis in the equatorial plane, $\theta = \frac{\pi}{2}$, the Lagrangian becomes:

$$\mathcal{L} = -F(r)\left(\frac{\mathrm{d}t}{\mathrm{d}\lambda}\right)^2 + G(r)\left(\frac{\mathrm{d}r}{\mathrm{d}\lambda}\right)^2 + r^2\left(\frac{\mathrm{d}\varphi}{\mathrm{d}\lambda}\right)^2. \tag{4}$$

By applying the Euler—Lagrange equations to the previous expression, we obtain the conserved quantities associated with the cyclic coordinates $t$ and $\varphi$, corresponding to the energy and angular momentum, respectively. Thereby, we can write:

$$L = r^2\frac{\mathrm{d}\varphi}{\mathrm{d}\lambda}, \qquad E = F(r)\frac{\mathrm{d}t}{\mathrm{d}\lambda}. \tag{5}$$

By substituting the conserved quantities, Eq. (5), in Eq. (4) and considering only null geodesics, $\mathcal{L} = 0$, leads (4) to

$$\left(\frac{\mathrm{d}r}{\mathrm{d}\lambda}\right)^2 = \frac{1}{\sigma^2}\left[E^2 - \frac{L^2F(r)}{r^2}\right]. \tag{6}$$

The expression above may be shown by analogy with the dynamics of a classical particle of unit mass, with energy $\bar{\mathcal{E}}$, subject to an effective potential $V_{\text{eff}}$, given by

$$\bar{\mathcal{E}} = E^2, \qquad \text{and} \qquad V_{\text{eff}} = \frac{L^2F(r)}{r^2}\ . \tag{7}$$

Radial motion is allowed only in regions where $\bar{\mathcal{E}} > V_{\text{eff}}$. In this context, we analyze the situation in which a photon begins its trajectory in a region sufficiently far from the center of the black hole and propagates until it reaches the so–called turning point, located outside the event horizon. Upon reaching this point, the photon reverses its radial motion due to the influence of the gravitational field, propagating again toward a distant region, generally associated with an asymptotically flat spacetime. However, it is important to emphasize that the spacetime considered in this analysis is not asymptotically flat due to the presence of the parameter associated with Lorentz symmetry violation. At the turning point $r_0$, the condition $\bar{\mathcal{E}} = V_{\text{eff}}(r_0)$ must be satisfied. Considering $\bar{\mathcal{E}} = E^2$, we obtain $V_{\text{eff}}(r_0) = E^2$, which, from Eq. (7), allows us to write:

$$\frac{1}{b^2} = \frac{F(r_0)}{r_0^2}\ , \tag{8}$$

where $b = \frac{L}{E}$ is the impact parameter of the light ray. At the photon sphere radius, $r_{ph}$, we have

$$\left.\frac{\mathrm{d}V_{\text{eff}}}{\mathrm{d}r}\right|_{r_{ph}} = 0\ . \tag{9}$$

The critical impact parameter, $b_c$, is defined by the condition $b_c = b(r_{ph})$. In addition, light rays with $b < b_c$ are completely absorbed by the black hole, while those with $b = b_c$ remain circling the photon sphere; in constrast, rays with $b > b_c$ are deflected and ultimately scattered. In the scattering regime, the deflection angle diverges in the limit $r_0 \to r_{ph}$, characterizing the so–called strong field regime. In the following sections, we will derive the approximation for the deflection angle in these two regimes, considering the geometry of a BH in Palatini Kalb-Ramond gravity.

## III. THE MODEL

The model under consideration was originally proposed in the framework of Kalb–Ramond gravity formulated within the Palatini formalism, in which a nonzero vacuum expectation value of the Kalb–Ramond field induces spontaneous breaking of local Lorentz symmetry. In this context, by treating the metric and the affine connection as independent variables, a new class of black hole solutions in the presence of a cosmological constant is obtained. However, the existence of these solutions does not necessarily depend on the presence of a cosmological constant; in this case, this allows us to investigate the geometric modifications induced by the vacuum expectation value of the Kalb–Ramond field in Schwarzschild spacetime.

In this view, considering the spherically symmetric spacetime described by the line element in the form of Eq. (1), we have that the metric function is defined as

$$F(r) = G(r)^{-1} = 1 - \frac{2M}{\rho\, r} - \frac{\sigma\, \Lambda\, r^2}{3\rho}, \tag{10}$$

where the parameters that characterize the violation of Lorentz symmetry are defined as

$$\sigma = \sqrt{\frac{\left(1 - l_1 - \frac{l_2}{2}\right)}{\left(1 - l_1\right)}}, \qquad \rho = \sqrt{\left(1 - l_1 - \frac{l_2}{2}\right)\left(1 - l_1\right)}. \tag{11}$$

We can observe that for the parameter $\sigma$ to be real, $l_1 < 1$. Another point is that in the limit where these parameters go to zero ($l_1 = l_2 = 0$), we must recover the results corresponding to the Schwarzschild–(A)dS geometry, as we should naturally expect.

For simplicity, in this work, we restrict our analysis to the modifications induced in the Schwarzschild geometry, considering the limit of a vanishing cosmological constant ($\Lambda = 0$), arising from the parameters associated with the violation of the Lorentz symmetry. In other words, using Eq. (9), we obtain the radius of the photon sphere, which is modified by the violation parameter: $r_{ph} = 3M/\rho$.

## IV. GEODESIC MOTION

The propagation of light and the motion of massive particles reveal how the Kalb–Ramond background changes the orbital structure of the Palatini solution. We examine both classes of trajectories for a vanishing cosmological constant, retaining the normalization of the Killing time and the areal radius used in the preceding sections. The physical geometry is shown in Eq. (10) [30]. We take $M > 0$, $1 - \ell_1 > 0$, and $1 - \ell_1 - \ell_2/2 > 0$, so that $\rho > 0$, $\sigma > 0$, and the event horizon lies at $r_h = 2\bar{M}$. All comparisons below hold $\bar{M}$ fixed. This choice isolates the residual radial deformation after the coefficient of the gravitational potential has been specified. The displayed values of $q$ illustrate the exact geometry and are independent of an observational fit.

### A. Equations of motion and conserved quantities

Let us consider a neutral test particle minimally coupled to $g_{\mu\nu}$. Its motion is governed by the Levi–Civita connection of the physical metric. The independent Palatini connection enters the gravitational field equations through the auxiliary metric. In geometric optics, minimally coupled electromagnetic rays obey the same metric equation with null normalization. For an affine parameter $\lambda$, we have

$$\frac{\mathrm{d}^2x^\mu}{\mathrm{d}\lambda^2}+\Gamma^\mu{}_{\alpha\beta}(g)\frac{\mathrm{d}x^\alpha}{\mathrm{d}\lambda}\frac{\mathrm{d}x^\beta}{\mathrm{d}\lambda}=0,\qquad g_{\mu\nu}\frac{\mathrm{d}x^\mu}{\mathrm{d}\lambda}\frac{\mathrm{d}x^\nu}{\mathrm{d}\lambda}=-\varepsilon. \tag{12}$$

The parameter $\varepsilon$ equals zero for photons and unity for massive particles; in the latter case, $\lambda$ is the proper time $\tau$. Writing $F'=\mathrm{d}F/\mathrm{d}r$, the four differential equations take the explicit form

$$\frac{\mathrm{d}^2t}{\mathrm{d}\lambda^2}+\frac{F'}{F}\frac{\mathrm{d}t}{\mathrm{d}\lambda}\frac{\mathrm{d}r}{\mathrm{d}\lambda}=0, \tag{13}$$

$$\frac{\mathrm{d}^2r}{\mathrm{d}\lambda^2}+\frac{FF'}{2\sigma^2}\left(\frac{\mathrm{d}t}{\mathrm{d}\lambda}\right)^2 \tag{14}$$

$$-\frac{F'}{2F}\left(\frac{\mathrm{d}r}{\mathrm{d}\lambda}\right)^2-\frac{rF}{\sigma^2}\left[\left(\frac{\mathrm{d}\theta}{\mathrm{d}\lambda}\right)^2+\sin^2\theta\left(\frac{\mathrm{d}\varphi}{\mathrm{d}\lambda}\right)^2\right]=0, \tag{15}$$

$$\frac{\mathrm{d}^2\theta}{\mathrm{d}\lambda^2}+\frac{2}{r}\frac{\mathrm{d}r}{\mathrm{d}\lambda}\frac{\mathrm{d}\theta}{\mathrm{d}\lambda}-\sin\theta\cos\theta\left(\frac{\mathrm{d}\varphi}{\mathrm{d}\lambda}\right)^2=0, \tag{16}$$

$$\frac{\mathrm{d}^2\varphi}{\mathrm{d}\lambda^2}+\frac{2}{r}\frac{\mathrm{d}r}{\mathrm{d}\lambda}\frac{\mathrm{d}\varphi}{\mathrm{d}\lambda}+2\cot\theta\frac{\mathrm{d}\theta}{\mathrm{d}\lambda}\frac{\mathrm{d}\varphi}{\mathrm{d}\lambda}=0. \tag{17}$$

Spherical symmetry allows the orbital plane to be chosen as $\theta=\pi/2$. Initial data with $\mathrm{d}\theta/\mathrm{d}\lambda=0$ preserve this plane throughout the motion. Time translations and azimuthal rotations supply the conserved quantities

$$E=F(r)\frac{\mathrm{d}t}{\mathrm{d}\lambda},\qquad L=r^2\frac{\mathrm{d}\varphi}{\mathrm{d}\lambda}. \tag{18}$$

For massive particles, $E$ and $L$ are measured per unit rest mass. We choose future directed motion with $E>0$ and orient the azimuth so that $L>0$ for the nonradial trajectories. For photons, the common normalization of $E$ and $L$ follows the choice of affine parameter, while $b_\gamma=L/E$ is invariant under an affine rescaling. Substitution into the normalization condition gives

$$\sigma^2\left(\frac{\mathrm{d}r}{\mathrm{d}\lambda}\right)^2+V_\varepsilon(r)=E^2,\qquad V_\varepsilon(r)=F(r)\left(\varepsilon+\frac{L^2}{r^2}\right). \tag{19}$$

The potential fixes the turning points at a prescribed $(\bar{M},E,L)$. The factor $\sigma$ controls the radial rate and the angular sweep between them. For numerical integration, an equivalent system is

$$\frac{\mathrm{d}r}{\mathrm{d}\lambda}=v_r,\quad \frac{\mathrm{d}v_r}{\mathrm{d}\lambda}=\frac{1}{\sigma^2}\left[-\frac{\varepsilon\bar{M}}{r^2}+\frac{L^2}{r^3}-\frac{3\bar{M}L^2}{r^4}\right],\quad \frac{\mathrm{d}\varphi}{\mathrm{d}\lambda}=\frac{L}{r^2}. \tag{20}$$

This system remains finite at $r_h$ and passes smoothly through radial turning points. The plotted exterior trajectories terminate at the horizon for capture or at a specified outer radius for scattering, as we have argued in the previous sections. Energy normalization is monitored through Eq. (19) along every solution.

For $L\neq 0$, the inverse radius $u=1/r$ obeys the common orbital equation

$$\sigma^2\frac{\mathrm{d}^2u}{\mathrm{d}\varphi^2}+u=\frac{\varepsilon\bar{M}}{L^2}+3\bar{M}u^2. \tag{21}$$

Table I. Impact parameters $b_\gamma/\bar{M}$ generated by Eqs. (24) and (25). The superscripts C and S denote capture and scattering, respectively. The common threshold is $b_c/\bar{M} = 3\sqrt{3} \simeq 5.1962$.

| $\lvert v_{r0}\rvert$ | $q=-0.6$ | $q=-0.3$ | $q=0$ | $q=0.3$ | $q=0.6$ |
|---|---|---|---|---|---|
| 0.18 | $5.1819^{\mathrm{C}}$ | $5.3928^{\mathrm{S}}$ | $5.6318^{\mathrm{S}}$ | $5.9057^{\mathrm{S}}$ | $6.2239^{\mathrm{S}}$ |
| 0.21 | $4.6534^{\mathrm{C}}$ | $4.8626^{\mathrm{C}}$ | $5.1028^{\mathrm{C}}$ | $5.3825^{\mathrm{S}}$ | $5.7139^{\mathrm{S}}$ |
| 0.25 | $4.0743^{\mathrm{C}}$ | $4.2745^{\mathrm{C}}$ | $4.5074^{\mathrm{C}}$ | $4.7831^{\mathrm{C}}$ | $5.1165^{\mathrm{C}}$ |
| 0.31 | $3.4117^{\mathrm{C}}$ | $3.5935^{\mathrm{C}}$ | $3.8078^{\mathrm{C}}$ | $4.0657^{\mathrm{C}}$ | $4.3843^{\mathrm{C}}$ |

Introducing $\psi = \varphi/\sigma$ puts Eq. (21) in its Schwarzschild–like form. At the same $(\bar{M}, E, L)$ and with an aligned angular origin, the exact relation is $r(\varphi) = r_{\mathrm{Sch}}(\varphi/\sigma)$. The physical azimuth retains its period $2\pi$, so the rescaling changes the angular accumulation in a full orbit. This property will organize the null and timelike trajectories considered in what follows.

### B. Null trajectories: scattering and capture

For photons, Eq. (19) reduces to

$$\sigma^2 \left(\frac{\mathrm{d}r}{\mathrm{d}\lambda}\right)^2 = E^2 \left[1 - b_\gamma^2 \frac{F(r)}{r^2}\right]. \tag{22}$$

The maximum of $F/r^2$ occurs at $r_{\mathrm{ph}} = 3\bar{M}$ and yields $b_c = 3\sqrt{3}\,\bar{M}$. For a scattered ray, the outer turning point satisfies

$$b_\gamma^2 = \frac{r_{\mathrm{min}}^2}{F(r_{\mathrm{min}})}, \qquad r_{\mathrm{min}} > 3\bar{M}. \tag{23}$$

To follow the trajectory construction used in our numerical light path analysis [20], we prescribe the initial position and spatial coordinate velocities and determine the future directed time component from the null condition. In the equatorial plane, the resulting initial integrals are

$$E = \sqrt{\sigma^2 v_{r0}^2 + F(r_0) r_0^2 \omega_0^2}, \;\; L = r_0^2 \omega_0, \;\; \left.\frac{\mathrm{d}t}{\mathrm{d}\lambda}\right|_0 = \frac{E}{F(r_0)}, \tag{24}$$

where $v_{r0} = (\mathrm{d}r/\mathrm{d}\lambda)_0 < 0$ and $\omega_0 = (\mathrm{d}\varphi/\mathrm{d}\lambda)_0 > 0$. The four trajectory families use

$$\begin{aligned} r_0 &= 7.7\bar{M}, &\qquad \varphi_0 &= \frac{3\pi}{2}, \\ \bar{M}\omega_0 &= 0.022, &\qquad v_{r0} &\in \{-0.18, -0.21, -0.25, -0.31\}. \end{aligned} \tag{25}$$

We vary $q$ over $\{-0.6, -0.3, 0, 0.3, 0.6\}$. Since $E$ contains $\sigma$ in Eq. (24), a common set of coordinate velocities produces a different impact parameter for each curve. The local launch direction also changes: the angle $\beta_0$ between the photon and the inward radial direction, measured by a static observer, satisfies

$$\sin\beta_0 = \frac{b_\gamma \sqrt{F(r_0)}}{r_0}, \qquad \cos\beta_0 = \frac{\sigma |v_{r0}|}{E}. \tag{26}$$

Table I displays the resulting $b_\gamma/\bar{M}$ and identifies capture or scattering. It makes the transition in the first two trajectory families explicit. For the two larger inward velocities, every displayed photon reaches the horizon.

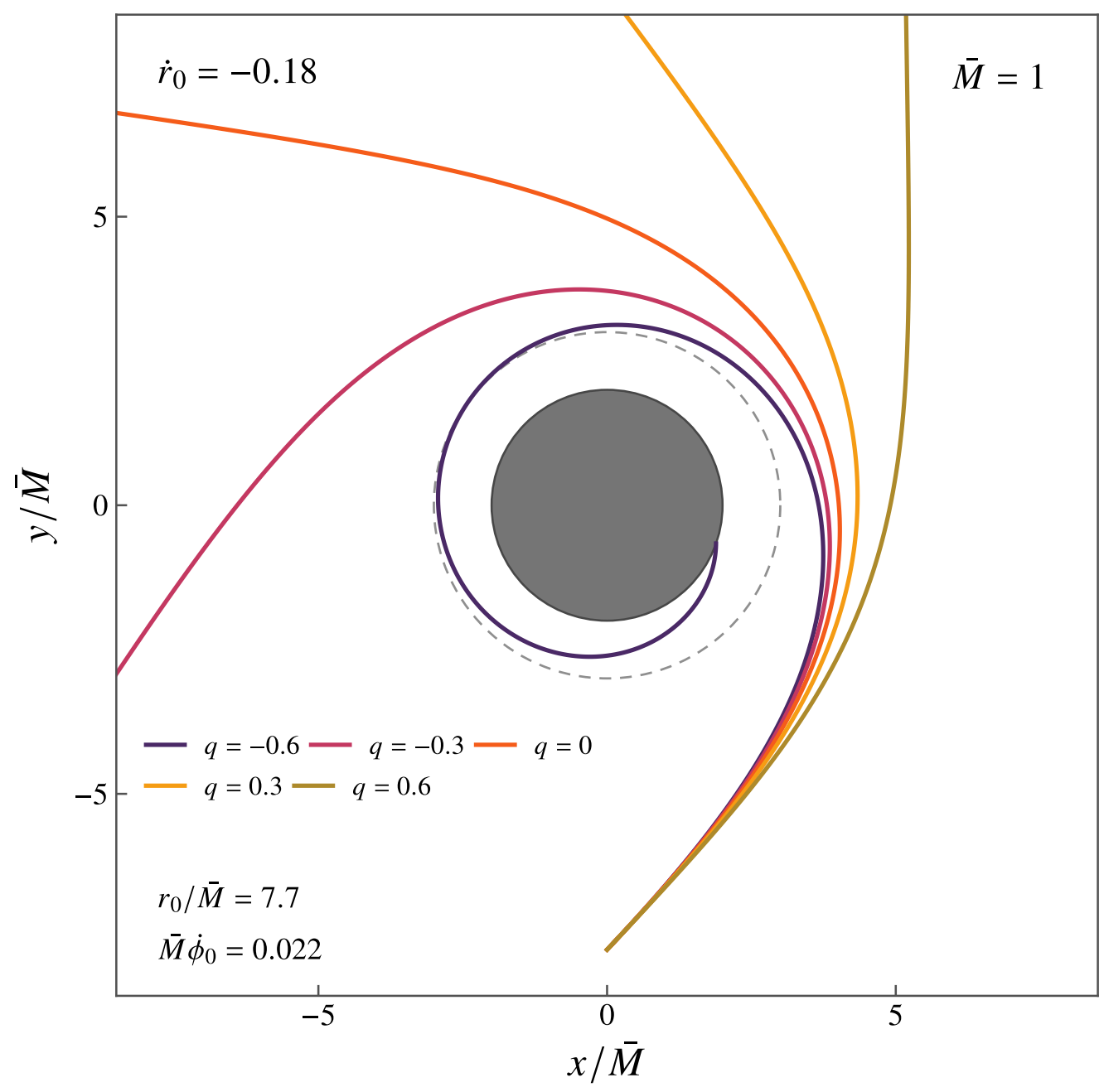

Figure 1. Null trajectories for $v_{r0} = -0.18$, $r_0 = 7.7\bar{M}$, $\varphi_0 = 3\pi/2$, and $\bar{M}\omega_0 = 0.022$, at fixed $\bar{M}$. The ray with $q = -0.6$ is captured after approaching the photon sphere; the remaining four rays scatter. The gray disk and dashed circle have radii $2\bar{M}$ and $3\bar{M}$, respectively.

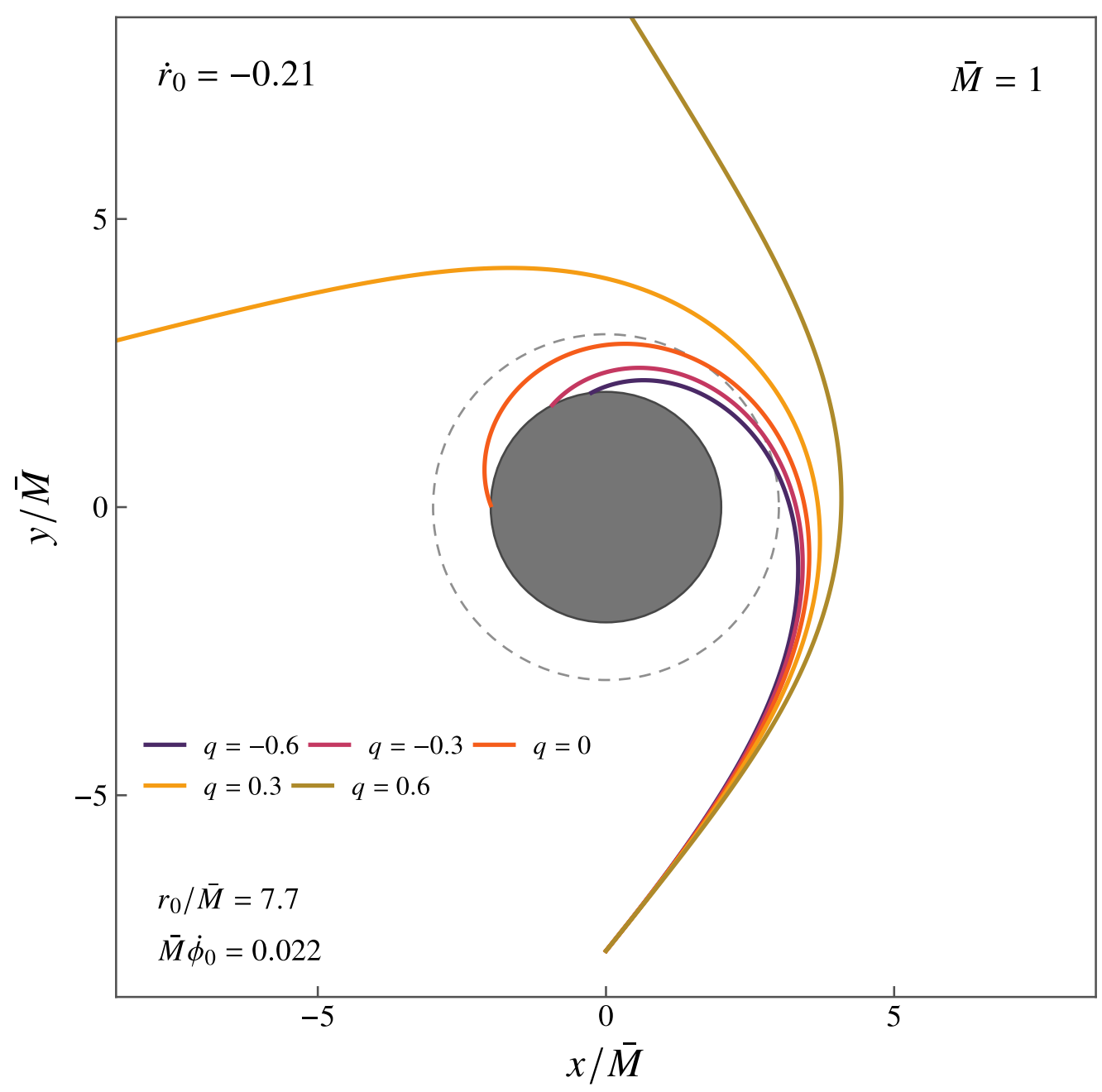

Figure 2. Null trajectories for $v_{r0} = -0.21$ and the remaining launch data of Fig. 1. The impact parameters place $q = -0.6, -0.3, 0$ in capture and $q = 0.3, 0.6$ in scattering. The transition follows the initial null normalization in Eq. (24).

Figures 1–4 present the four families individually. The coordinates $x = r\cos\varphi$ and $y = r\sin\varphi$ provide a planar representation of the areal radius and physical azimuth. The central disk indicates the horizon, and the dashed circle marks the common photon sphere. Proper radial lengths follow from $\sigma\,\mathrm{d}r/\sqrt{F}$; notice that the displayed plane is a coordinate representation of the trajectories.

We now specify a common incident ray by fixing $E = 1$ and $b_\gamma = 6\bar{M}$. Equation (26) gives the same local launch angle at a common $r_0$, and Eq. (23) supplies a common closest approach. The ingoing radial velocity becomes

$$v_{r0} = -\frac{1}{\sigma}\sqrt{1 - b_\gamma^2 \frac{F(r_0)}{r_0^2}}. \tag{27}$$

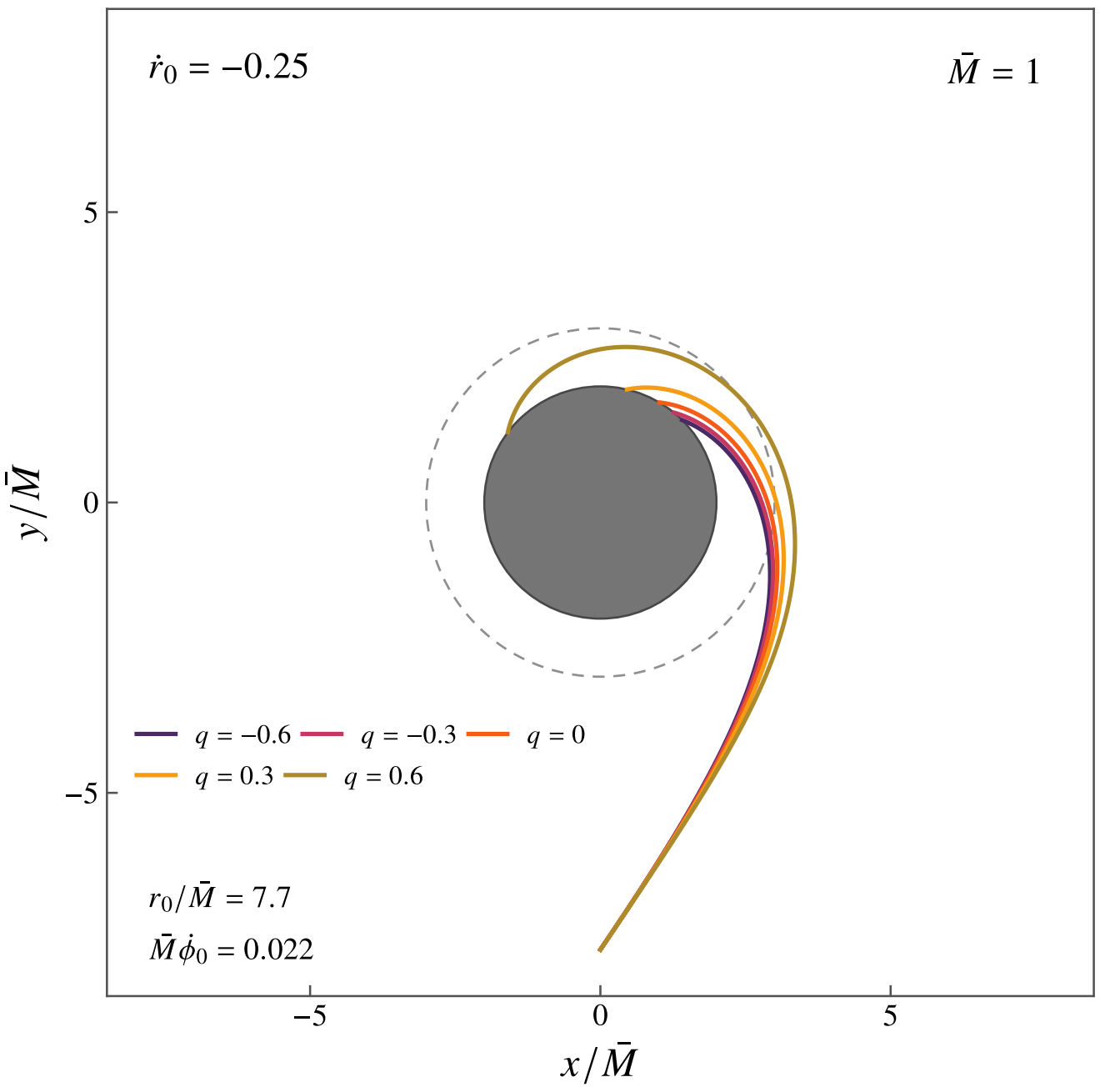

Figure 3. Captured null trajectories for $v_{r0} = -0.25$, with the same initial position and angular velocity. Every impact parameter lies below $b_c$, while the angular accumulation before horizon arrival depends on $q$.

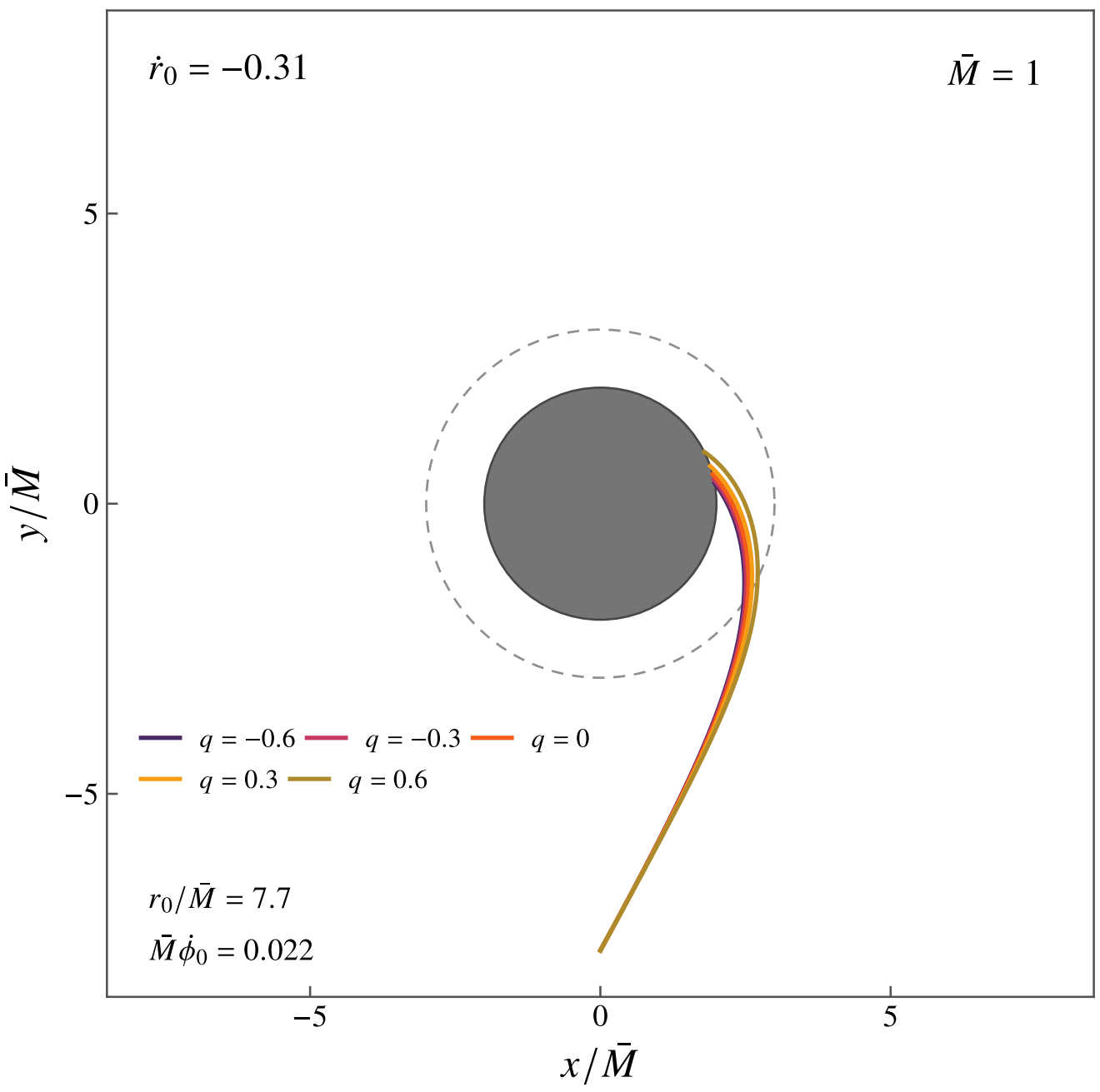

Figure 4. Captured null trajectories for $v_{r0} = -0.31$. The increased inward coordinate velocity lowers the impact parameters in all five cases. Each curve terminates at $r = 2\bar{M}$.

Figure 5 isolates the angular deformation for this choice. Each ray reaches the same turning radius and returns to the outer sphere, while its azimuthal sweep scales with $\sigma$. Increasing $q$ decreases $\sigma$ and reduces this sweep. An important remark which deserves being pointed out, at fixed $M$, changing the couplings would also change $\bar{M} = M/\rho$ and rescale the horizon, the photon sphere, and the turning radii.

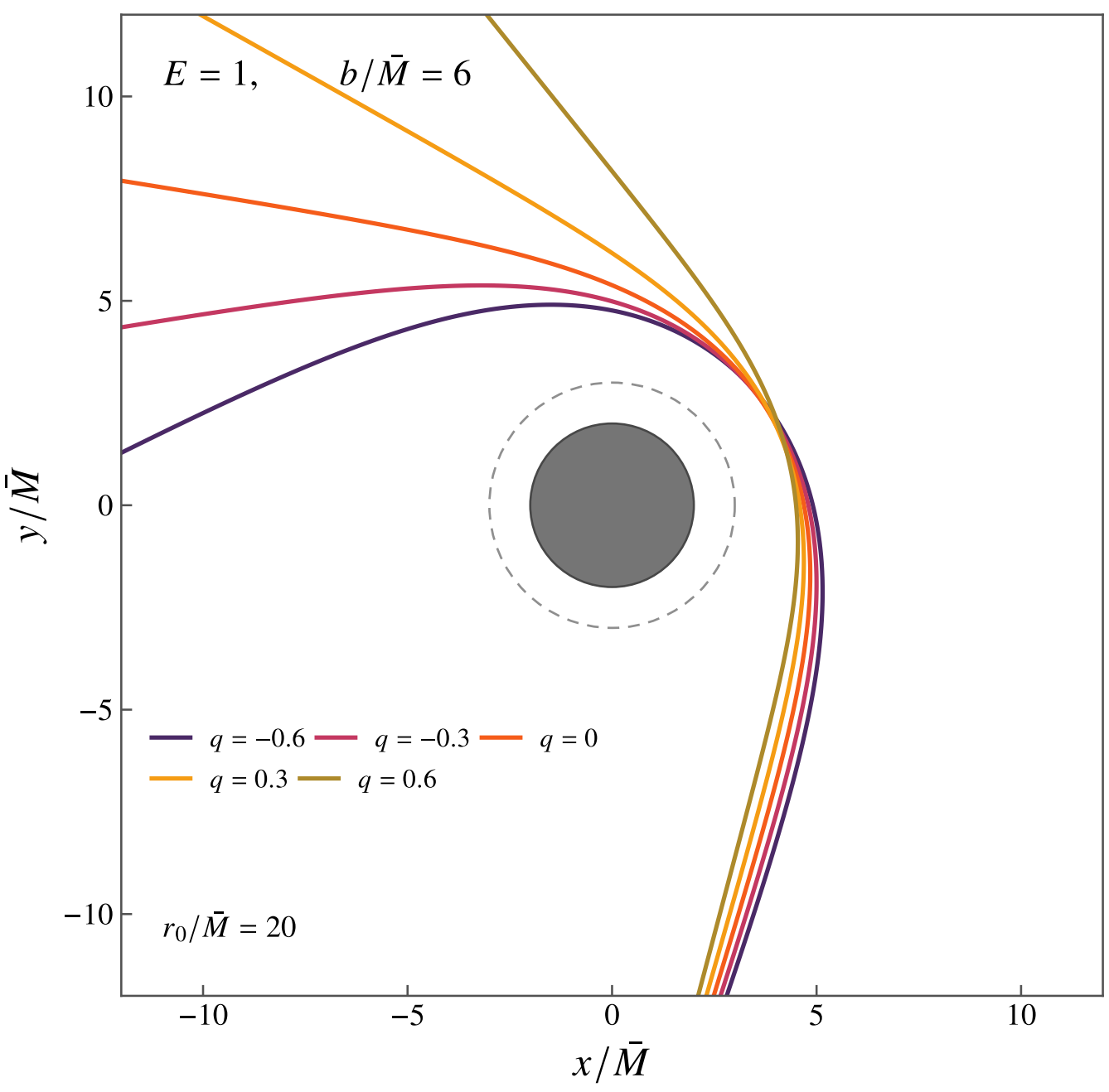

Figure 5. Scattering trajectories at fixed $E = 1$ and $b_\gamma = 6\bar{M}$, launched from $r_0 = 20\bar{M}$ and $\varphi_0 = 3\pi/2$. The common closest approach is $r_{\min} \simeq 4.4534\bar{M}$; differences between the curves arise from the angular factor $\sigma$. The same $q$ values and color assignment as in Figs. 1–4 are used.

### C. Angular bending from weak to strong deflection

For a scattered photon arriving from and returning to large radius, the total azimuthal sweep is

$$\Delta\varphi(b_\gamma) = 2\sigma \int_{r_{\min}}^{\infty} \frac{b_\gamma\,\mathrm{d}r}{r^2\sqrt{1 - b_\gamma^2 F(r)/r^2}} = \sigma\,\Delta\varphi_{\rm Sch}(b_\gamma). \tag{28}$$

Its reference angle follows from the asymptotic spatial geometry. On the equatorial plane, the $\bar{M} = 0$ metric has radial distance $\sigma r$ and circumference $2\pi r$. A complete ray in this conical geometry sweeps $\pi\sigma$. We define, in this case, the excess in the physical azimuth by subtracting the ray in the same zero mass geometry:

$$\alpha_{\rm cone}(b_\gamma) \equiv \Delta\varphi(b_\gamma) - \pi\sigma = \sigma\,\alpha_{\rm Sch}(b_\gamma), \qquad \alpha_{\rm Sch} \equiv \Delta\varphi_{\rm Sch} - \pi. \tag{29}$$

Using $\pi$ as the Euclidean angular reference gives $\Delta\varphi - \pi = \pi(\sigma - 1) + \alpha_{\rm cone}$. These definitions keep the asymptotic geometry explicit. In the locally unfolded azimuth $\psi = \varphi/\sigma$, the corresponding excess is $\Delta\psi - \pi = \alpha_{\rm Sch}$. In this manner, $\alpha_{\rm cone}$ measures accumulated coordinate azimuth. At an exterior radius $r$, a static observer measures $\sin\beta = b_\gamma\sqrt{F(r)}/r$.

The exact factorization in Eq. (29) gives the weak deflection expansion

$$\alpha_{\rm cone} = \sigma\left[\frac{4\bar{M}}{b_\gamma} + \frac{15\pi}{4}\left(\frac{\bar{M}}{b_\gamma}\right)^2 + \mathcal{O}\left(\frac{\bar{M}^3}{b_\gamma^3}\right)\right]. \tag{30}$$

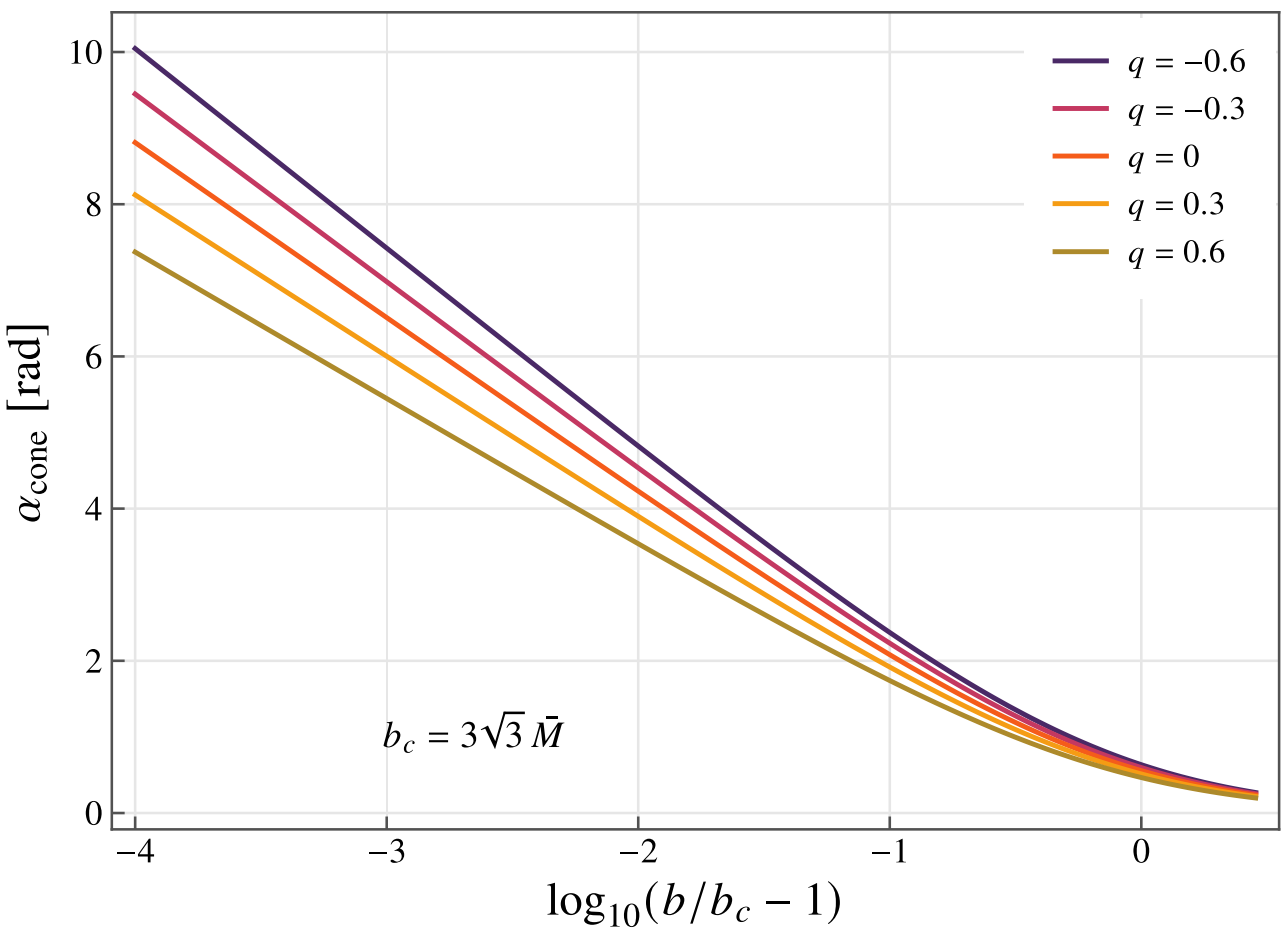

Figure 6. Coordinate angular excess $\alpha_{\rm cone}$ relative to the zero mass conical geometry, computed from Eq. (28). The distance from the critical impact parameter is shown on a logarithmic scale. The curves retain the exact dependence on $\sigma$; the divergence has the coefficient given in Eq. (31).

Near the capture, let us set $\delta = b_\gamma/b_c - 1 > 0$. The Schwarzschild strong deflection expansion [? ] yields

$$\alpha_{\rm cone} = \sigma\left[-\ln\delta + \ln\Big(216[7-4\sqrt{3}]\Big) - \pi\right] + \mathcal{O}(\delta\ln\delta). \tag{31}$$

The coefficient of the logarithm is $\sigma$, while the critical impact parameter stays fixed at a given $\bar{M}$. Figure 6 shows the exact numerical integral across the strong and moderate bending regimes. The increased angular accumulation near $b_c$ accounts for the long excursions around the photon sphere in the trajectory plots.

## D. Instability of critical photon motion

The location of the photon sphere and the rate at which a nearby ray departs from it encode different aspects of the geometry. Linearizing the radial equation about $r = 3\bar{M}$ gives

$$\frac{\mathrm{d}^2\delta r}{\mathrm{d}\lambda^2} = \frac{E^2}{3\sigma^2\bar{M}^2}\,\delta r. \tag{32}$$

On the circular orbit, $\mathrm{d}t/\mathrm{d}\lambda = 3E$. The instability exponent in normalized Killing time and the orbital angular frequency are [54]

$$\delta r \propto e^{\lambda_t t}, \qquad \lambda_t = \frac{1}{3\sqrt{3}\,\sigma\bar{M}}, \qquad \Omega_{\rm ph} = \frac{\mathrm{d}\varphi}{\mathrm{d}t} = \frac{1}{3\sqrt{3}\,\bar{M}}. \tag{33}$$

The growing perturbation mode is amplified by $\exp(2\pi/\sigma)$ during one full azimuthal revolution. Equivalently, the instability time is $\lambda_t^{-1} = 3\sqrt{3}\,\sigma\bar{M}$. Larger $\sigma$ permits a longer residence near the critical orbit, in agreement with the stronger angular accumulation in Eq. (31). The two quantities satisfy the exact relation

$$\frac{\Omega_{\rm ph}}{\lambda_t} = \sigma. \tag{34}$$

Figure 7 displays their different dependence on $q$. At fixed orbital mass, the same photon sphere radius can accompany different instability times. The calculation follows directly from geodesic motion and the specified normalization of time.

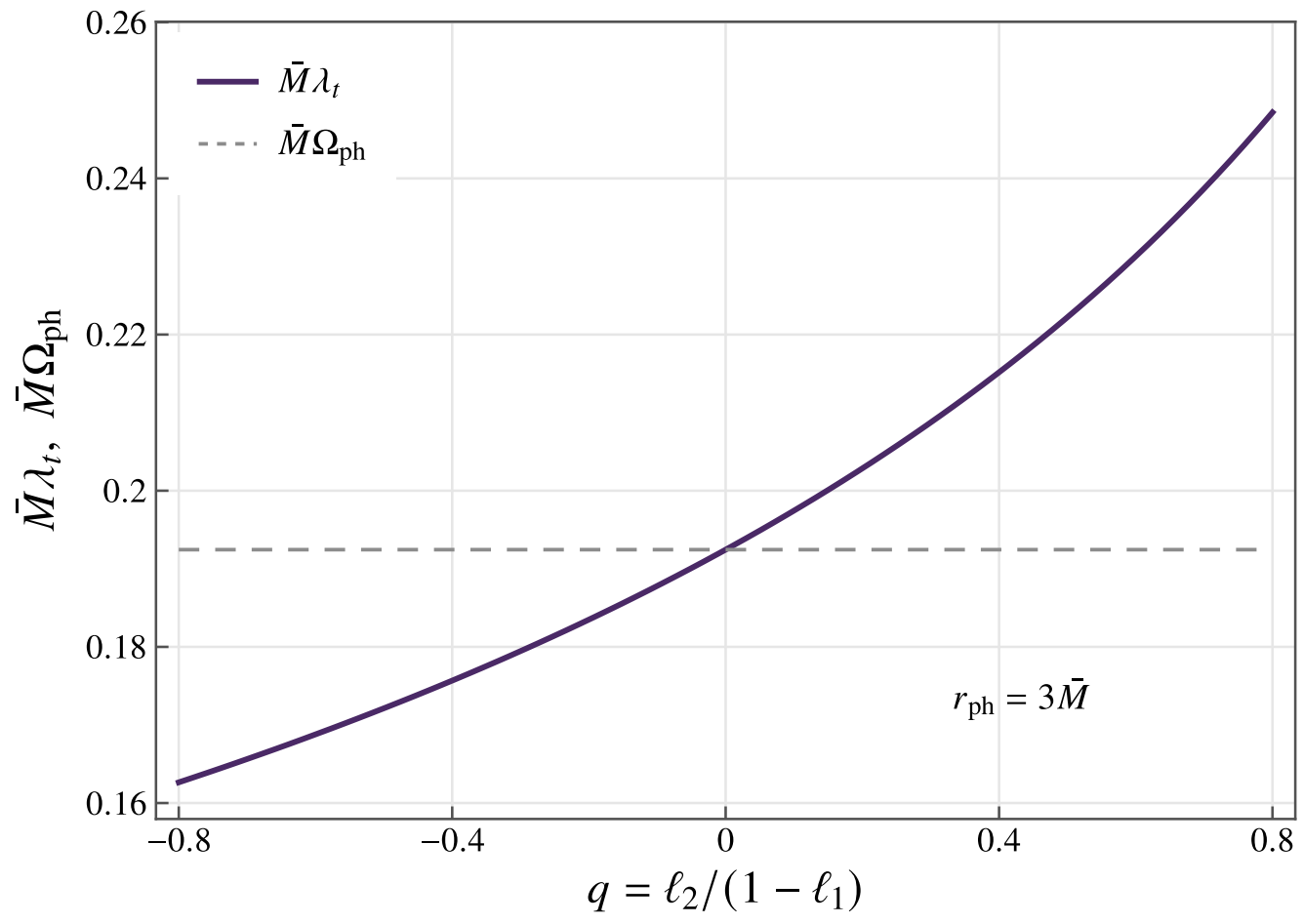

Figure 7. Dimensionless instability exponent $\bar{M}\lambda_t$ and photon orbit frequency $\bar{M}\Omega_{\rm ph}$ as functions of $q$, at fixed $\bar{M}$. Increasing $q$ lowers $\sigma = \sqrt{1-q/2}$ and increases the instability exponent. The orbital frequency remains constant.

### E. Bound timelike motion and orbital stability

For a massive particle, the effective potential is $V_1 = F(1 + L^2/r^2)$. A circular orbit at $r = r_c$ requires $V_1(r_c) = E_c^2$ and $V_1'(r_c) = 0$, giving

$$E_c^2 = \frac{(r_c - 2\bar{M})^2}{r_c(r_c - 3\bar{M})}, \qquad L_c^2 = \frac{\bar{M} r_c^2}{r_c - 3\bar{M}}. \tag{35}$$

Circular timelike trajectories exist for $r_c > 3\bar{M}$. Their radial stability follows from

$$V_1''(r_c) = \frac{2\bar{M}(r_c - 6\bar{M})}{r_c^3(r_c - 3\bar{M})}, \qquad r_{\rm ISCO} = 6\bar{M}, \qquad r_{\rm mb} = 4\bar{M}. \tag{36}$$

The marginally bound orbit has $E_c = 1$ at $r_{\rm mb}$. The sign of $V_1''$ selects stable circular motion for $r_c > 6\bar{M}$, while $3\bar{M} < r_c < 6\bar{M}$ gives radial instability. For a small radial oscillation about a stable circular orbit, the frequencies in Killing time obey

$$\Omega_c^2 = \frac{\bar{M}}{r_c^3}, \qquad \kappa_t^2 = \frac{\bar{M}(r_c - 6\bar{M})}{\sigma^2 r_c^4}, \qquad \frac{\Omega_c}{\kappa_t} = \frac{\sigma}{\sqrt{1 - 6\bar{M}/r_c}}. \tag{37}$$

The critical radii keeps their Schwarzschild values in units of $\bar{M}$, and the radial frequency carries the deformation. In the small eccentricity limit, the azimuthal sweep between successive periapses is $2\pi\Omega_c/\kappa_t$, and the coordinate periapsis advance relative to one full revolution is $2\pi(\Omega_c/\kappa_t - 1)$.

For finite eccentricity, an exact description of a bound orbit is obtained with a radial anomaly $\chi$:

$$\frac{r(\chi)}{\bar{M}} = \frac{p}{1 + e\cos\chi}, \qquad \frac{{\rm d}\varphi}{{\rm d}\chi} = \sigma\sqrt{\frac{p}{p - 6 - 2e\cos\chi}}, \tag{38}$$

with $0 \le e < 1$, and $p > 6 + 2e$. Here, $p$ is the dimensionless semilatus rectum, and the bound motion constants are

$$E^2 = \frac{(p-2)^2 - 4e^2}{p(p - 3 - e^2)}, \qquad \frac{L^2}{\bar{M}^2} = \frac{p^2}{p - 3 - e^2}. \tag{39}$$

The periapsis and apoapsis occur at $r_p = p\bar{M}/(1+e)$ and $r_a = p\bar{M}/(1-e)$. Integrating Eq. (38) over a radial period gives

$$\Phi_r = 4\sigma\sqrt{\frac{p}{p - 6 + 2e}}\, K\!\left(\frac{4e}{p - 6 + 2e}\right), \qquad \Delta\varphi_p = \Phi_r - 2\pi, \tag{40}$$

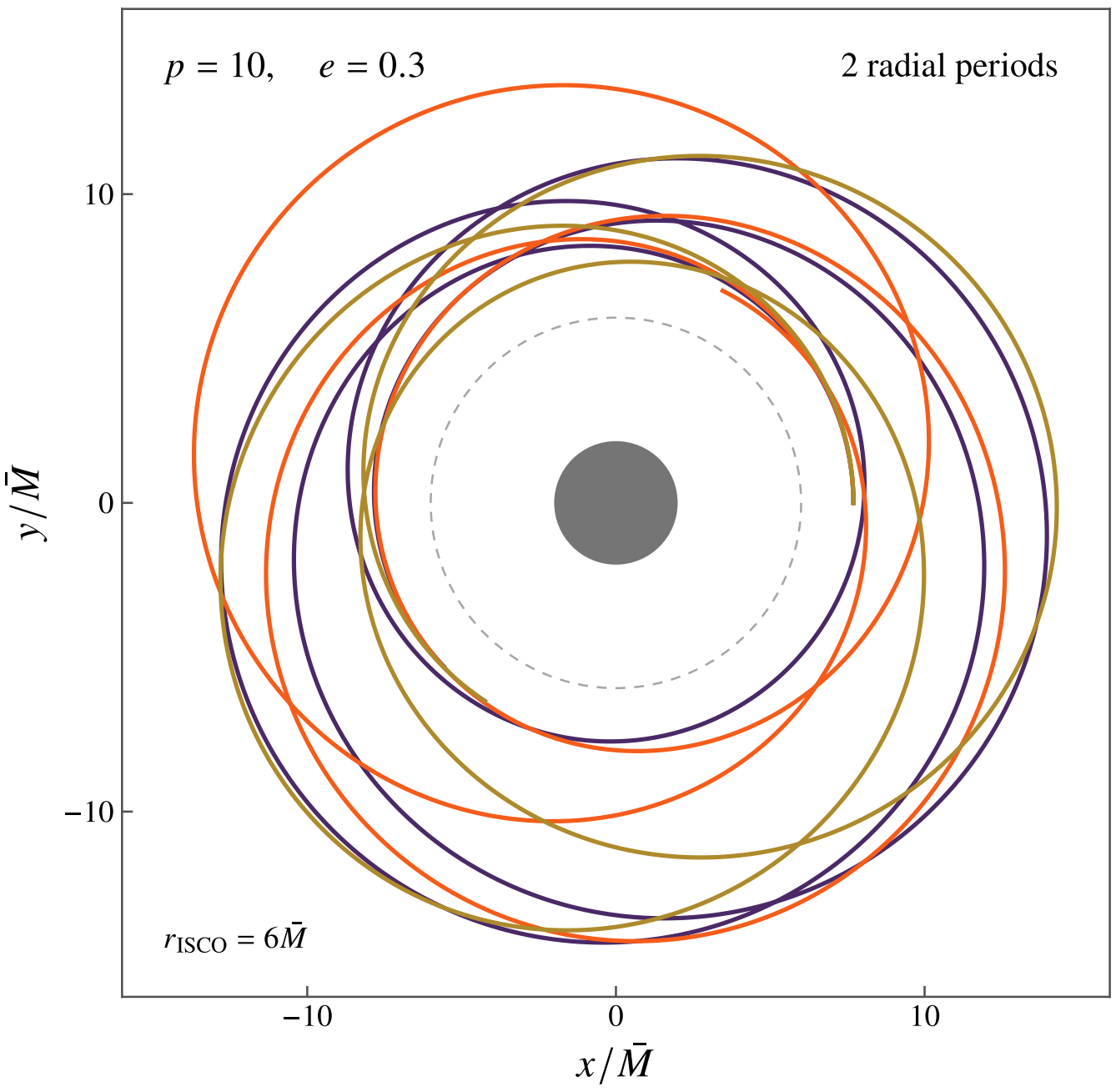

Figure 8. Bound timelike trajectories over two radial periods, with $p = 10$, $e = 0.3$, and $q = -0.6, 0, 0.6$. Each starts at periapsis with $\varphi = 0$. The common turning radii are $r_p = 10\bar{M}/1.3$ and $r_a = 10\bar{M}/0.7$, while the azimuthal period follows Eq. (40). The central disk marks the event horizon, and the dashed circle indicates $r_{\rm ISCO} = 6\bar{M}$.

where $K(z) = \int_0^{\pi/2}(1 - z\sin^2\zeta)^{-1/2}\,d\zeta$ is the complete elliptic integral with parameter $z$. In this sense, $\Phi_r = \sigma\Phi_r^{\rm Sch}$ holds for arbitrary eccentricity in the stated domain. In the weak field limit, this relation yields

$$\Delta\varphi_p = 2\pi(\sigma - 1) + \frac{6\pi\sigma}{p} + \mathcal{O}(p^{-2}). \tag{41}$$

Notice that the first term reflects the angular periodicity of the asymptotic geometry, while the second is the relativistic contribution to the coordinate periapsis advance. Figure 8 shows two radial periods for $p = 10$ and $e = 0.3$. All curves share $r_p$, $r_a$, $E$, and $L$, and differ in the accumulated azimuth.

## V. GRAVITATIONAL DOPPLER EFFECT

Considering the static solution of Palatini Kalb-–Ramond gravity with a vanishing cosmological constant [30], along with the definitions previously established at the beginning of Sec. IV, we can determine the physical quantities associated with the Doppler effect. Let us remember that, $\ell_1 = \xi_1 b^2$ and $\ell_2 = \xi_2 b^2$ are the dimensionless KR couplings, $M$ is the conserved Iyer–Wald mass [31], and $\bar{M}$ fixes the Newtonian potential and the areal horizon, $r_h = 2\bar{M}$. We work in the domain $M > 0$, $1 - \ell_1 > 0$, and $1 - \ell_1 - \ell_2/2 > 0$, with $\sigma > 0$. Neutral test particles and electromagnetic radiation couple minimally to $g_{\mu\nu}$. Their trajectories, measured frequencies, and proper times follow from the physical metric [23].

For a massive particle moving radially, conservation of the specific Killing energy and normalization of its four velocity give

$$\mathcal{E} = F(r)\dot{t}, \qquad \sigma^2\dot{r}^2 + F(r) = \mathcal{E}^2, \tag{42}$$

where the dot denotes differentiation with respect to proper time. Release from rest at $r = B > r_h$ fixes

$$\mathcal{E} = \sqrt{F(B)}, \qquad \dot{r} = -\frac{\sqrt{\mathcal{E}^2 - F(r)}}{\sigma}. \tag{43}$$

The static orthonormal frame measures the inward speed

$$v(r;B) = \frac{\sigma}{F(r)}\left|\frac{\mathrm{d}r}{\mathrm{d}t}\right| = \sqrt{1 - \frac{F(r)}{F(B)}} = \sqrt{\frac{2\bar{M}(B-r)}{r(B-2\bar{M})}}. \tag{44}$$

In this manenr, $v$ increases from zero at release to unity as $r \to r_h^+$. The latter value is the limit measured by static observers in the exterior [62]. The coordinate velocity is $\mathrm{d}r/\mathrm{d}t = -F(r)v/\sigma$.

Let a static observer remain at $B$ while exchanging radial photons with the infalling particle. For each photon, the conserved Killing frequency $\omega_\infty = -k_t > 0$ determines

$$k^\mu = \left(\frac{\omega_\infty}{F(r)}, \epsilon\frac{\omega_\infty}{\sigma}, 0, 0\right), \qquad \epsilon = +1 \text{ (outgoing)}, \\ \epsilon = -1 \text{ (ingoing)}. \tag{45}$$

The two observer velocities are

$$u^\mu_{\mathrm{ff}} = \left(\frac{\mathcal{E}}{F(r)}, -\frac{\sqrt{\mathcal{E}^2 - F(r)}}{\sigma}, 0, 0\right), \\ u^\mu_B = \left(\frac{1}{\sqrt{F(B)}}, 0, 0, 0\right). \tag{46}$$

Using $\omega = -k_\mu u^\mu$ [49], the local photon frequency along the falling trajectory becomes

$$\omega_{\mathrm{ff}} = \frac{\omega_\infty}{F(r)}\left[\mathcal{E} + \epsilon\sqrt{\mathcal{E}^2 - F(r)}\right]. \tag{47}$$

For an outgoing signal emitted by the particle and received at $B$, this gives

$$g_{\mathrm{out}}(r;B) \equiv \frac{\omega_B^{(r)}}{\omega_{\mathrm{ff}}^{(s)}} = \frac{F(r)}{\mathcal{E}\left[\mathcal{E} + \sqrt{\mathcal{E}^2 - F(r)}\right]} = 1 - v(r;B). \tag{48}$$

For an ingoing signal emitted at $B$ and received by the particle, the frequency ratio is

$$g_{\mathrm{in}}(r;B) \equiv \frac{\omega_{\mathrm{ff}}^{(r)}}{\omega_B^{(s)}} = \frac{\mathcal{E}\left[\mathcal{E} - \sqrt{\mathcal{E}^2 - F(r)}\right]}{F(r)} = \frac{1}{1 + v(r;B)}. \tag{49}$$

The superscripts $(s)$ and $(r)$ identify the emitted and received frequencies. Both ratios start at unity, while

$$g_{\mathrm{out}} \longrightarrow 0, \qquad g_{\mathrm{in}} \longrightarrow \frac{1}{2}, \qquad r \longrightarrow 2\bar{M}^+. \tag{50}$$

The outgoing signal accumulates an unbounded redshift near the horizon. For the ingoing signal, the gravitational frequency increase and the motion of the receiver combine into the finite value 1/2 [48, 62]. Release from infinity gives $\mathcal{E} = 1$ and

$$g_{\mathrm{out}}^{(\infty)} = 1 - \sqrt{\frac{2\bar{M}}{r}}, \qquad g_{\mathrm{in}}^{(\infty)} = \left(1 + \sqrt{\frac{2\bar{M}}{r}}\right)^{-1}. \tag{51}$$

At fixed $(\bar{M}, r, B)$, the factors of $\sigma$ in the radial motion and the photon wave vector cancel from the measured ratios. Figures 9 and 10 show the resulting profiles for $B = 10\bar{M}$. A comparison at fixed conserved mass $M$ retains the coupling dependence through $\bar{M} = M/\rho$; both the lapse and the areal horizon then vary with $\rho$. This distinction specifies the physical mass scale held fixed in each comparison.

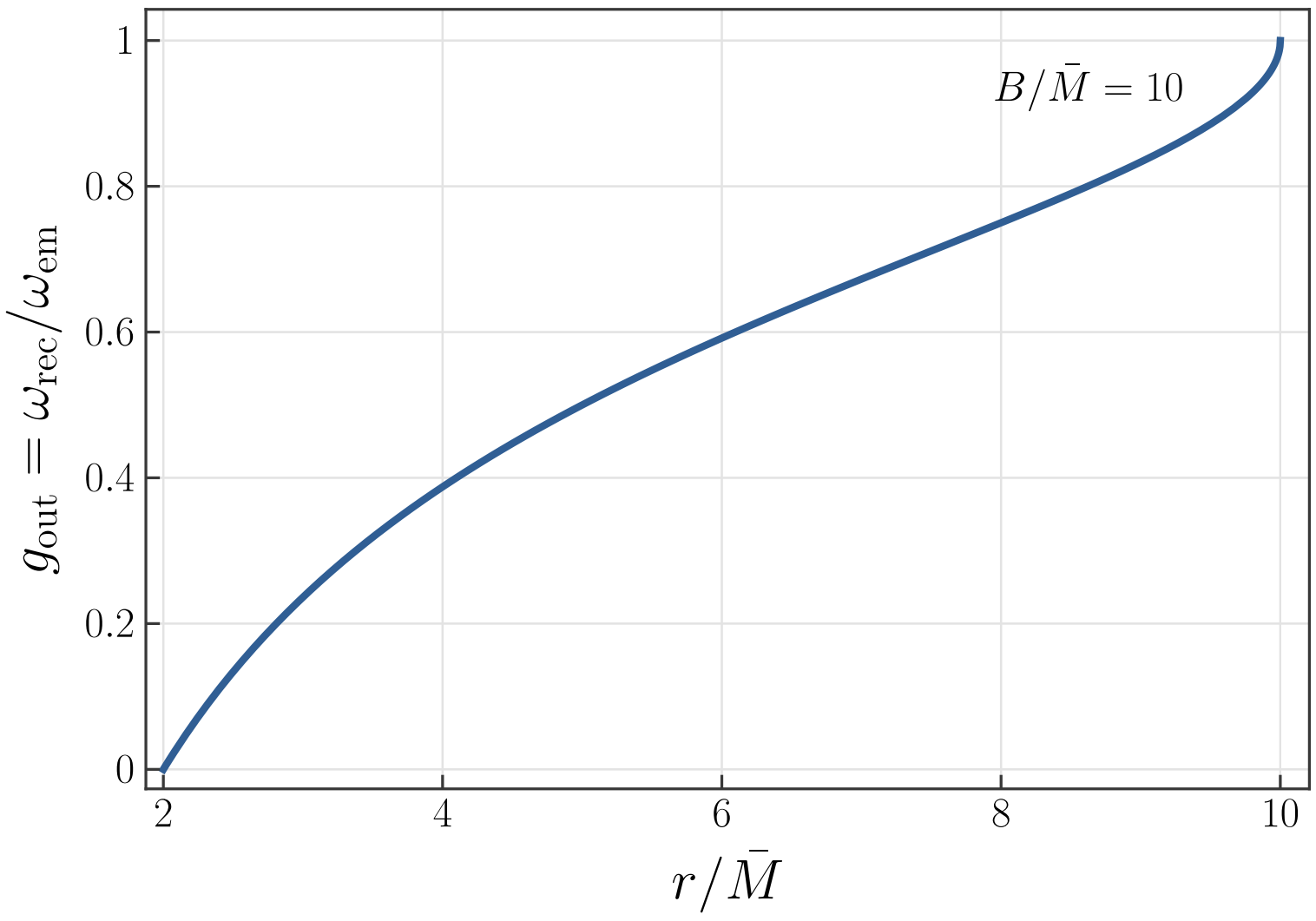

Figure 9. Outgoing frequency ratio for release from rest at $B = 10\bar{M}$. The horizontal coordinate is the areal radius in units of $\bar{M}$. The profile applies to every admissible $q$ at this fixed mass scale, with $g_{\rm out} = 1$ at release and $g_{\rm out} \to 0$ at the horizon.

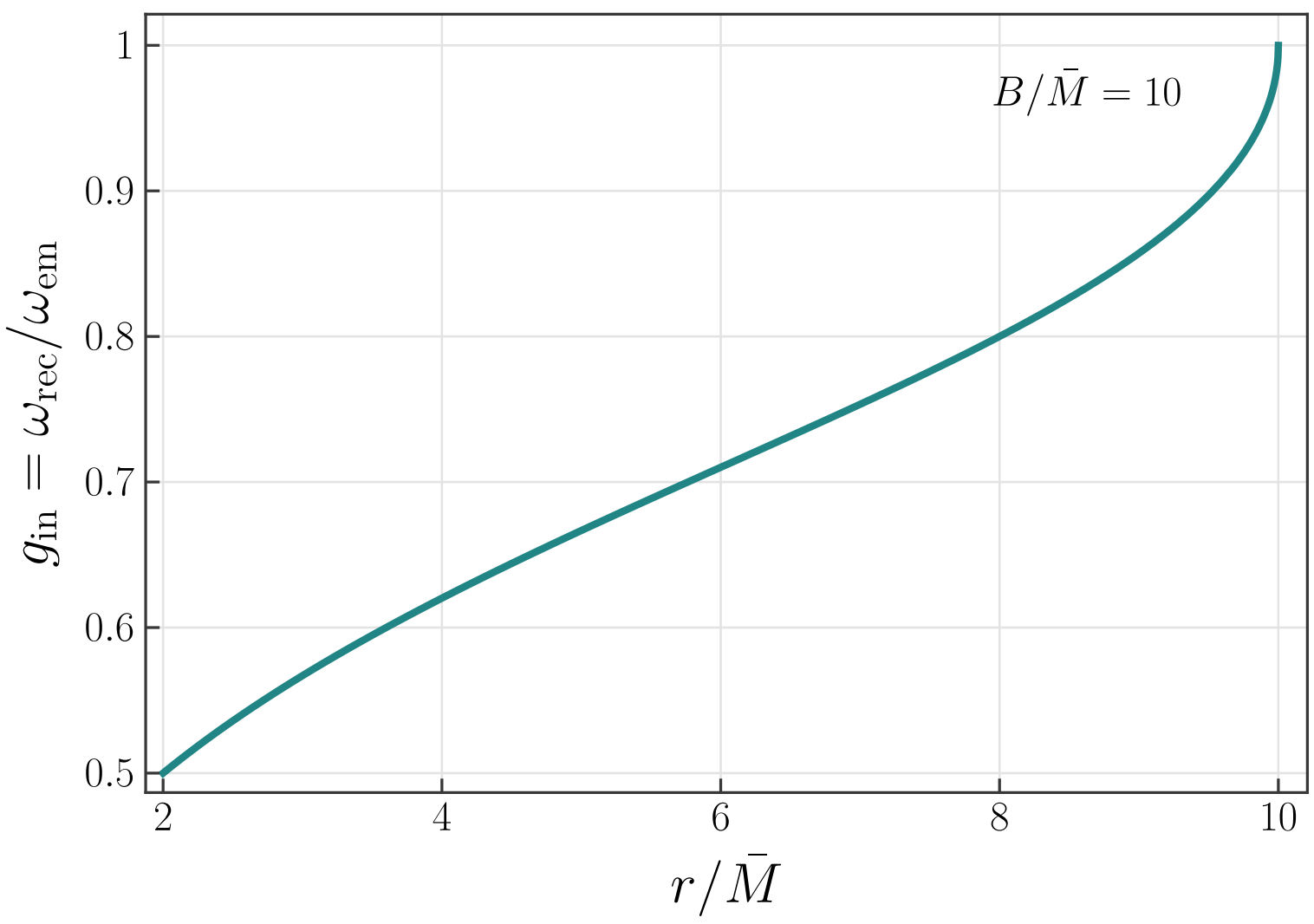

Figure 10. Ingoing frequency ratio for the same release radius. During the fall, $g_{\rm in}$ decreases from unity to its horizon limit 1/2. The curves for all admissible $q$ coincide at fixed $r/\bar{M}$ and $B/\bar{M}$.

The KR deformation enters directly when the position is expressed through proper radial distance. On a hypersurface of constant $t$, the distance between $r$ and $B$ is

$$\mathcal{L}(B, r) = \sigma \int_r^B \frac{\mathrm{d}r'}{\sqrt{F(r')}} = \sigma\big[\mathcal{P}(B) - \mathcal{P}(r)\big], \tag{52}$$

with

$$\mathcal{P}(r) = \sqrt{r(r - 2\bar{M})} + 2\bar{M}\ \mathrm{arcosh}\sqrt{\frac{r}{2\bar{M}}}. \tag{53}$$

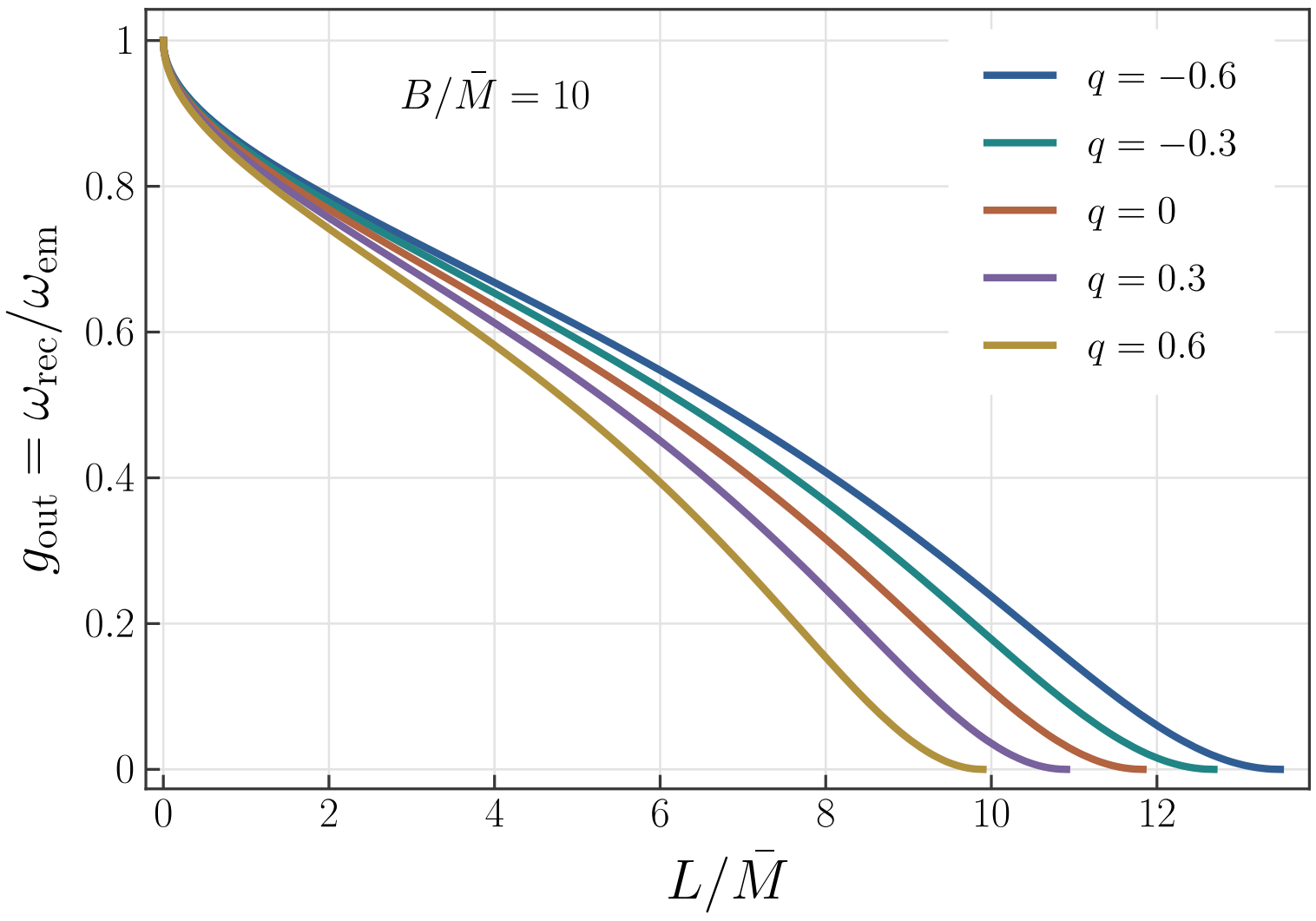

Figure 11. Outgoing frequency ratio as a function of the proper radial distance from the release point, $B = 10\bar{M}$. The illustrated values $q = -0.6, -0.3, 0, 0.3, 0.6$ show how $\sigma = \sqrt{1 - q/2}$ rescales the distance axis. The endpoint of each curve corresponds to $r \to 2\bar{M}^+$.

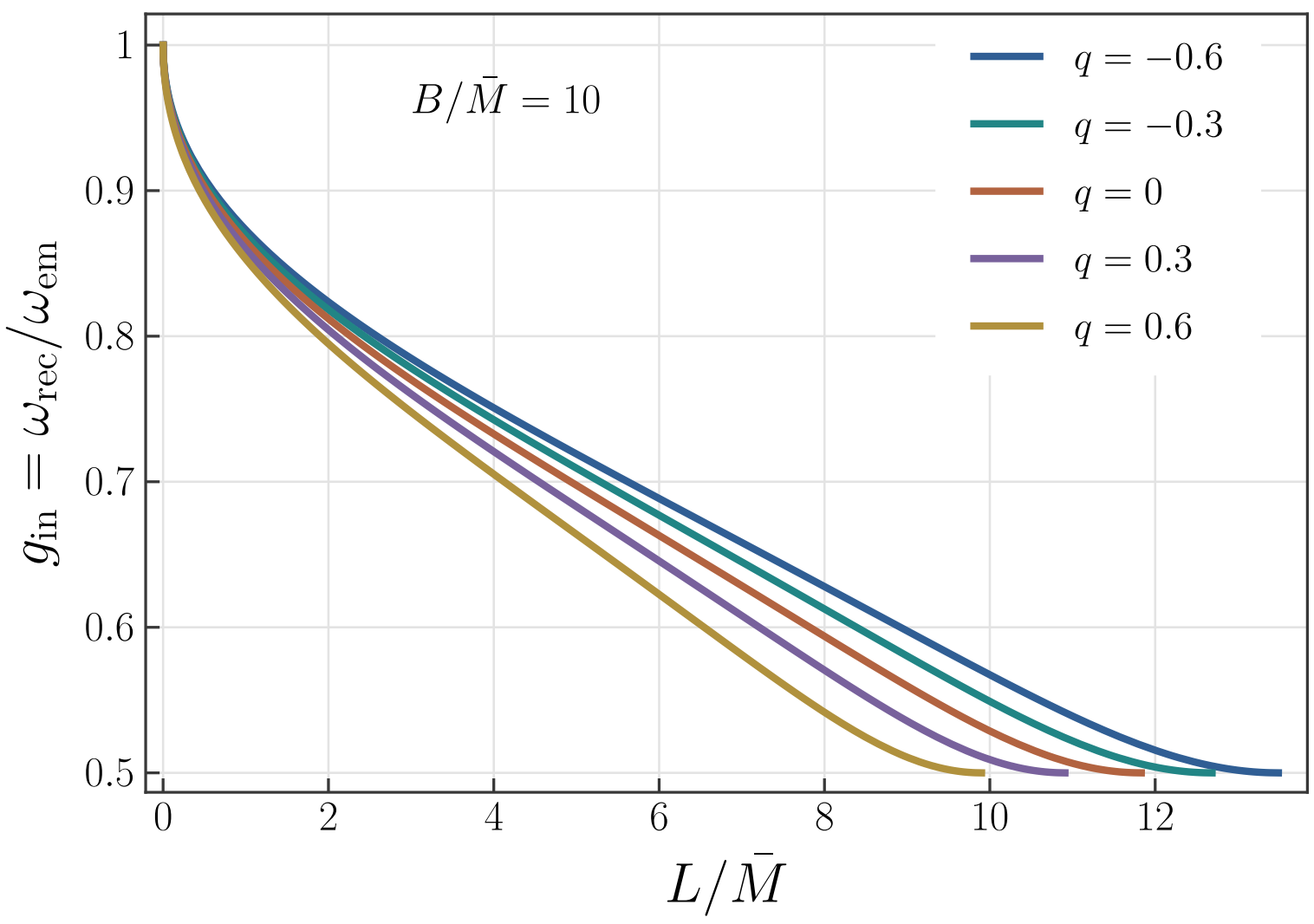

Figure 12. Ingoing frequency ratio versus proper radial distance for the same configurations. Every curve begins at unity and reaches 1/2 at its horizon endpoint. Positive $q$ reduces the proper distance required to attain a given ratio.

This distance is defined by the static spatial geometry. For $q > 0$, $\sigma < 1$, and a specified frequency ratio is reached after a smaller proper radial interval. For $q < 0$, $\sigma > 1$, and the corresponding interval increases. The separated curves in Figs. 11 and 12 describe this spatial dependence. Each curve ends at its own horizon distance, $\mathcal{L}_h = \sigma\mathcal{P}(B)$.

The elapsed proper time of the particle and the Killing time of an outgoing radial signal also carry the factor

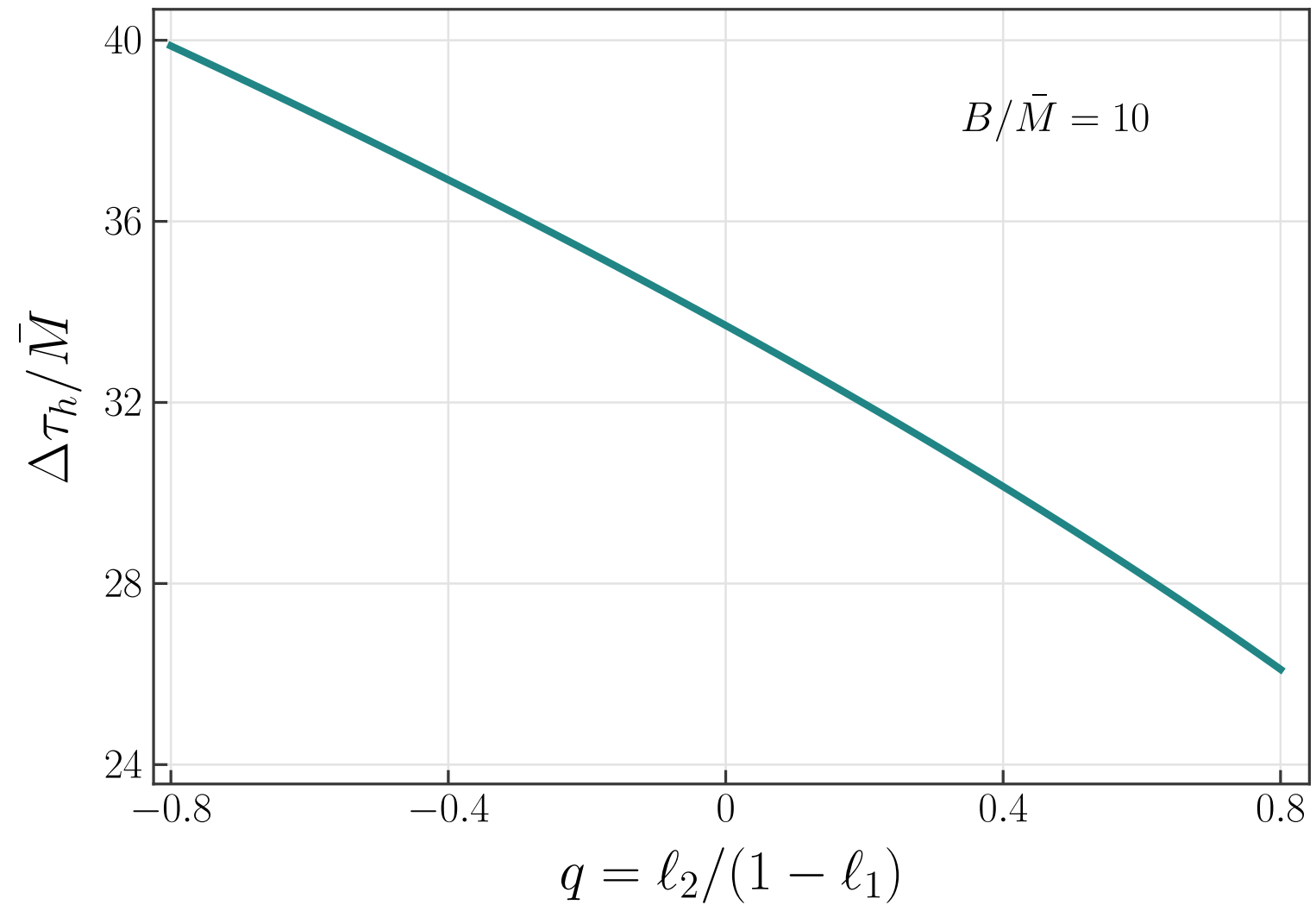

Figure 13. Proper time from release at $B = 10\bar{M}$ to the horizon, in units of $\bar{M}$. The dependence on the KR couplings is $\Delta\tau_h(q) = \sqrt{1-q/2}\,\Delta\tau_h(0)$ at fixed $\bar{M}$ and $B/\bar{M}$.

$\sigma$:

$$\Delta\tau(B \to r) = \sigma\sqrt{\frac{B^3}{2\bar{M}}}\left[\arccos\sqrt{\frac{r}{B}} + \frac{\sqrt{r(B-r)}}{B}\right], \tag{54}$$

$$\Delta t_\gamma(r \to B) = \sigma\left[B - r + 2\bar{M}\ln\left(\frac{B-2\bar{M}}{r-2\bar{M}}\right)\right]. \tag{55}$$

The static observer at $B$ measures a signal propagation interval $\sqrt{F(B)}\,\Delta t_\gamma$. The particle crosses the horizon after a finite proper time, while the outgoing propagation interval grows logarithmically as the emission radius approaches $r_h$. Figure 13 displays the horizon arrival time for $B = 10\bar{M}$. At fixed $\bar{M}$, the fractional variation of each distance or duration proportional to $\sigma$ is $\sigma - 1 = -q/4 + O(q^2)$.

For a nonzero cosmological constant, the same frequency construction holds locally with

$$F_\Lambda(r) = 1 - \frac{2\bar{M}}{r} - \frac{\Lambda r^2}{3(1-\ell_1)}, \qquad v^2 = 1 - \frac{F_\Lambda(r)}{F_\Lambda(B)}. \tag{56}$$

Equations (48) and (49) retain the forms $1 - v$ and $(1+v)^{-1}$ where $F_\Lambda > 0$ and $F_\Lambda(B) \geq F_\Lambda(r)$ along the inward trajectory. Release from rest initiates inward motion when $F'_\Lambda(B) > 0$; for $\Lambda > 0$, this requires $B^3 < 3(1-\ell_1)\bar{M}/\Lambda$ within the static patch. The observer at $B$ fixes the reference clock in this case. The figures above use $\Lambda = 0$.

## VI. TIDAL FORCES

The deformation of a freely falling body is determined by the curvature of the physical metric projected onto its comoving frame [51]. For the minimally coupled matter considered here, the relevant covariant derivative and Riemann tensor are those of $g_{\mu\nu}$ [23]. Let us remember that the independent Palatini connection enters the gravitational field equations through the auxiliary geometry, as we have argued in the previous sections.

### A. Tidal tensor in the freely falling frame

Let $\eta^\mu$ connect neighboring radial timelike geodesics, with $u_\mu\eta^\mu = 0$. Their relative acceleration satisfies [50, 63]

$$\frac{D^2\eta^\mu}{D\tau^2} + R^\mu{}_{\nu\alpha\beta}(g)u^\nu\eta^\alpha u^\beta = 0. \tag{57}$$

We adopt the convention

$$\begin{aligned} R^\mu{}_{\nu\alpha\beta} &= \partial_\alpha\Gamma^\mu_{\beta\nu}(g) - \partial_\beta\Gamma^\mu_{\alpha\nu}(g) \\ &\quad + \Gamma^\mu_{\alpha\lambda}(g)\Gamma^\lambda_{\beta\nu}(g) - \Gamma^\mu_{\beta\lambda}(g)\Gamma^\lambda_{\alpha\nu}(g). \end{aligned} \tag{58}$$

Writing $w(r) = \sqrt{\mathcal{E}^2 - F(r)}$, an orthonormal tetrad adapted to inward motion is [50, 51]

$$\begin{aligned} e^\mu_{\hat 0} &= \left(\frac{\mathcal{E}}{F}, -\frac{w}{\sigma}, 0, 0\right), & e^\mu_{\hat 1} &= \left(-\frac{w}{F}, \frac{\mathcal{E}}{\sigma}, 0, 0\right), \\ e^\mu_{\hat 2} &= \left(0, 0, \frac{1}{r}, 0\right), & e^\mu_{\hat 3} &= \left(0, 0, 0, \frac{1}{r\sin\theta}\right). \end{aligned} \tag{59}$$

It obeys $e^\mu_{\hat 0} = u^\mu$, $g_{\mu\nu}e^\mu_{\hat a}e^\nu_{\hat b} = \mathrm{diag}(-1,1,1,1)_{\hat a\hat b}$, and $De^\mu_{\hat a}/D\tau = 0$. In this parallelly transported frame, the spatial components of Eq. (57) become

$$\frac{\mathrm{d}^2\eta^{\hat a}}{\mathrm{d}\tau^2} = \mathcal{T}^{\hat a}{}_{\hat b}\eta^{\hat b}, \qquad \mathcal{T}_{\hat a\hat b} = -R_{\hat 0\hat a\hat 0\hat b}, \qquad \hat a, \hat b = \hat 1, \hat 2, \hat 3. \tag{60}$$

Direct projection gives

$$\begin{aligned} R_{\hat 0\hat 1\hat 0\hat 1} &= \frac{F''}{2\sigma^2}, & R_{\hat 0\hat\imath\hat 0\hat\imath} &= \frac{F'}{2\sigma^2 r}, \\ R_{\hat 1\hat\imath\hat 1\hat\imath} &= -\frac{F'}{2\sigma^2 r}, & R_{\hat 2\hat 3\hat 2\hat 3} &= \frac{\sigma^2 - F}{\sigma^2 r^2}, \qquad \hat\imath = \hat 2, \hat 3, \end{aligned} \tag{61}$$

with each repeated $\hat\imath$ identifying a single component. The tidal tensor is diagonal,

$$\mathcal{T}^{\hat a}{}_{\hat b} = \mathrm{diag}(\lambda_r, \lambda_\perp, \lambda_\perp), \quad \lambda_r = -\frac{F''}{2\sigma^2}, \quad \lambda_\perp = -\frac{F'}{2\sigma^2 r}. \tag{62}$$

The relation $R_{\hat 0\hat\imath\hat 0\hat\imath} = -R_{\hat 1\hat\imath\hat 1\hat\imath}$ makes these transverse components invariant under a radial Lorentz boost. Together with the radial component, it removes $\mathcal{E}$ from the tidal eigenvalues. Every radial freely falling observer measures the same tidal strengths at a given areal radius.

### B. Radial tidal stretching

For $F(r) = 1 - 2\bar M/r$, the radial relative acceleration is

$$\frac{\mathrm{d}^2\eta^{\hat 1}}{\mathrm{d}\tau^2} = \lambda_r(r)\eta^{\hat 1}, \qquad \lambda_r(r) = \frac{2\bar M}{\sigma^2 r^3} = \frac{2\bar M}{(1-q/2)r^3}. \tag{63}$$

The positive eigenvalue stretches neighboring worldlines along the direction of the fall. Its magnitude grows as $r^{-3}$ toward the central singularity and has the finite horizon value

$$\lambda_r(r_h) = \frac{1}{4\sigma^2\bar M^2}. \tag{64}$$

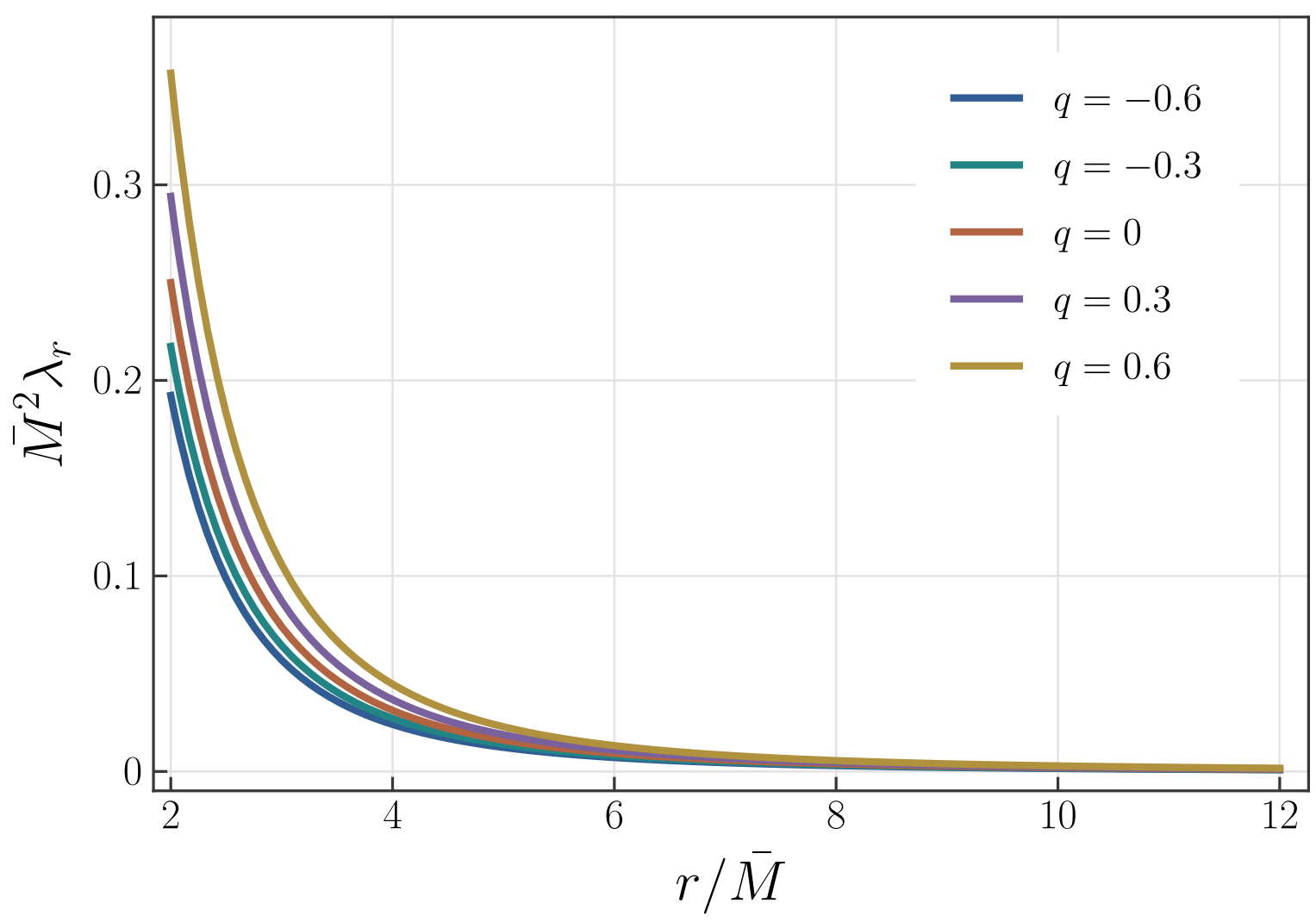

Figure 14. Dimensionless radial tidal eigenvalue $\bar{M}^2\lambda_r$ versus $r/\bar{M}$ for representative $q = -0.6, -0.3, 0, 0.3, 0.6$. Each exterior profile begins at the common areal horizon $r_h/\bar{M} = 2$. Positive $q$ increases the radial stretching at fixed $r/\bar{M}$.

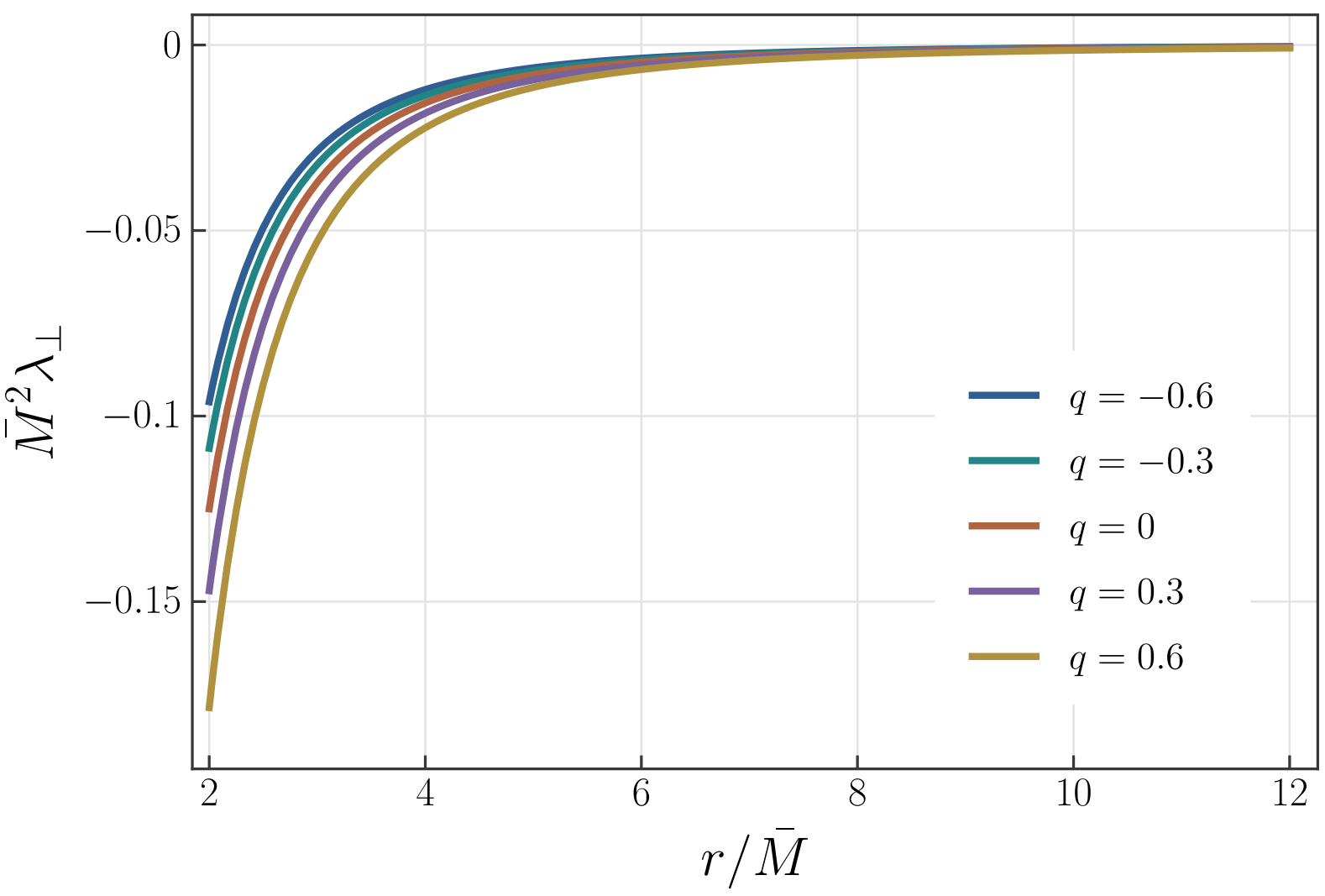

Figure 15. Radial tidal strength at the horizon in units of $\bar{M}^{-2}$. The value at $q = 0$ is $1/4$, and the KR dependence follows $\bar{M}^2\lambda_r(r_h) = [4(1 - q/2)]^{-1}$. The mass scale $\bar{M}$ is held fixed.

At fixed $(\bar{M}, r)$, positive $q$ strengthens this acceleration, while negative $q$ weakens it. Relative to the Schwarzschild value,

$$\frac{\lambda_r}{\lambda_r^{\mathrm{Schw}}} = \frac{1}{1 - q/2} = 1 + \frac{q}{2} + O(q^2). \tag{65}$$

Figures 14 and 15 display the exterior profile and the horizon dependence at this fixed mass scale.

The accumulated radial deformation can also be exactly obtained [50, 63]. For release from rest at $B$, impose $\eta^{\hat{1}}(B) = \eta_{r0}$ and $\dot{\eta}^{\hat{1}}(B) = 0$. Combining Eqs. (42) and (63) yields

$$(\mathcal{E}^2 - F)\frac{\mathrm{d}^2\eta^{\hat{1}}}{\mathrm{d}r^2} - \frac{F'}{2}\frac{\mathrm{d}\eta^{\hat{1}}}{\mathrm{d}r} + \frac{F''}{2}\eta^{\hat{1}} = 0. \tag{66}$$

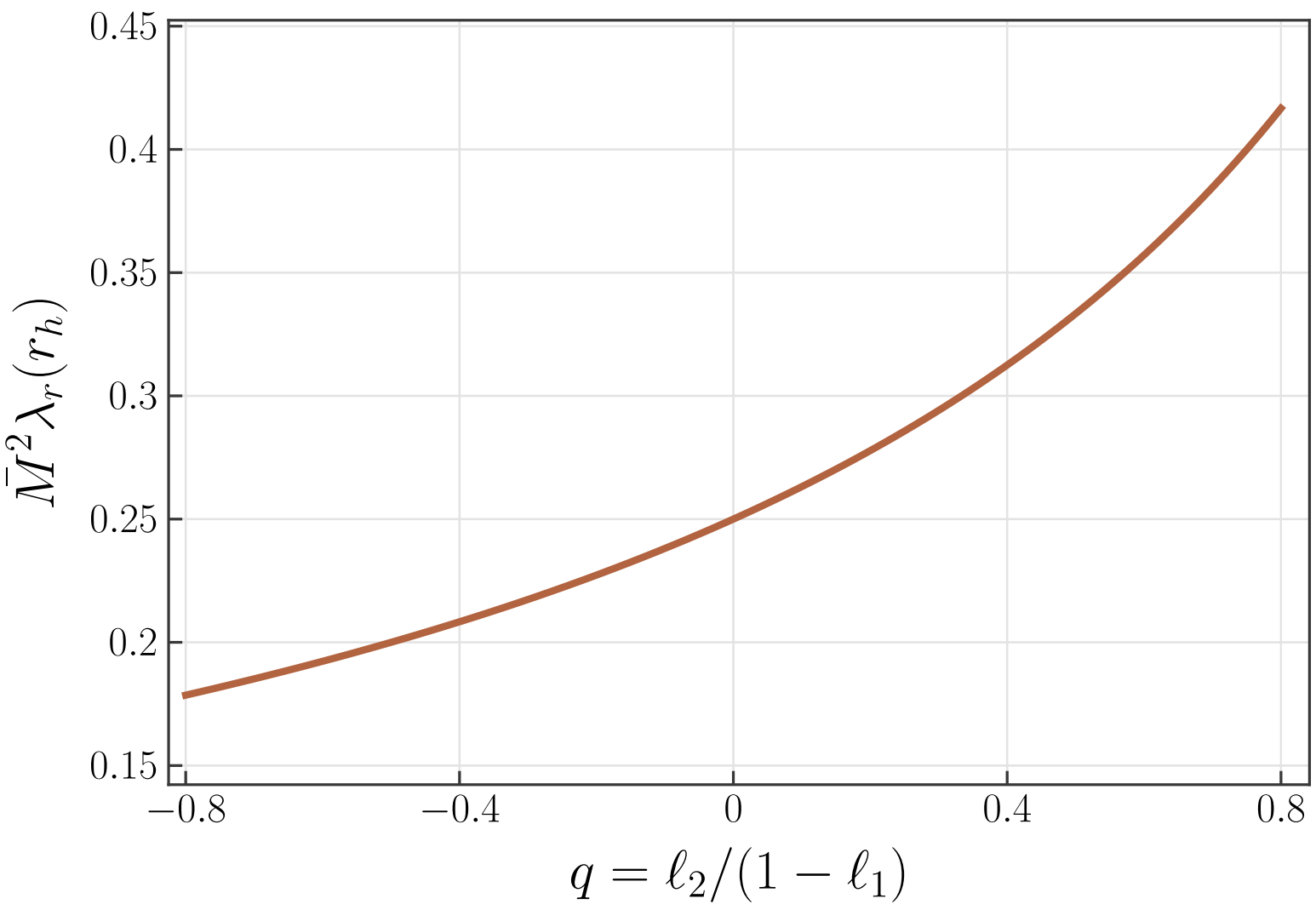

Figure 16. Dimensionless transverse eigenvalue $\bar{M}^2\lambda_\perp$ for the configurations used in Fig. 14. The curves retain a negative sign throughout the exterior and approach zero at large radius. Positive $q$ increases the magnitude of the compression.

With $y = r/B$, its regular solution is

$$\frac{\eta^{\hat{1}}(r)}{\eta_{r0}} = \frac{3-y}{2} + \frac{3}{2}\sqrt{\frac{1-y}{y}}\ \arccos\sqrt{y}, \qquad 0 < y \leq 1. \tag{67}$$

This expression describes a separation that increases during infall. At a fixed value of $r/B$, its profile is shared by all $q$; notice that the coupling dependence remains in the proper time required to reach that radius, given by Eq. (54).

The choice of mass held fixed also affects the horizon comparison. Using $\bar{M} = M/[(1-\ell_1)\sigma]$ in Eq. (64) gives

$$\lambda_r(r_h) = \frac{(1-\ell_1)^2}{4M^2}. \tag{68}$$

It is important to mention that, at fixed $(M, \ell_1)$, the change in horizon radius compensates the $\ell_2$ dependence of the radial curvature coefficient.

### C. Transverse tidal compression

The two angular directions have equal eigenvalues,

$$\frac{\mathrm{d}^2\eta^{\hat{i}}}{\mathrm{d}\tau^2} = \lambda_\perp(r)\eta^{\hat{i}}, \qquad \lambda_\perp(r) = -\frac{\bar{M}}{\sigma^2 r^3}, \qquad \hat{i} = \hat{2}, \hat{3}. \tag{69}$$

The negative sign gives compression across the direction of motion. Its magnitude has the same $r^{-3}$ dependence and the same multiplicative KR factor as the radial stretching. At the horizon,

$$\lambda_\perp(r_h) = -\frac{1}{8\sigma^2\bar{M}^2} = -\frac{(1-\ell_1)^2}{8M^2}. \tag{70}$$

Figure 16 displays the signed eigenvalue. For positive $q$, its larger absolute value corresponds to stronger transverse compression at each fixed $r/\bar{M}$.

For an initially comoving transverse separation, $\eta^{\hat{i}}(B) = \eta_{\perp 0}$ and $\dot{\eta}^{\hat{i}}(B) = 0$, the exact solution is

$$\eta^{\hat{i}}(r) = \eta_{\perp 0}\frac{r}{B}. \tag{71}$$

Together with Eq. (67), this gives the shape evolution of an infinitesimal body released with vanishing internal relative velocity. A fixed $r/B$ corresponds to the same radial and transverse separations for every $q$. At fixed elapsed proper time, $(\bar{M}, B)$, and initial separations, a positive $q$ advances the body further along the fall and increases its deformation, as long as the compared trajectories keep in their common radial interval.

The stretching and compression obey

$$\lambda_r = -2\lambda_\perp, \qquad \operatorname{tr}\mathcal{T} = \lambda_r + 2\lambda_\perp = 0. \tag{72}$$

In this manner, the KR couplings preserve the relative factor of two and the tracefree electric tidal tensor for radial fall. The conical part of the physical curvature appears in

$$R_{\hat{2}\hat{3}\hat{2}\hat{3}} = \frac{\sigma^2 - 1}{\sigma^2 r^2} + \frac{2\bar{M}}{\sigma^2 r^3}, \qquad R(g) = \frac{2(\sigma^2-1)}{\sigma^2 r^2}. \tag{73}$$

The $r^{-2}$ term records the angular curvature of the asymptotically conical geometry, while the radial tidal eigenvalues follow the mass dependent $r^{-3}$ profile.

For completeness, retaining the cosmological term in Eq. (56) gives

$$\begin{aligned} \lambda_r^{(\Lambda)} &= \frac{2\bar{M}}{\sigma^2 r^3} + \frac{\Lambda}{3(1-\ell_1-\ell_2/2)}, \\ \lambda_\perp^{(\Lambda)} &= -\frac{\bar{M}}{\sigma^2 r^3} + \frac{\Lambda}{3(1-\ell_1-\ell_2/2)}. \end{aligned} \tag{74}$$

The cosmological contribution is isotropic, as in the Kottler geometry [64], with $\operatorname{tr}\mathcal{T} = \Lambda/(1-\ell_1-\ell_2/2)$. Positive $\Lambda$ adds stretching in all three directions, and negative $\Lambda$ adds compression. For $\Lambda > 0$, the transverse eigenvalue changes sign at $r^3 = 3(1-\ell_1)\bar{M}/\Lambda$ whenever this radius lies in the physical domain under consideration.

## VII. PHOTON SPHERES, SHADOWS, AND OBSERVATIONAL CONSTRAINTS

The photon sphere determines the critical null trajectories that delimit the shadow. We examine this structure for the physical Palatini Kalb–Ramond metric, using the areal radius and the Killing time normalized at infinity [6, 30]

$$\begin{aligned} \mathrm{d}s^2 &= -F(r)\,\mathrm{d}t^2 + \frac{\sigma^2}{F(r)}\,\mathrm{d}r^2 + r^2\mathrm{d}\Omega^2, \\ F(r) &= 1 - \frac{2\bar{M}}{r}, \quad \bar{M} = \frac{M}{\rho}. \end{aligned}$$

The deformation parameters obey

$$\begin{aligned} &\sigma^2 = 1 - \frac{q}{2}, \quad q = \frac{\ell_2}{1-\ell_1}, \\ &\rho = (1-\ell_1)\sigma, \quad \rho^2 = (1-\ell_1)\left(1 - \ell_1 - \frac{\ell_2}{2}\right). \end{aligned}$$

We take $M > 0$, $1 - \ell_1 > 0$, and $1 - \ell_1 - \ell_2/2 > 0$, with $\sigma > 0$. The mass $M$ is the conserved Iyer–Wald charge, while $\bar{M}$ is the coefficient measured through the Newtonian potential. Electromagnetic rays follow null geodesics of $g_{\mu\nu}$, and the exterior begins at $r_h = 2\bar{M}$.

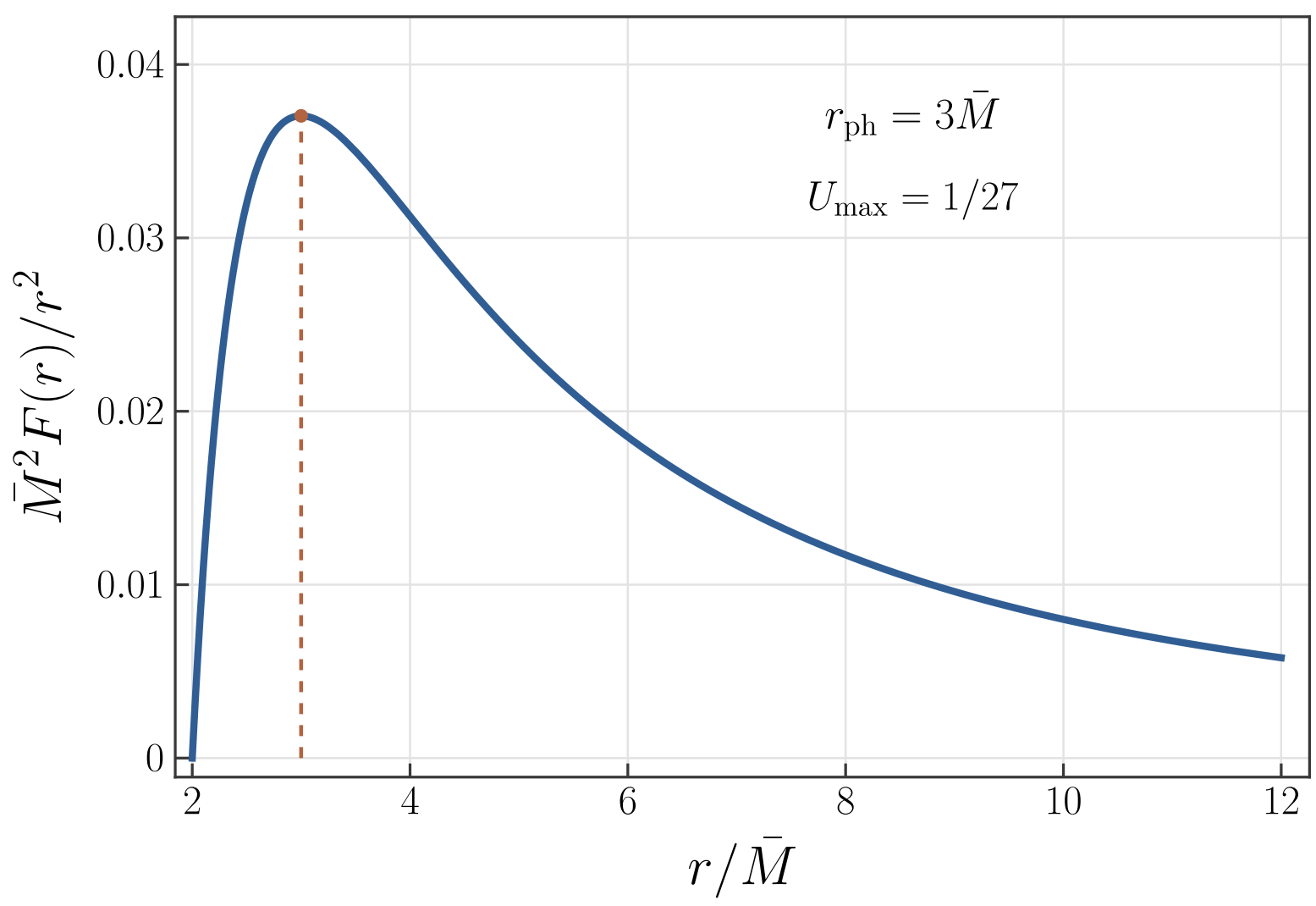

Figure 17. Normalized null effective potential $\mathcal{U} = (1 - 2/x)/x^2$. The maximum at $r/\bar{M} = 3$ determines the photon sphere. The horizon is at $r/\bar{M} = 2$, where the potential vanishes. The mass scale $\bar{M}$ is held fixed.

### A. Null geodesics and the photon sphere

Spherical symmetry allows the photon trajectory to be placed in the equatorial plane. For an affine parameter $\lambda$, the conserved photon energy and angular momentum are

$$E_\gamma = F(r)\frac{\mathrm{d}t}{\mathrm{d}\lambda}, \qquad L_\gamma = r^2\frac{\mathrm{d}\varphi}{\mathrm{d}\lambda}, \qquad b_\gamma = \frac{L_\gamma}{E_\gamma}. \tag{75}$$

The null condition gives

$$\sigma^2\left(\frac{\mathrm{d}r}{\mathrm{d}\lambda}\right)^2 + V_\gamma(r) = E_\gamma^2, \qquad V_\gamma(r) = L_\gamma^2\frac{F(r)}{r^2}. \tag{76}$$

A circular photon orbit satisfies $V'_\gamma(r_{\rm ph}) = 0$ and $V_\gamma(r_{\rm ph}) = E_\gamma^2$ [5]. The radial condition reduces to

$$r_{\rm ph}F'(r_{\rm ph}) - 2F(r_{\rm ph}) = 0, \qquad r_{\rm ph} = 3\bar{M} = \frac{3M}{\rho}. \tag{77}$$

The second derivative at the circular orbit is

$$V''_\gamma(r_{\rm ph}) = -\frac{2L_\gamma^2}{81\bar{M}^4}, \tag{78}$$

which identifies an unstable maximum. The critical impact parameter follows directly:

$$b_{\rm c}^2 = \frac{r_{\rm ph}^2}{F(r_{\rm ph})} = 27\bar{M}^2, \qquad b_{\rm c} = 3\sqrt{3}\,\bar{M} = \frac{3\sqrt{3}\,M}{\rho}. \tag{79}$$

For photons incident from large radius, $b_\gamma < b_{\rm c}$ gives capture, $b_\gamma > b_{\rm c}$ permits scattering, and $b_\gamma = b_{\rm c}$ approaches the circular orbit. Figure 17 displays the dimensionless potential $\mathcal{U}(x) = \bar{M}^2 V_\gamma/L_\gamma^2$, with $x = r/\bar{M}$. Its maximum occurs at $x = 3$ and $\mathcal{U} = 1/27$ for every admissible $q$ at fixed $\bar{M}$.

The radial deformation enters the growth rate of a perturbation about the circular orbit. In terms of the normalized Killing time, the orbital frequency and radial Lyapunov exponent are [54]

$$\Omega_{\rm ph} = \frac{\sqrt{F(r_{\rm ph})}}{r_{\rm ph}} = \frac{1}{3\sqrt{3}\,\bar{M}}, \qquad \gamma_t = \frac{\Omega_{\rm ph}}{\sigma} = \frac{1}{3\sqrt{3}\,\sigma\bar{M}}. \tag{80}$$

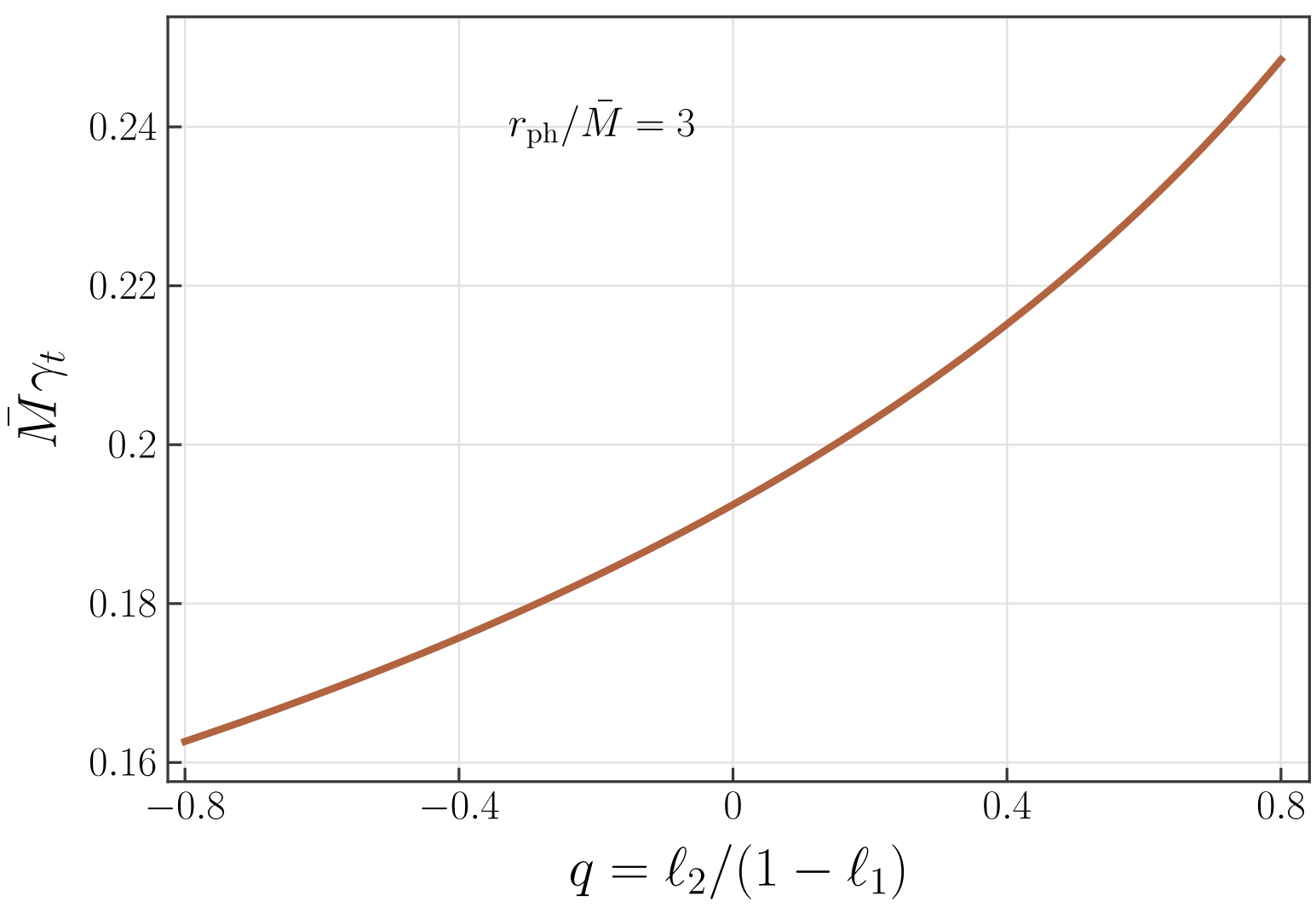

Figure 18. Dimensionless radial Lyapunov exponent of the photon orbit. The KR dependence is $\bar{M}\gamma_t = [3\sqrt{3}\sqrt{1-q/2}]^{-1}$, with $\bar{M}\gamma_t = 1/(3\sqrt{3})$ at $q = 0$.

Table II. Photon orbit quantities expressed using the two mass normalizations. The columns distinguish the fixed lapse mass $\bar{M}$ from the conserved mass $M = \rho\bar{M}$.

| Quantity | Units of $\bar{M}$ | Units of $M$ |
|---|---|---|
| Horizon radius | $r_h/\bar{M} = 2$ | $r_h/M = 2/\rho$ |
| Photon radius | $r_{\rm ph}/\bar{M} = 3$ | $r_{\rm ph}/M = 3/\rho$ |
| Critical impact parameter | $b_{\rm c}/\bar{M} = 3\sqrt{3}$ | $b_{\rm c}/M = 3\sqrt{3}/\rho$ |
| Orbital frequency | $\bar{M}\Omega_{\rm ph} = 1/(3\sqrt{3})$ | $M\Omega_{\rm ph} = \rho/(3\sqrt{3})$ |
| Radial instability | $\bar{M}\gamma_t = 1/(3\sqrt{3}\sigma)$ | $M\gamma_t = (1-\ell_1)/(3\sqrt{3})$ |

In this sense, $\gamma_t/\Omega_{\rm ph} = (1-q/2)^{-1/2}$. Positive $q$ increases the radial instability at fixed $\bar{M}$, while negative $q$ decreases it. The critical curve and the radial escape rate therefore probe distinct features of the geometry. Figure 18 shows this dependence.

### B. Shadow in the static observer's frame

The apparent angular size follows from the photon momentum projected onto the observer's orthonormal frame [6]. For a static observer at $r = r_o > r_{\rm ph}$, the temporal and radial frame vectors are

$$\begin{aligned} e_{\hat{t}} &= F(r_o)^{-1/2}\partial_t, \qquad & e_{\hat{r}} &= \frac{\sqrt{F(r_o)}}{\sigma}\partial_r, \\ e_{\hat{\theta}} &= \frac{1}{r_o}\partial_\theta, \qquad & e_{\hat{\varphi}} &= \frac{1}{r_o\sin\theta}\partial_\varphi. \end{aligned} \tag{81}$$

The measured photon energy is $E_\gamma/\sqrt{F(r_o)}$ and the transverse momentum is $L_\gamma/r_o$. Their ratio determines the angular radius of the shadow:

$$\sin\psi_{\rm sh} = \frac{b_{\rm c}\sqrt{F(r_o)}}{r_o}, \quad \psi_{\rm sh} = \arcsin\left[\frac{3\sqrt{3}\,\bar{M}}{r_o}\sqrt{1-\frac{2\bar{M}}{r_o}}\right]. \tag{82}$$

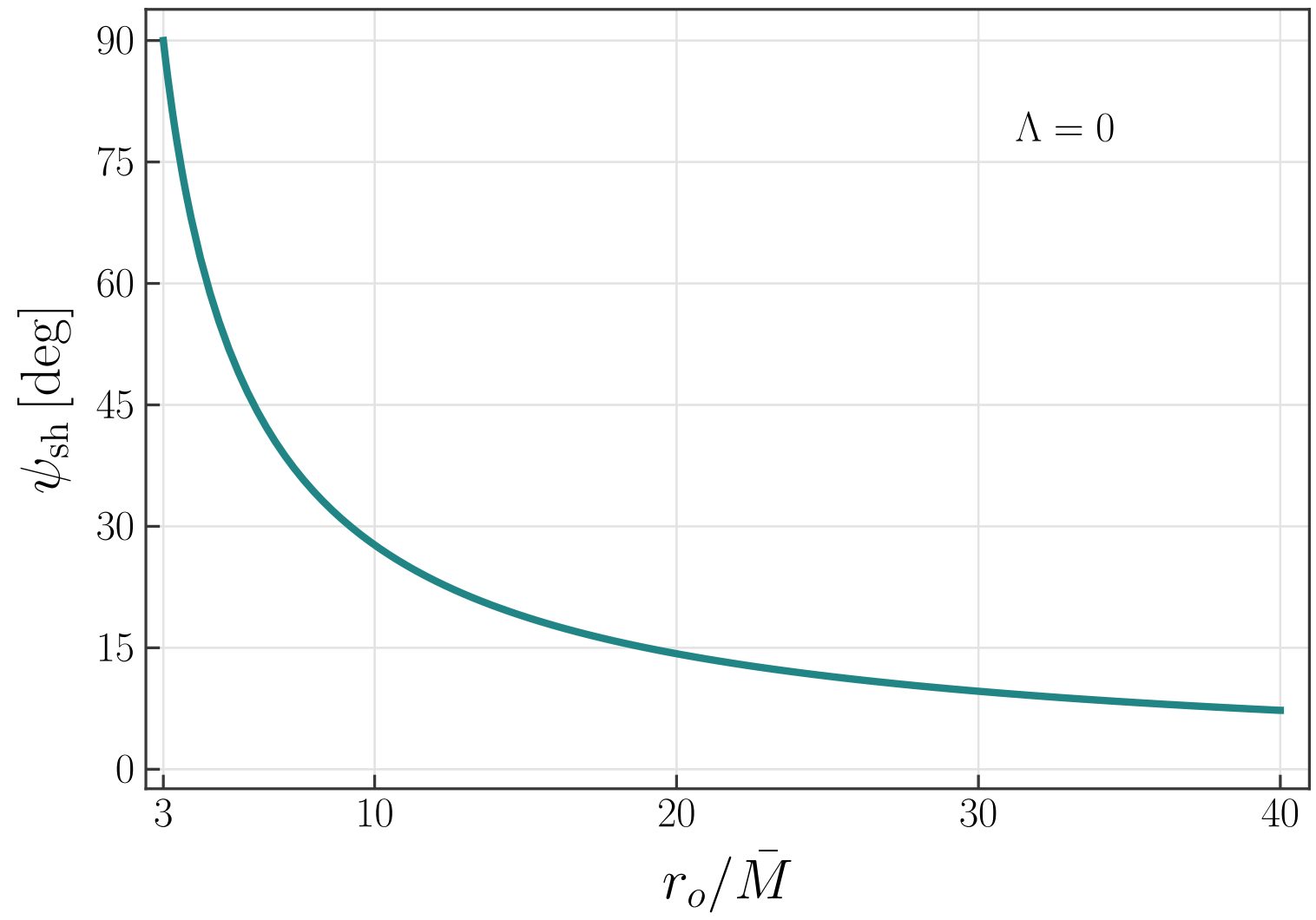

Figure 19. Angular radius of the shadow measured by a static observer outside the photon sphere. The angle is $\pi/2$ at $r_o = 3\bar{M}$ and approaches $3\sqrt{3}\bar{M}/r_o$ at large radius.

The shadow fills a hemisphere at $r_o = 3\bar{M}$ and subtends a decreasing angle as the observer moves outward. Figure 19 gives the finite distance profile. On an areal screen with azimuthal angle $\chi$, define $\alpha = r_o \sin\psi_{\rm sh}\cos\chi$ and $\beta = r_o \sin\psi_{\rm sh}\sin\chi$. The critical curve satisfies

$$\begin{aligned} &\alpha^2 + \beta^2 = \mathcal{R}_{\rm sh}^2, \qquad \mathcal{R}_{\rm sh} = b_{\rm c}\sqrt{F(r_o)}, \\ &\lim_{r_o\to\infty} \mathcal{R}_{\rm sh} = 3\sqrt{3}\,\bar{M}. \end{aligned} \tag{83}$$

Spherical symmetry fixes the circular shape, while the lapse mass determines its size. Figure 20 shows the common normalized contour at large observer radius. The factors of $\sigma$ in the radial photon momentum and the radial frame vector cancel in the angular measurement.

At large observer radius, the angular diameter and the angular mass scale are

$$\theta_{\rm sh} = 2\psi_{\rm sh} = 6\sqrt{3}\,\vartheta_K + O(\vartheta_K^2), \qquad \vartheta_K = \frac{\bar{M}}{D_A}. \tag{84}$$

Here, $D_A$ ($D_A = r_o$) specifies the areal distance convention of the observer screen. The asymptotic spatial geometry is $\mathrm{d}\ell^2 \simeq \sigma^2\mathrm{d}r^2 + r^2\mathrm{d}\Omega^2$, so a proper radial distance $D_{\rm prop}$ obeys $D_{\rm prop}/r_o \to \sigma$. With an independently specified $D_{\rm prop}$, the same angular measurement reads $\theta_{\rm sh} \simeq 6\sqrt{3}\sigma\bar{M}/D_{\rm prop}$.

For completeness, the cosmological solution has $F_\Lambda(r) = 1 - 2\bar{M}/r - \Lambda r^2/[3(1-\ell_1)]$. Its photon sphere remains at $3\bar{M}$, while

$$\begin{aligned} &\mathcal{H} = 1 - \frac{9\Lambda\bar{M}^2}{1-\ell_1}, \quad b_{\rm c}^{(\Lambda)} = \frac{3\sqrt{3}\,\bar{M}}{\sqrt{\mathcal{H}}}, \\ &\sin\psi_{\rm sh}^{(\Lambda)} = \frac{b_{\rm c}^{(\Lambda)}\sqrt{F_\Lambda(r_o)}}{r_o}. \end{aligned} \tag{85}$$

Notice that the static photon orbit needs $\mathcal{H} > 0$, and the observer lies in a region with $F_\Lambda(r_o) > 0$.

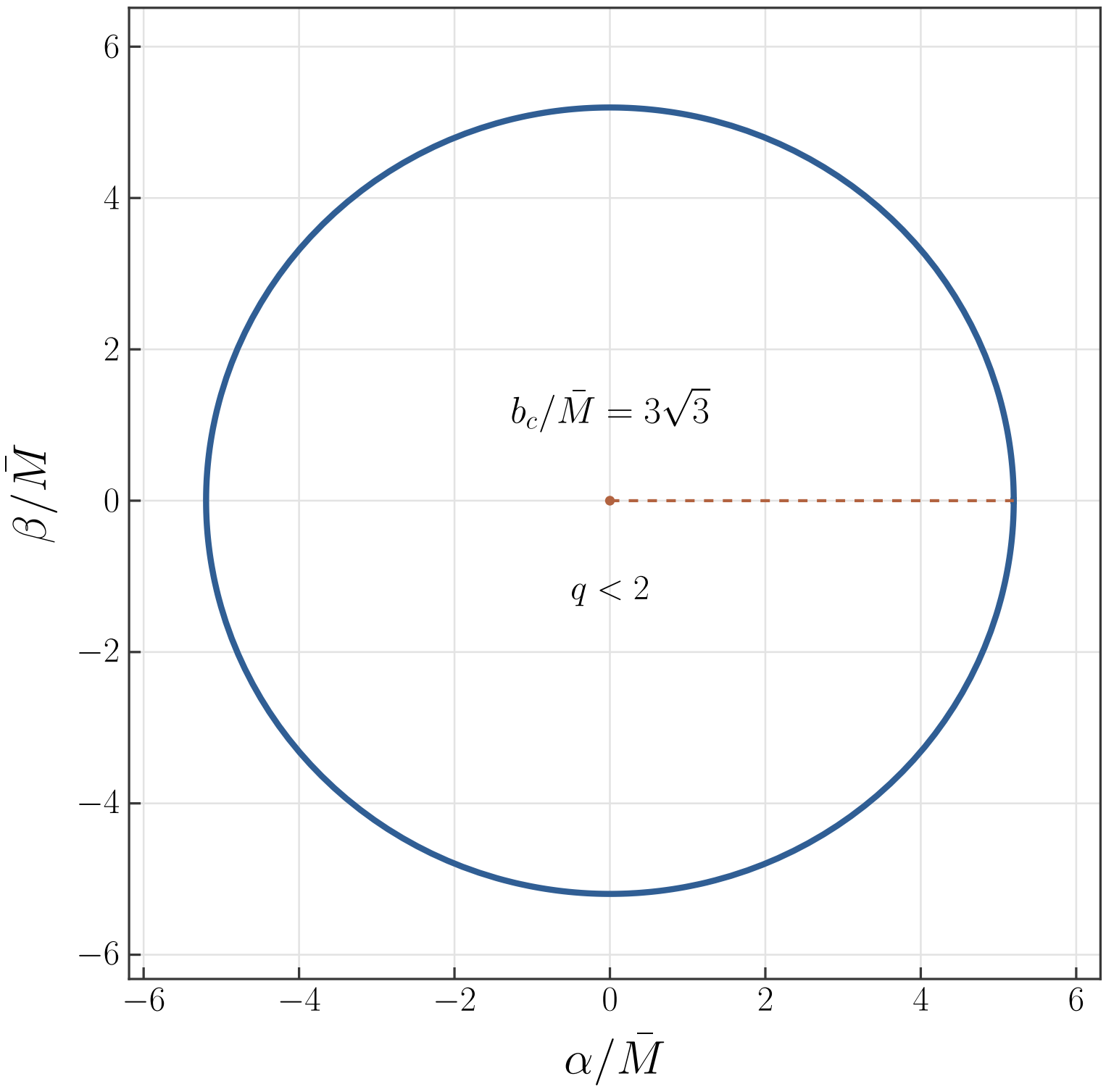

Figure 20. Shadow contour on the normalized areal screen for a distant observer. The radius is $3\sqrt{3}$ in units of $\bar{M}$. Every admissible KR configuration shares this contour at fixed $\bar{M}$, while its null trajectories retain the radial dependence on $\sigma$.

### C. Angular size constraints from Sgr A* and M87*

For $\Lambda = 0$, the frequency of a circular massive orbit satisfies $\Omega^2_{\rm orb} = \bar{M}/r^3$. Stellar dynamics sets the lapse mass entering Eq. (84) [30]. We use the published dynamical angular scales as inputs to this geometric comparison.

For Sgr A*, the EHT shadow statistic is [2]

$$\delta_{\rm sh} = \frac{\theta_{\rm sh}}{6\sqrt{3}\,\vartheta_{\rm dyn}} - 1. \tag{86}$$

The VLTI and Keck mass distance configuration give the alternative 68% credible intervals in Table III. The fiducial `eht-imaging`+GRMHD analysis gives $\theta_{\rm sh} = 48.9^{+5.2}_{-5.1}\,\mu$as [2].

For M87*, the published statistic compares the EHT angular mass scale with the stellar-dynamical prior [1]:

$$\delta_{\theta_g} = \frac{\vartheta_{\rm EHT}}{\vartheta_{\rm dyn}} - 1. \tag{87}$$

The reported value $\delta_{\theta_g} = -0.01 \pm 0.17$ uses the image setting and its uncertainty. In the static KR calculation, the geometric angular scale is $\bar{M}/D_A$. The M87* interval supplies a comparison at the accuracy of the Kerr/GRMHD.

With $\vartheta_{\rm dyn} = \bar{M}/D_A$, the model predicts

$$\delta^{\rm KR}_{\rm sh} = 0, \;\; \delta^{\rm KR}_{\theta_g} = 0, \;\; \left(\frac{\partial \delta^{\rm KR}_{\rm sh}}{\partial q}\right)_{\vartheta_K} = \left(\frac{\partial \delta^{\rm KR}_{\theta_g}}{\partial q}\right)_{\vartheta_K} = 0, \tag{88}$$

Table III. Angular scales and 68% intervals used in the geometric shadow comparison [1, 2]. The two uncertainties on the Sgr A* angular scales are statistical and systematic. VLTI and Keck provide alternative priors for the same EHT data. The M87* row uses the stellar prior and the angular mass statistic.

| Source and prior | $\vartheta_{\rm dyn}$ [$\mu$as] | Published deviation | 68% interval |
|---|---|---|---|
| Sgr A*, VLTI | $5.125 \pm 0.009 \pm 0.020$ | $\delta_{\rm sh} = -0.08 \pm 0.09$ | $[-0.17, 0.01]$ |
| Sgr A*, Keck | $4.92 \pm 0.03 \pm 0.01$ | $\delta_{\rm sh} = -0.04^{+0.09}_{-0.10}$ | $[-0.14, 0.05]$ |
| M87*, stellar | $3.62^{+0.60}_{-0.34}$ | $\delta_{\theta_g} = -0.01 \pm 0.17$ | $[-0.18, 0.16]$ |

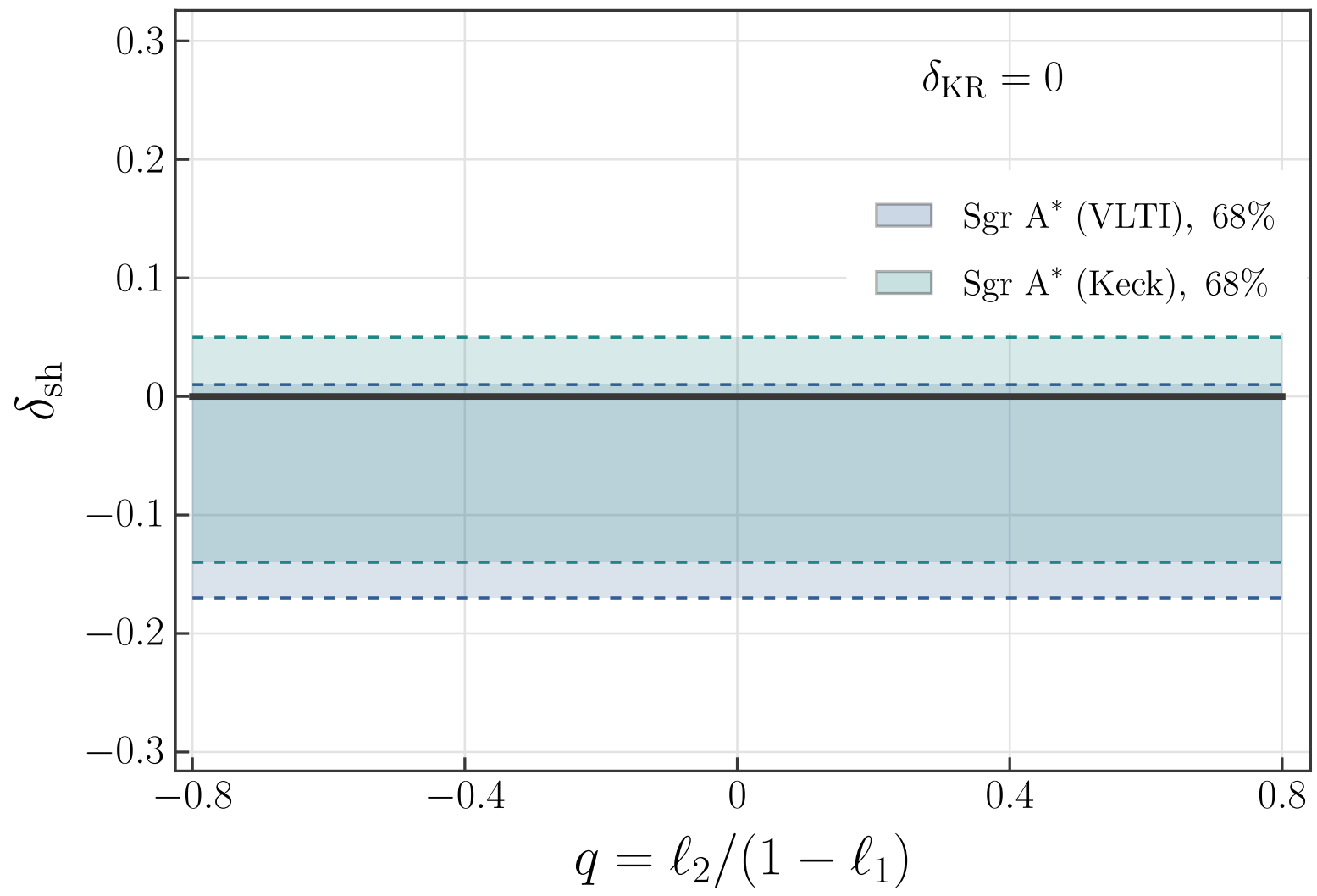

Figure 21. Sgr A* shadow-deviation intervals from the VLTI and Keck priors, shown with the KR prediction $\delta_{\rm sh} = 0$ at fixed $\bar{M}/D_A$. The shaded ranges are the separate published 68% intervals. The horizontal parameter range illustrates the exact geometric degeneracy in $q$.

to the distant observer accuracy of Eq. (84). Each interval in Table III contains this prediction. The diameter information fixes the angular lapse mass scale, and the couplings remain free within the admissible metric domain at this level of inference. Figures 21 and 22 display the constant prediction against the published intervals; Fig. 23 compares the three configurations directly.

The emission ring measurements show additional context for this comparison. Sgr A* has a ring diameter $51.8 \pm 2.3\,\mu$as [3], and the multi epoch M87* analysis gives a mean ring diameter $43.9 \pm 0.6\,\mu$as [4]. Their relation to the geometric critical curve depends on the emission and transfer model. In the KR metric, the photon orbit equation keeps the factor $\sigma$ through

$$\frac{{\rm d}\varphi}{{\rm d}r} = \pm\frac{\sigma b_\gamma}{r^2\sqrt{1 - b_\gamma^2 F(r)/r^2}}. \tag{89}$$

This factor changes the accumulated azimuth and the time spent near the photon sphere.

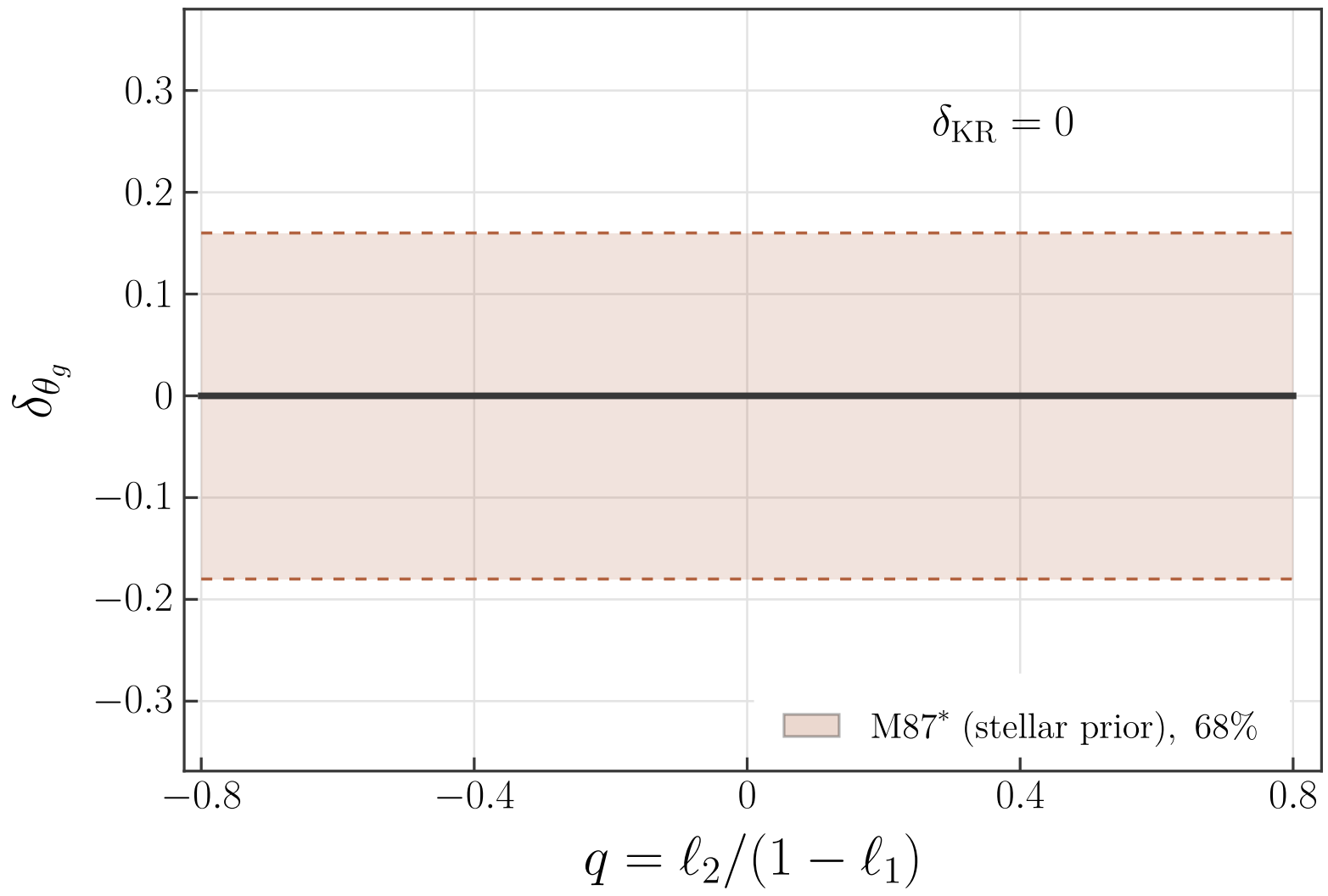

Figure 22. M87* angular mass consistency interval and the static KR prediction. The 68% range is $-0.18 \le \delta_{\theta_g} \le 0.16$.

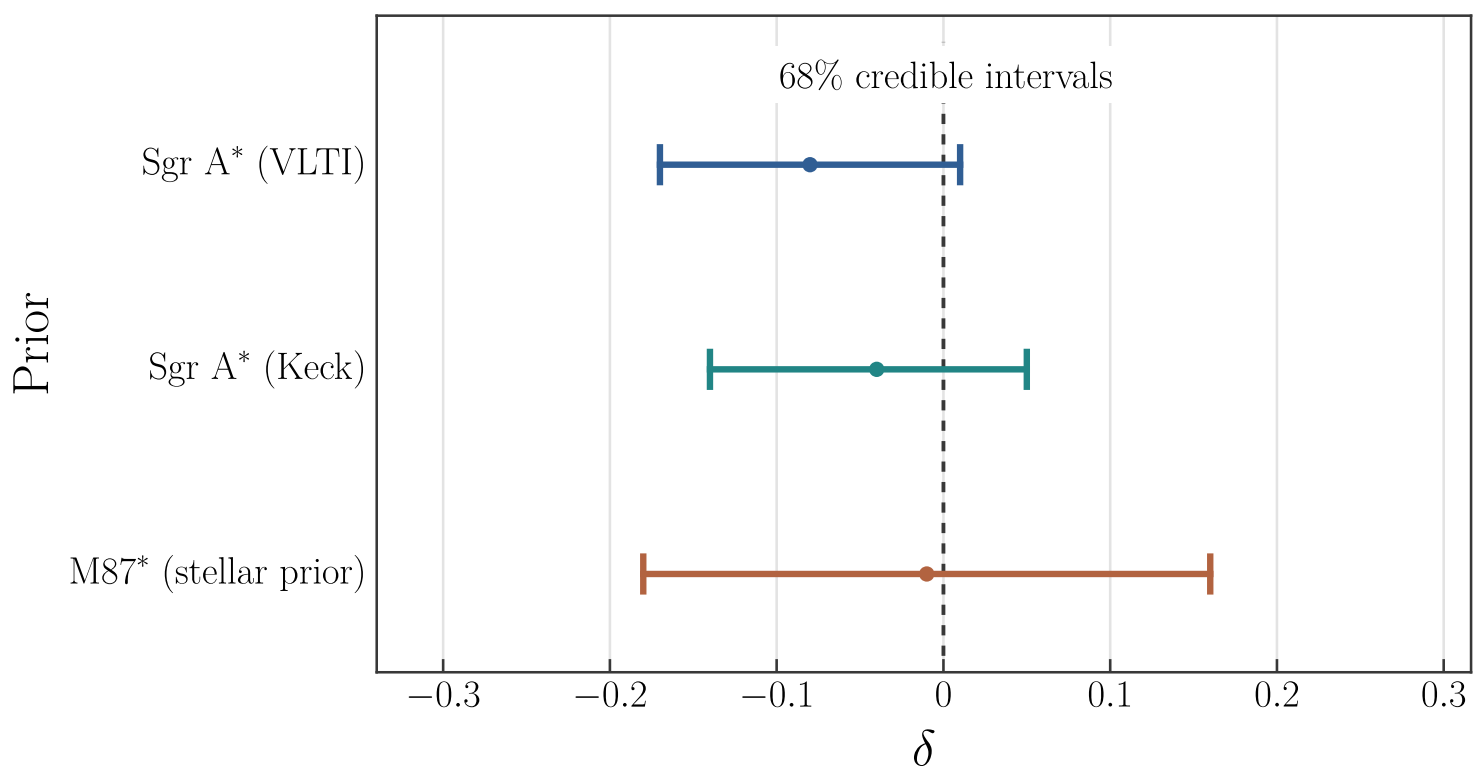

Figure 23. Sgr A* and M87* statistics with their 68% intervals. The vertical line marks the static KR prediction. Sgr A* uses the shadow diameter statistic, while M87* uses the angular mass statistic defined in Eq. (87).

### D. Conditional sensitivities in the coupling plane

An independent measurement of the conserved mass $M$, together with an areal distance, would give us a second angular scale $\vartheta_M = M/D_A$. The shadow could determine $\rho$ through

$$\Delta_M = \frac{\theta_{\rm sh}}{6\sqrt{3}\,\vartheta_M} - 1 = \frac{1}{\rho} - 1. \tag{90}$$

If such a measurement supplied a fractional interval $\Delta_- \le \Delta_M \le \Delta_+$, with $\Delta_- > -1$, its transformation would be

$$\rho_- = \frac{1}{1+\Delta_+} \le \rho \le \frac{1}{1+\Delta_-} = \rho_+. \tag{91}$$

The corresponding strip in the $(\ell_1, \ell_2)$ plane is

$$2(1-\ell_1) - \frac{2\rho_+^2}{1-\ell_1} \le \ell_2 \le 2(1-\ell_1) - \frac{2\rho_-^2}{1-\ell_1}, \qquad 1-\ell_1 > 0. \tag{92}$$

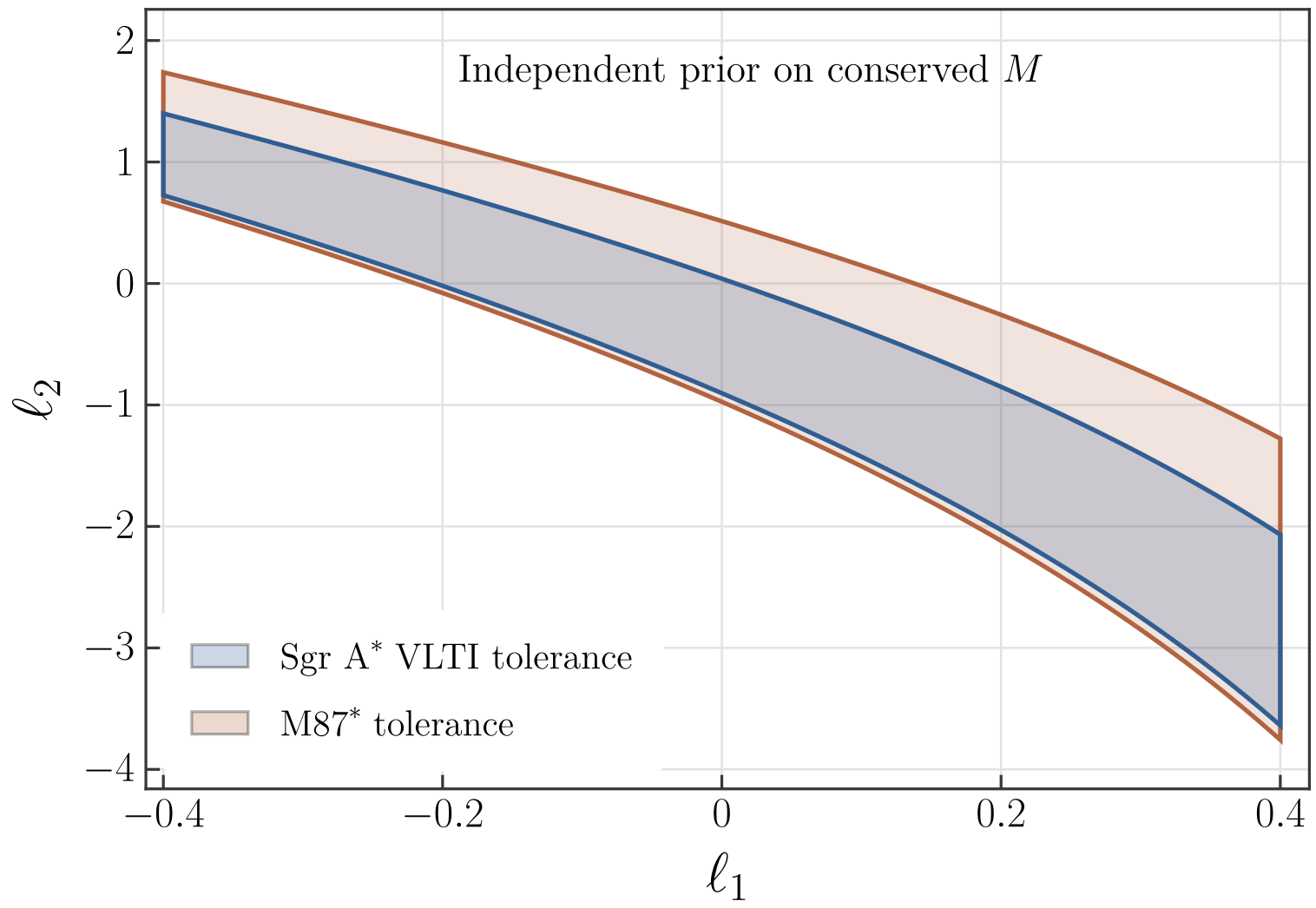

Figure 24. Sensitivity strips in the KR coupling plane, obtained from Eq. (92). The plot assumes an independently conserved mass and uses the Sgr A* VLTI and M87* fractional tolerances separately.

Table IV adopts the fractional ranges in Table III as illustrative tolerances for this prospective independent mass comparison. It displays the sensitivity that those ranges would provide under the additional normalization in Eq. (90). The dynamical priors are set $\bar{M}/D_A$; a mass measurement is the additional input required to turn these conditional sensitivities into bounds on $\rho$.

Table IV. Conditional sensitivities for an independently conserved mass angular scale. The fractional tolerances are taken from Table III; the entries describe the additional assumption in Eq. (90). The $q$ column also fixes $\ell_1 = 0$. Each row is transformed separately.

| Tolerance | $[\Delta_-, \Delta_+]$ | $[\rho_-, \rho_+]$ | $q$ interval at $\ell_1 = 0$ |
|---|---|---|---|
| Sgr A*, VLTI | $[-0.17, 0.01]$ | $[0.9901, 1.2048]$ | $[-0.9032, 0.0394]$ |
| Sgr A*, Keck | $[-0.14, 0.05]$ | $[0.9524, 1.1628]$ | $[-0.7042, 0.1859]$ |
| M87*, stellar | $[-0.18, 0.16]$ | $[0.8621, 1.2195]$ | $[-0.9744, 0.5137]$ |

At $\ell_1 = 0$, the conversion used in the last column is $q = 2(1-\rho^2)$. Figure 24 displays the separate VLTI and M87* tolerance strips. Their overlap describes the common algebraic region under the stated mass assumption. The quoted source intervals retain their individual statistical definitions.

An independent proper radial distance would define a different comparison, with $\vartheta_{\text{prop}} = \bar{M}/D_{\text{prop}}$ and $\Delta_{\text{prop}} = \sigma - 1$. Its conversion is

$$2\left[1-(1+\Delta_+)^2\right] \leq q \leq 2\left[1-(1+\Delta_-)^2\right]. \tag{93}$$

## VIII. LENSING PHENOMENA

### A. Expansion for light deflection in the weak-field limit

To analyze the deflection of light in the model under consideration, we employ the line element and the metric function defined in Eqs. (1) and (10), respectively. From these expressions, the impact parameter can be obtained using the relation established in Eq. (8), which yields:

$$\frac{1}{b^2(r_0)} = \frac{1}{r_0^2}\left[1 - \frac{2M}{\rho\, r} - \frac{\sigma\,\Lambda\, r^2}{3\rho}\right]. \tag{94}$$

By rewriting the radial equation in terms of the conserved quantities defined in Eqs. (5) and (6) and the metric tensor components given in Eq. (10), we can exploit the symmetry of the problem to determine the total angular deflection. In this context, the contributions from the photon trajectories before and after the point of closest approach are identical. Thus, the total angular deflection can be obtained by multiplying the contribution from either of these regions by a factor of 2. Thus, we have that:

$$\Delta\varphi = 2\sigma \int_{r_0}^{\infty} \frac{\mathrm{d}r}{r^2}\left[\frac{1}{b^2} - \frac{F(r)}{r^2}\right]^{-1/2}. \tag{95}$$

Due to the symmetry of the problem, it suffices to consider only the solution corresponding to the positive sign of the expression above. Thus, we will use the following coordinate transformation in Eq. (95) $u = \frac{1}{r}$. Note that this coordinate transformation causes the limits of integration to become $u \to 0$ as $r \to \infty$, and $u \to u_0 = \frac{1}{r_0}$ as $r \to r_0$. Consequently, the equation for the angular deviation in the new coordinates takes the form

$$\Delta\varphi = 2\sigma \int_{0}^{u_0} \mathrm{d}u \left[\frac{1}{b^2(r_0)} - u^2 + \left(\frac{2M}{\rho}\right)u^3\right]^{-1/2}. \tag{96}$$

It is worth highlighting that the above expression was obtained by considering the case of a vanishing cosmological constant ($\Lambda = 0$). To obtain the expansion of the light deflection angle in the weak-field regime, we return to the expression for the impact parameter given in Eq. (94) and apply the previously introduced coordinate transformation. Thus, the relation can be rewritten as $\frac{1}{b^2} = u_0^2\left[1 - \left(\frac{2M}{\rho}\right)u_0\right]$. Using this new expression for the impact parameter and considering the regime in which the photon is sufficiently far from the BH, we can adopt the weak-field approximation, where $M/b \ll 1$. In this context, we expand $u_0$ in inverse powers of the impact parameter $b$, retaining terms up to the second order in $1/b$. Thus, we obtain $u_0 \approx 1/b + M/\rho\, b^2$.

In this sense, by expanding the orbital equation Eq. (96) to second order in $M$, we find that the light deflection is given by $\delta\varphi = \Delta\varphi - \sigma\,\pi$ [1] and, for the purpose of graphical analysis, we discard terms higher than second order in mass. Thereby, the expression for angular deviation is given by

$$\delta\varphi = \pi\Big(\sigma - 1\Big) + \frac{4\,M\,\sigma}{\rho\, b} + \frac{15\,\pi\,\sigma\,M^2}{4\,\rho^2\, b^2}. \tag{97}$$

The expression above describes the gravitational deflection experienced by a light ray in the geometry of a BH within the framework of Kalb-Ramond gravity, described in the Palatini formalism. First, it is important to emphasize that, due to the non-asymptotically flat nature of the spacetime, the first term of Eq. (97) exhibits a modification associated with the presence of the Lorentz symmetry violation parameter $\sigma$. Similarly, the next two terms can be interpreted as corrections up to second order in the mass parameter $M$ for the Schwarzschild BH, arising from the modifications introduced by the parameters associated with the nonzero vacuum expectation value of the Kalb-Ramond field, which is responsible for the spontaneous breaking of Lorentz symmetry. The result corresponding to the gravitational deflection of light in the Schwarzschild BH spacetime is fully recovered by taking the limits $\sigma \to 1$ and $\rho \to 1$ [65–70].

[1] Here, $\Delta\varphi$ represents the total variation of the angular coordinate along the photon trajectory, from its initial position at infinity to its return to infinity. In flat spacetime, a straight-line trajectory corresponds to an angular variation of $\Delta\varphi = \pi$. However, since the geometry under consideration is not asymptotically flat, this angular reference is modified by the presence of the parameter $\sigma$, which introduces a deformation in the geometric structure of spacetime. Thus, the light deflection angle is defined as $\delta\varphi = \Delta\varphi - \sigma\pi$.

### B. The Gauss–Bonnet approach

In this section, we will analyze light deflection in the weak-field regime using the Gauss–Bonnet method [39], with the aim of comparing the results obtained with those derived from the approach adopted in the previous subsection, and investigating the contributions associated with the different orders of expansion of the parameters involved. In this context, the application of the Gauss–Bonnet method allows us to determine the light deflection angle by establishing a relationship between the global geometric properties of the optical geometry and the trajectories followed by light rays.

Initially, we will analyze the properties of the photon spheres determined by Eq. (9), with particular emphasis on the stability conditions of the corresponding orbits. For this investigation, we will employ the Gaussian curvature associated with the optical geometry of the spacetime under consideration, whose sign provides a geometric criterion for classifying these orbits. In this context, positive values of the Gaussian curvature indicate the presence of stable photon orbits, whereas negative values are associated with unstable photon orbits.

#### 1. *Stability of the critical orbits*

The trajectories corresponding to critical photon orbits are closely related to the geometric properties of the optical manifold associated with the spacetime under consideration. In this context, the stability of these orbits depends not only on the underlying geometric symmetries, but also on the behavior of neighboring null geodesics under small perturbations around the circular trajectory. In this case, a circular photon orbit can be characterized as a critical null trajectory whose stability is determined by the local geometric properties of spacetime. For unstable photon orbits, arbitrarily small perturbations are sufficient to cause photons to deviate from their circular trajectories, allowing them to propagate toward other regions of spacetime. On the other hand, in the case of stable photon orbits, small perturbations do not produce a significant deviation from the critical radius, so that photon trajectories may remain confined to its vicinity.

This behavior can be understood in terms of the geometric properties of the optical manifold, particularly through the analysis of the Gaussian curvature $\mathcal{K}(r)$. In this context, the Cartan–Hadamard theorem establishes that, under its conditions of applicability, geometries with nonpositive Gaussian curvature, $\mathcal{K}(r) \leq 0$, have no conjugate points, providing a geometric basis for analyzing the instability of circular photon orbits. On the other hand, regions characterized by positive Gaussian curvature, $\mathcal{K}(r) > 0$, may admit conjugate points and are therefore compatible with the existence of stable circular photon orbits. To investigate these properties, we consider a generic metric and impose the null condition $\mathrm{d}s^2 = 0$, allowing photon trajectories to be described in terms of an equivalent optical geometry, as discussed in Refs. [71, 72].

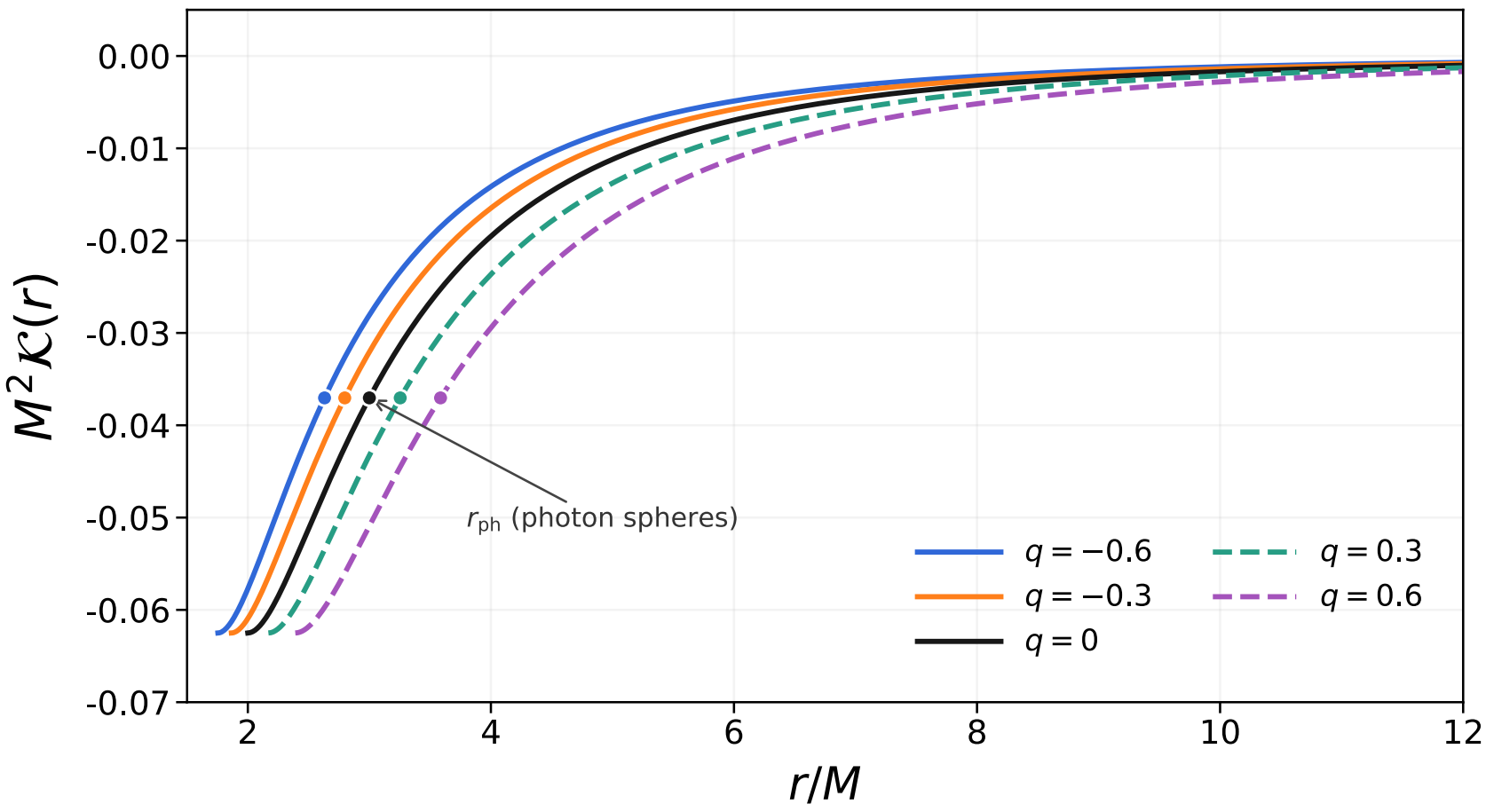

Figure 25. Gaussian curvature of the optical geometry as a function of $r/M$.

Then, we present below the line element defining the optical metric associated with the geometry of interest, which describes the propagation of a photon in the equatorial plane:

$$\mathrm{d}t^2 = \tilde{\gamma}_{ij}\mathrm{d}x^i\mathrm{d}x^j = \frac{\sigma^2\,\mathrm{d}r^2}{F^2(r)} + \frac{r^2\mathrm{d}\varphi^2}{F(r)}. \tag{98}$$

Here, the indices $i$ and $j$ denote exclusively the spatial directions, while $\tilde{\gamma}_{ij}$ represents the components of the induced metric in the optical geometry. Once the optical metric is defined, the geometric content of the optical manifold can be characterized in terms of its Gaussian curvature. Below, we explicitly present the expression for this quantity in terms of the metric functions under consideration, whose complete derivation can be found in Ref. [39, 44, 73]:

$$\mathcal{K}(r) = \frac{R}{2} = -\frac{F^{3/2}(r)}{\sigma^2\,r}\frac{\mathrm{d}}{\mathrm{d}r}\left[F(r)\frac{\mathrm{d}}{\mathrm{d}r}\left(\frac{r}{\sqrt{F(r)}}\right)\right], \tag{99}$$

where $R$ denotes the Ricci scalar associated with the two-dimensional optical metric.

A substantial body of previous research has demonstrated that the stability of circular photon orbits under small perturbations can be geometrically analyzed through the Gaussian curvature $\mathcal{K}(r)$ associated with the optical manifold of spacetime. In regions where $\mathcal{K}(r) > 0$, initially neighboring null geodesics tend to converge, favoring the confinement of photon trajectories and, in this manner, the stability of circular photon orbits. Conversely, in regions characterized by $\mathcal{K}(r) < 0$, neighboring null geodesics tend to diverge, indicating unstable behavior under small perturbations. In the latter case, photons may leave their circular orbits, following trajectories that lead them either toward the compact object or allow them to escape to infinity [18, 74–76].

As discussed previously, the sign of the Gaussian curvature can be used as a geometric criterion to investigate the possible stability and instability regimes of circular photon orbits. This is precisely the criterion adopted in Fig. 25, where we analyze the optical geometry of the BH considering different values of the parameter $q$, associated with the violation of Lorentz symmetry. It is observed that, for positive values of $q$ (dotted curves), the orbits are located at larger radii compared to the case of the Schwarzschild BH ($q = 0$), which is represented by the black curve. In contrast, negative values of $q$ (solid colored curves) shift these trajectories toward smaller radii. Furthermore, the Gaussian curvature remains negative at the positions of the photon sphere for all values of $q$ considered, indicating that these orbits retain their unstable nature, despite the geometric modifications induced by the violation of Lorentz symmetry. Although the parameter $q$ induces deformations in the optical geometry relative to the Schwarzschild case, these modifications are not sufficient to change the unstable nature of the photon spheres within the range of parameters considered.

### 2. *Weak deflection angle*

In the weak-field regime, the light deflection angle can be determined through a geometric approach based on the Gauss-Bonnet theorem [39]. Within this framework, the Gaussian curvature, defined in Eq. (99), plays a fundamental role in determining the deflection of photon trajectories. To apply this formalism, we restrict the optical geometry to the equatorial plane, $\theta = \pi/2$, such that the problem can be described by a two-dimensional optical manifold. Accordingly, the surface element associated with this geometry can be expressed as:

$$\mathrm{d}S = \sqrt{\tilde{\gamma}}\,\mathrm{d}r\mathrm{d}\varphi = \frac{\sigma\, r}{F^{3/2}(r)}\,\mathrm{d}r\mathrm{d}\varphi. \tag{100}$$

The integral determining the light deflection angle is evaluated under approximations consistent with the geometric properties of the spacetime under investigation. To this end, we consider the weak-field regime, in which the impact parameter is sufficiently larger than the characteristic gravitational scale of the BH $b \gg 2M$. This condition allows the application of the Gauss–Bonnet formalism to the study of gravitational lensing, as discussed in Ref. [19, 44–47, 77].

Furthermore, assuming that the mass parameter $M$ is sufficiently small in the weak-field regime, we adopt a perturbative approach to the exact orbital expression, retaining contributions up to second order in $M$. This approximation is commonly employed in studies of gravitational light deflection and has been widely used in related investigations [20, 78–80]. It is also worth emphasizing that the geometry under consideration is not asymptotically flat, since the parameter associated with Lorentz symmetry violation introduces a deformation in the angular structure of spacetime. This behavior can be demonstrated by taking the limit $r \to \infty$ of the metric presented in Eq. (10), in the absence of the cosmological constant ($\Lambda = 0$), and then performing the coordinate transformation $\bar{r} = \sigma\, r$, making it $d^2s = -d^2t + d^2\bar{r} + \frac{\bar{r}^2}{\sigma^2}d^2\varphi$.

Therefore, in the expression for the deflection angle presented below, we consider perturbations to the exact photon orbit, retaining terms up to order $1/b^2$, following the approaches detailed in Refs. [44]. We obtain:

$$\delta\varphi = \pi(\sigma-1) - \sigma\int_0^{\pi}\int_{r(\varphi/\sigma)}^{\infty}\mathcal{K}\mathrm{d}S = \pi(\sigma-1) + \frac{4\,M\,\sigma}{\rho\,b} + \frac{15\,\pi\,M^2\sigma}{4\,\rho^2\,b^2}, \tag{101}$$

where $r(\varphi/\sigma) = u^{-1}(\varphi/\sigma)$ is defined below

$$u(\varphi/\sigma) = \left[\frac{\sin(\varphi/\sigma)}{b} + \frac{M(1-\cos(\varphi/\sigma))^2}{\rho\, b^2} + \frac{M^2}{\rho^2\, b^3}\left(-\frac{15\varphi}{4\sigma}\cos\left(\frac{\varphi}{\sigma}\right) + \frac{5}{16}\sin\left(\frac{\varphi}{\sigma}\right)\right)\right.$$
$$\left. + \frac{M^2}{\rho^2\, b^3}\left(2\sin\left(\frac{2\,\varphi}{\sigma}\right) - \frac{2}{16}\sin\left(\frac{3\,\varphi}{\sigma}\right)\right)\right]^{-1}. \tag{102}$$

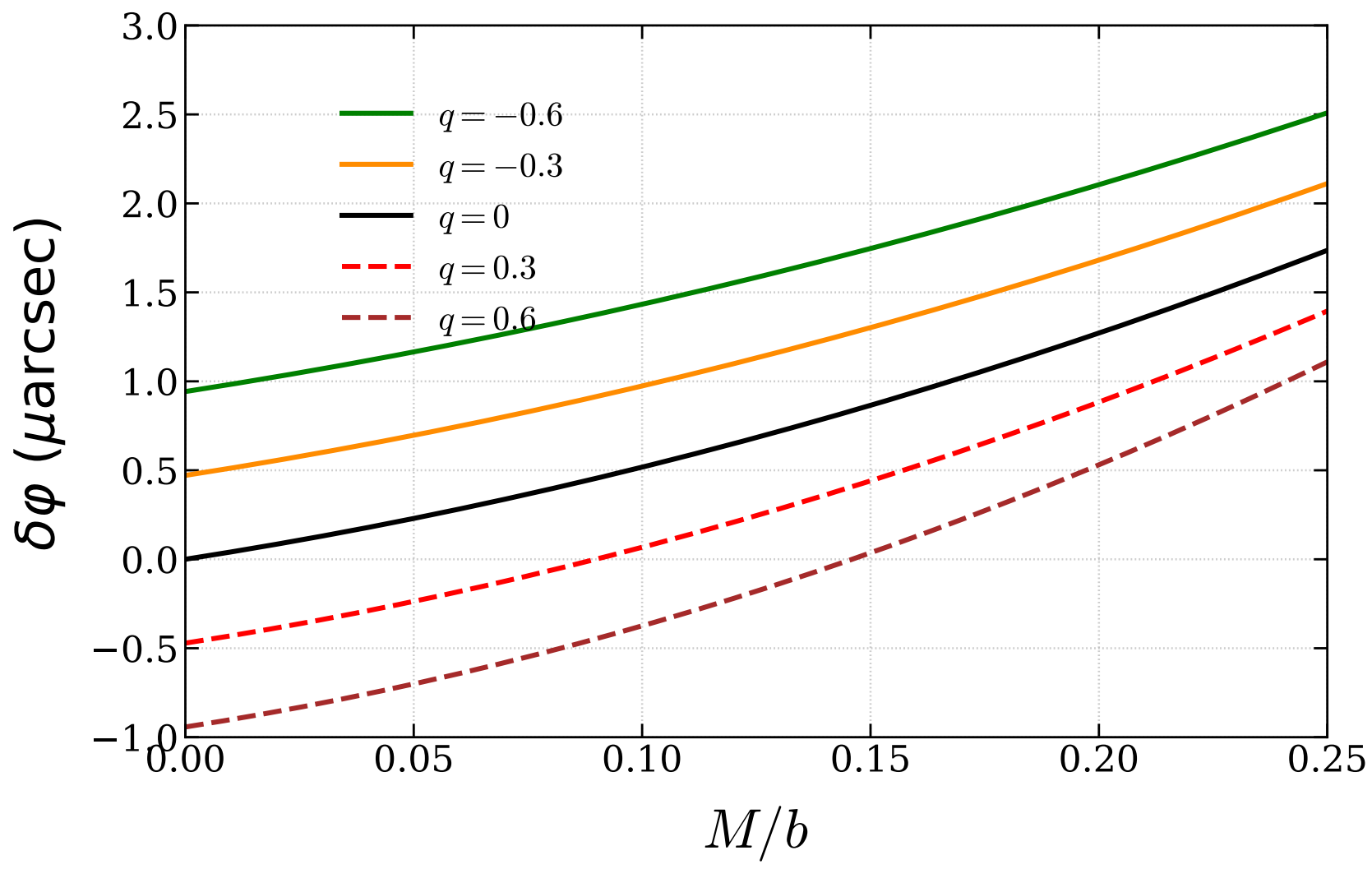

Figure 26. Deflection angle in the weak-field limit as a function of $M/b$ for different values of the Lorentz-symmetry violation parameter

As can be seen in the expression above, Eq. (101), we calculate the deflection of light in the weak-field regime using the Gauss-Bonnet method, which is applicable to both asymptotically flat and non-asymptotically flat geometries. In this context, our primary objective is to establish a proper comparison between the results obtained via this formalism and those derived from the method traditionally employed in the literature, as presented in Section VIII. It is worth noting that, by carefully considering the same orders of expansion for the parameter assumed to be small ($M \ll 1$), we arrive at the same expressions regardless of the method used, as demonstrated in equations (97) and (101).

In Fig. (26), we present a graphical representation of the light deflection angle in the context of Kalb–Ramond gravity, formulated within the Palatini formalism, as a function of the dimensionless ratio $M/b$. The black curve represents the light deflection angle for the Schwarzschild black hole, corresponding to the case where effects associated with Lorentz symmetry violation are absent. On the other hand, the solid curves illustrate the behavior of the deflection angle for two negative values of $q$ corresponding to $l_1 = 0$, revealing a progressive positive increase in the magnitude of the angular deflection.

On the other hand, the dotted curves represent positive values of the Lorentz symmetry violation parameter, with $\ell_1 = 0$. In this case, the deflection angle tends to become negative, suggesting the emergence of more intense gravitational effects. This phenomenon is not exclusive to the geometry under consideration, as it has also been reported in other gravitational scenarios in the literature.

### C. Deflection of light in the strong-field limit

In this subsection, we investigate the gravitational deflection of light in the strong-field regime, employing the formalism originally proposed by Bozza [36] and subsequently refined by Tsukamoto [37]. To carry out this analysis, we introduce the variable transformation $z = 1 - \frac{r_0}{r}$ into the orbital equation given in Eq. (95), allowing it to be rewritten as follows:

$$\Delta\varphi(r_0) = \pm \int_0^1 \frac{2\,\sigma\, r_0 \mathrm{d}z}{\sqrt{G(z,r_0)}}, \tag{103}$$

where

$$G(z,r_0) = \frac{r_0^4}{b^2} - (1-z)^2\, r_0^2 + \frac{2\,M\, r_0\,(1-z)^3}{\rho}. \tag{104}$$

Performing the series expansion of the function $G(z, r_0)$ at the point $z = 0$, which corresponds to the limit $r \to r_0$, and considering $b \to b(r_0)$ given by Eq. (94), we obtain that:

$$G(z,r_0) \approx \Lambda_1(r_0) z + \Lambda_2(r_0) z^2, \tag{105}$$

where the expansion parameters $\Lambda_1$ and $\Lambda_2$ are defined as

$$\Lambda_1(r_0) = .2\, r_0\,(r_0 - r_{ph}), \qquad \Lambda_2(r_0) = r_0 \Big( -r_0 + \frac{6\,M}{\rho}\Big), \tag{106}$$

$$\Lambda_1(r_0 \;\to\; r_{ph}) = 0, \qquad \Lambda_2(r_0 \;\to\; r_{ph}) = \frac{9\,M^2}{\rho^2}. \tag{107}$$

In the preceding expressions, the expansion coefficients $\Lambda_1$ and $\Lambda_2$ are presented separately, considering the limit $r_0 \to r_{\rm ph}$. To verify the consistency of the results obtained, we examine the particular case corresponding to the Schwarzschild BH, which is recovered by setting $\rho = 1$. Although the expression presented in Eq. (105) allows us to investigate the geometric properties associated with an arbitrary turning point $r_0$, the analysis of gravitational light deflection in the strong-field regime requires considering the critical limit in which the impact parameter approaches its critical value $b \to b_c$, corresponding to the approach of the turning point to the photon sphere radius $r_0 \to r_{\rm ph}$.

In this critical limit, the turning point $r_0$ approaches the radius of the outer photon sphere, $r_{\rm ph}$, causing the expansion coefficient $\Lambda_1$ to tend to zero. Consequently, integrating the expression given in Eq. (103), considering the limit $r_0 \to r_{\rm ph}$ and the expansion coefficients defined in Eqs. (106) and (107), reveals the emergence of a logarithmic divergence characteristic of the strong-field regime. To properly handle this divergent behavior, following the Bozza and Tsukamoto formalism, we decompose the integral into two distinct contributions: a regular part, $\Delta\varphi_R(r_0)$, which remains finite in the critical limit, and a divergent part, $\Delta\varphi_D(r_0)$, responsible for the logarithmic singularity. In this sense, we can write:

$$\Delta\varphi(r_0) = \Delta\varphi_R(r_0) + \Delta\varphi_D(r_0). \tag{108}$$

$$\begin{aligned}
\Delta\varphi_D(r_0) \;&=\; \int_0^1 \frac{2\,\sigma\, r_0 \mathrm{d}z}{\sqrt{\Lambda_1(r_0)z + \Lambda_2(r_0)z^2}} = -\frac{4\,\sigma\, r_0}{\sqrt{\Lambda_2(r_0)}} \log\Big(\sqrt{\Lambda_1(r_0)}\Big) \\
&+\; \frac{4\,\sigma\, r_0}{\sqrt{\Lambda_2(r_0)}} \log\Big(\sqrt{\Lambda_2(r_0)} + \sqrt{\Lambda_1(r_0) + \Lambda_2(r_0)}\Big), \qquad \text{when} \qquad r_0 \to r_{ph} \\
\Delta\varphi_D(r_{ph}) \;&=\; -\frac{2\,\sigma\, r_{ph}}{\sqrt{\Lambda_2(r_{ph})}} \log\left(\Lambda_1(r_{ph})\right) + \frac{2\,\sigma\, r_{ph}}{\sqrt{\Lambda_2(r_{ph})}} \log\left(4\Lambda_2(r_{ph})\right).
\end{aligned} \tag{109}$$

To control the divergent contribution to the integral, it is necessary to expand the coefficient $\Lambda_1(r_0)$ and the impact parameter $b(r_0)$ around the radius of the photon sphere $r_{ph}$. In this limit, we can improve the expression in Eq. (106), which will be important for rewriting the approximate form of the impact parameter in Eq. (94). So, we have

$$\Lambda_1(r_0) = \frac{6\,M^2\sqrt{6}}{\rho^2}\left(\frac{\rho\,b}{3\sqrt{3}M} - 1\right)^{1/2}. \tag{110}$$

After performing some algebraic manipulations and considering the expansion coefficients $\Lambda_1$ and $\Lambda_2$ in the critical limit $r_0 \to r_{\rm ph}$, we explicitly obtain the expression corresponding to the divergent part of the integral, given in Eq. (109):

$$\Delta\varphi_D(r_{ph}) = \sigma\,\log\left(\frac{\rho\,b}{3\,\sqrt{3}\,M} - 1\right) + \sigma\,\log(6). \tag{111}$$

The expression obtained previously corresponds to the divergent contribution to the integral describing the gravitational deflection of light. In turn, the regular contribution, which remains finite in the limit where the turning point approaches the photon sphere radius, is defined by the following expression:

$$\Delta\varphi_R = \int_0^1 \frac{2\,\sigma\,r_{ph}}{\sqrt{G(z,r_{ph})}}dz - \int_0^1 \frac{2\,\sigma\,r_{ph}}{\sqrt{\Lambda_2(r_{ph})}}\frac{dz}{z} = 2\,\sigma\,\log\left(12 - 6\,\sqrt{3}\right). \tag{112}$$

The total light deflection angle induced by the gravitational field of a BH described within Kalb–Ramond gravity in the Palatini formalism, taking into account the effects associated with Lorentz symmetry violation, can be expressed as $\Delta\varphi = \Delta\varphi_D + \Delta\varphi_R - \sigma\pi$, where $\Delta\varphi_D$ and $\Delta\varphi_R$ represent the divergent and regular contributions, respectively, while the term $\sigma\pi$ corresponds to the reference angular contribution arising from the non-asymptotically flat nature of the geometry. The results obtained in this work are in agreement with those reported in the literature, since the corresponding expressions are consistently recovered by taking the appropriate limits of the model parameters [65, 66, 69], particularly when the effects of the holonomy correction are switched off ($a \to 0$) and the contribution associated with the global monopole vanishes ($\alpha \to 0$).

### D. Lens equation and observables

In this subsection, we investigate the connection between the gravitational deflection of light in the weak- and strong-field regimes, whose results are presented in Eq. (101) and Eqs. (111) and (112), respectively. In the strong-field regime, the deflection angle is expressed as $\delta\varphi = \Delta\varphi_R + \Delta\varphi_D - \sigma\pi$. Based on these results, we employ the gravitational lensing formalism to analyze light propagation in the spacetime of a BH under the effects of Lorentz symmetry violation. This approach allows us to determine observables associated with gravitational lensing, providing a means to investigate how the modifications induced by Lorentz symmetry violation affect the optical properties of the geometry and, in principle, to establish observational constraints on the model parameters. We begin by focusing our analysis on the strong field regime.

The Fig. (27) presents a schematic representation of the geometric configuration of the gravitational lens system. In this setup, a light beam emitted by the source **S** undergoes gravitational deflection due to the presence of the BH, located at **L**, and subsequently travels toward the observer **O**. The deflection angle of the light path is denoted by $\alpha$, while $\psi$ and $\theta$ represent, respectively, the angular positions of the source and the image with respect to the optical axis $\overline{LO}$. Considering a configuration of near alignment between the source **S**, the lens **L**, and the observer **O** a scenario in which effects associated with the formation of relativistic images become particularly significant [34, 35] we can establish the gravitational lens equation that relates the angular positions $\psi$ and $\theta$, expressed as:

$$\psi = \theta - \frac{D_{LS}}{D_{OS}}\Delta\alpha_n, \tag{113}$$

where $\Delta\alpha_n$ is the deflection angle with all the loops completed by the photons before reaching the observer subtracted out; that is, $\Delta\alpha_n = \alpha - 2n\pi$. In this approach, we use the following approximation for the impact parameter: $b \approx \theta D_{OL}$. Thus, the contributions to the angular deflection, Eqs. (111) and (112), are rewritten as:

$$\alpha(\theta) = -\bar{a}\log\left(\frac{\theta D_{OL}}{b_c} - 1\right) + \bar{b}, \tag{114}$$

where we have

$$b_c = \frac{3\sqrt{3}\,M}{\rho}, \qquad \bar{a} = \sigma \tag{115}$$

$$\bar{b} = \sigma\,\log(6) + 2\,\sigma\log\left(12 - 6\sqrt{3}\right) - \sigma\,\pi. \tag{116}$$

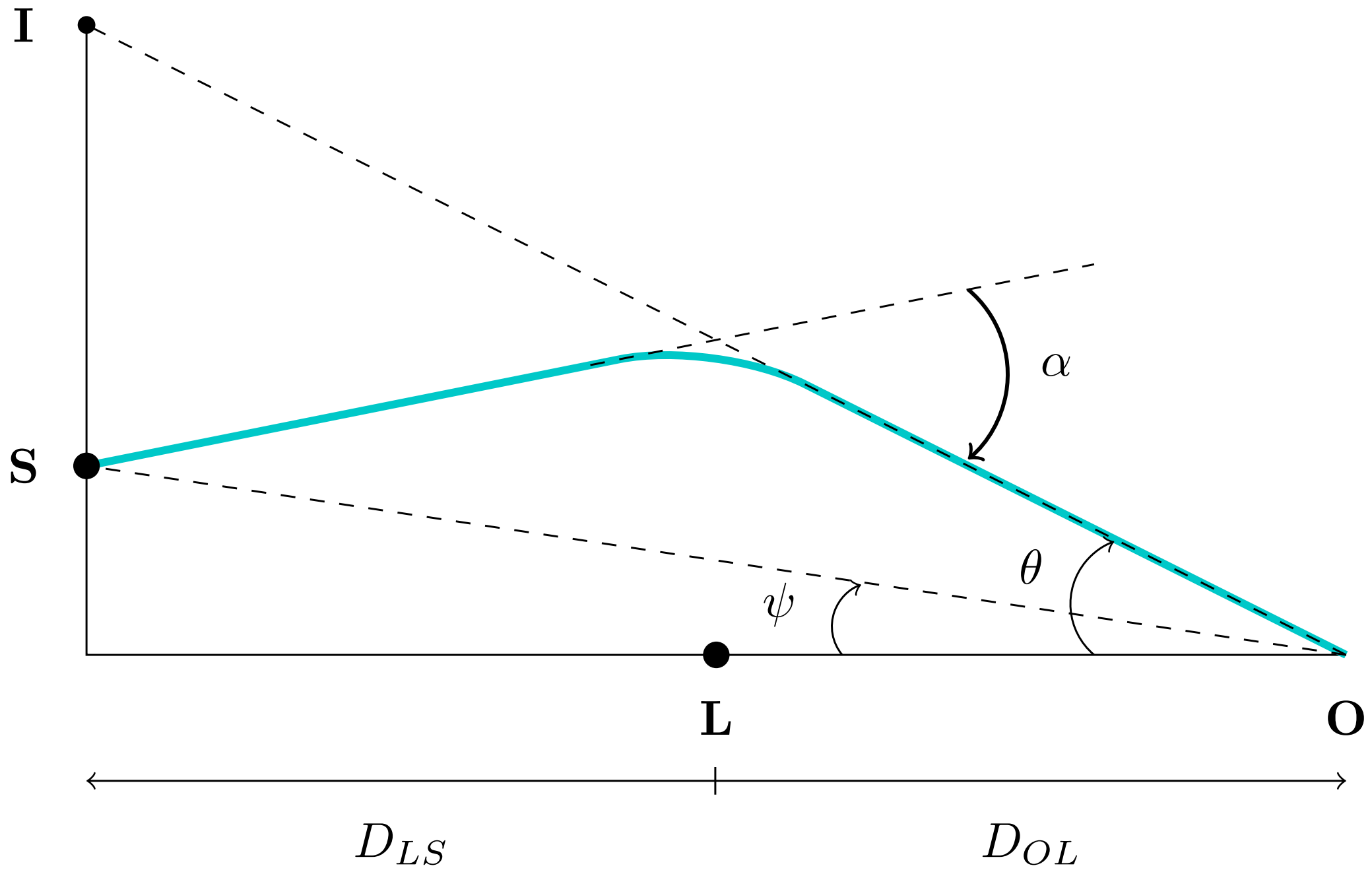

Figure 27. Light angular deflection diagram.

To obtain $\Delta\alpha_n$, we expand $\alpha(\theta)$ close to $\theta = \theta_n^0$, where $\alpha(\theta_n^0) = 2n\pi$, and find that

$$\Delta\alpha_n = \frac{\partial\alpha}{\partial\theta}\Big|_{\theta=\theta_n^0}\left(\theta - \theta_n^0\right). \tag{117}$$

Evaluating Eq. (114) on $\theta = \theta_n^0$, we have

$$\theta_n^0 = \frac{b_c}{D_{OL}}(1 + e_n), \qquad \text{with} \quad e_n = e^{\frac{\bar{b} - 2n\pi}{\bar{a}}}. \tag{118}$$

Substituting Eq. (114) and (118) in Eq. (117), we have

$$\Delta\alpha_n = -\frac{\bar{a}D_{OL}}{b_c e_n}\left(\theta - \theta_n^0\right). \tag{119}$$

B substituting Eq. (119) and Eq. (113), we have

$$\theta_n \approx \theta_n^0 + \left(\frac{e_n b_c}{\bar{a}}\right) \frac{D_{OS}(\psi - \theta_n^0)}{D_{OL} D_{LS}}. \tag{120}$$

Although light deflection preserves the surface brightness of the source, the gravitational lensing effect modifies the solid angle subtended by the observed image. In this regard, the total received flux associated with each lensed image is modulated by its corresponding magnification factor $\mu_n$, which is defined by $\mu_n = \left|\frac{\psi}{\theta}\frac{\partial\psi}{\partial\theta}\left|_{\theta=\theta_n^0}\right|^{-1}$. Therefore, substituting Eq. (113) into Eq. (119), we have

$$\mu_n = \frac{e_n(1+e_n)}{\psi \bar{a}} \left(\frac{b_c}{D_{OL}}\right)^2 \frac{D_{OS}}{D_{LS}}. \tag{121}$$

It is observed that the amplification decreases sharply as $n$ increases, such that the first relativistic image, $\theta_1$, provides the dominant contribution to the observed brightness, while higher-order images become progressively less intense. Moreover, the presence of the factor $\left(\frac{b_c}{D_{OL}}\right)^2$ indicates that the amplification is, in general, significantly reduced. However, in the limit $\psi \to 0$, corresponding to a perfect alignment between the source, the lens, and the observer, the amplification tends to diverge, which, in principle, enhances the detectability of relativistic images.

### E. Observables in the strong-field limit

In the previous sections, we derived expressions for the angular positions of the relativistic images and their corresponding fluxes in terms of the expansion coefficients $\bar{a}$, $\bar{b}$, and the critical impact parameter $b_c$. At this stage, we adopt an inverse approach, in which these parameters are determined from observational quantities associated with gravitational lensing. This procedure allows us to investigate the physical and geometrical properties of the object responsible for light deflection as well as to assess possible modifications introduced into the spacetime geometry. In this way, a connection can be established between gravitational lensing observables and the model parameters, enabling the theoretical predictions to be confronted with observational results. The impact parameter may be written in terms of $\theta_\infty$ [36],

$$b_c = D_{OL}\theta_\infty. \tag{122}$$

We will follow Bozza [37] and assume that only the outermost image $\theta_1$ is resolved as a single image while the others are encapsulated in $\theta_\infty$. Thus, Bozza defined the following observables,

$$s = \theta_1 - \theta_\infty = \theta_\infty e^{\frac{\bar{b}-2\pi}{\bar{a}}}, \tag{123}$$

$$\tilde{r} = \frac{\mu_1}{\sum_{n=2}^{\infty}\mu_n} = e^{\frac{2\pi}{\bar{a}}}, \tag{124}$$

where the parameters $\bar{a}$ and $\bar{b}$ are defined in the expressions (115) and (116).

In the preceding expressions, **s** represents the angular separation between the first relativistic image and the remaining images, while $\tilde{r}$ characterizes the ratio between the flux of the first image and the sum of the fluxes of the subsequent images. From these observational quantities, it is possible to invert the corresponding relations and determine the expansion coefficients associated with the strong-field regime. For our analysis, we consider a compact object with an estimated mass of $4.4 \times 10^6 M_\odot$, located at an approximate distance of $D_{OL} = 8.5\,\text{kpc}$, adopting values comparable to those attributed to the supermassive black hole at the center of the Galaxy [81]. In geometrized units, the mass is expressed through rescaling $M \to MG/c^2$. Using these parameters, we obtain the limiting angular position $\theta_\infty = 26.5473\,\mu$as, which corresponds to the value predicted for the Schwarzschild geometry.

Table V. Observable quantities varying as a function of the Lorentz symmetry violation parameter $q$.

| $l_1 = 0$ | $q = l_2$ | $\theta_\infty$ ($\mu$ arcsecs) | $s$ ($\mu$ arcsecs) | $r = 2.5 \log_{10} \bar{r}$ (**magnitudes**) |
|---|---|---|---|---|
| 0 | -0.8 | 18.9690 | 0.1429 | 4.8728 |
| 0 | -0.6 | 20.4282 | 0.1090 | 5.2476 |
| 0 | -0.3 | 23.0927 | 0.0656 | 5.9321 |
| 0 | 0.0 | 26.5566 | 0.0332 | 6.8219 |
| 0 | 0.3 | 31.2431 | 0.0129 | 8.0257 |
| 0 | 0.6 | 37.9380 | 0.0032 | 9.7456 |
| 0 | 0.8 | 44.2610 | 0.0008 | 11.3698 |

Table V presents the numerical results for the observables associated with gravitational lensing in the strong field regime, considering different values of the Lorentz symmetry violation parameter $q = l_2$, with $l_1 = 0$. It can be observed that increasing $q$ leads to an increase in the limiting angular position $\theta_\infty$, accompanied by a significant decrease in the angular separation $s$, indicating that the relativistic images tend to cluster closer to $\theta_\infty$. Furthermore, the relative magnitude $r$ progressively increases, revealing a greater dominance of the brightness of the first relativistic image over the subsequent ones. Compared to the Schwarzschild case ($q = 0$), negative values of $q$ produce effects opposite to those observed for positive values.

### F. Observables in the weak-field limit

Next, we turn our attention to gravitational lensing observables in the weak field regime, characterized by the condition $b \gg M$, where the impact parameter is significantly larger than the characteristic mass scale of the compact object. In this regime, light rays undergo only small gravitational deflections, without completing multiple loops around the lens. Furthermore, to simplify the analytical treatment, we restrict our investigation to the geometry of a Schwarzschild-type BH modified by the effects of Lorentz symmetry violation, retaining only first-order contributions in the mass parameter $M$. Under these approximations, the expression for the gravitational deflection angle given in Eq. (101) reduces to:

$$\delta\varphi \approx \pi\,(\sigma - 1) + \frac{4\,\sigma\,M}{\rho\,b}. \tag{125}$$

Starting from the gravitational lens equation presented in Eq. (113), which establishes the relationship between the angular positions of the source and the image, we consider a configuration of perfect alignment between the source, the compact object responsible for light deflection, and the observer. This condition, characterized by $\psi = 0$, corresponds to the formation of an Einstein ring, whose angular position can be determined by the

Table VI. Observable quantities as functions of the violation parameter $q$.

| $l_1 = 0$ | $q = l_2$ | $R_E$ (km) | $\theta_E$ (arcsecs) |
|---|---|---|---|
| 0 | -0.8 | $7.76 \times 10^{16}$ | 129600.00 |
| 0 | -0.6 | $5.82 \times 10^{16}$ | 97200.00 |
| 0 | -0.3 | $2.91 \times 10^{16}$ | 48600.00 |
| 0 | 0.0 | $1.27 \times 10^{12}$ | 2.12 |
| 0 | 0.3 | $5.52 \times 10^{7}$ | $9.22 \times 10^{-5}$ |
| 0 | 0.6 | $2.76 \times 10^{7}$ | $4.68 \times 10^{-5}$ |
| 0 | 0.8 | $2.07 \times 10^{7}$ | $3.46 \times 10^{-5}$ |

following expression:

$$\theta = \frac{D_{LS}}{D_{OS}} \Delta\alpha_n, \tag{126}$$

where $\Delta\alpha_n$ is given by Eq. (125). Therefore, substituting Eq. (125) into Eq. (126) and keeping only terms up to first order in the mass $M$, we obtain a second-degree algebraic equation, which when solved gives us the angular position $\theta_E$ for the Einstein ring:

$$\theta = \theta_E = \frac{D_{LS}\,\pi\,(\sigma-1)}{2\,D_{OS}} + \frac{1}{2}\sqrt{\left(\frac{D_{LS}\,\pi\,(\sigma-1)}{D_{OS}}\right)^2 + \frac{16\,\sigma\,M\,D_{LS}}{\rho\,D_{OL}\,D_{OS}}}. \tag{127}$$

The above expression shows that the angular position of the Einstein ring is modified by the presence of parameters associated with Lorentz symmetry violation, even when only first-order contributions in the mass parameter $M$ are considered. However, by taking the limit $\sigma = \rho \to 1$, corresponding to the absence of the effects introduced by Kalb–Ramond gravity, we exactly recover the result predicted for the Schwarzschild BH. Thus, considering only the positive solution for the angular position of the Einstein ring, we obtain the same expression as that presented in Eq. (50) of Ref. [65]:

$$\theta = \theta_E = \sqrt{\frac{4M\,D_{LS}}{D_{OS}D_{OL}}}. \tag{128}$$

Furthermore, we can determine the Einstein ring radius using the small-angle approximation, $b \approx \theta D_{OL}$, together with the expression for the angular position of the Einstein ring given in Eq. (127) [65, 66]. Thus, the Einstein ring radius can be expressed in terms of the distance between the observer and the gravitational lens, $D_{OL}$, and the angular position $\theta_E$, yielding $R_E = D_{OL}\theta_E$.

Based on the expressions obtained previously, we can determine the observable gravitational lensing $R_E$ and $\theta_E$, investigating their dependence on the parameter $\alpha$, associated with modifications to the gravitational theory, and on the mass $M$ of the compact object. For this analysis, performed in the weak-field regime, we consider a gravitational lensing configuration involving a star located in the Galactic bulge [82]. In this scenario, we adopt the distances $D_{OL} = 4\,\text{kpc}$ and $D_{OS} = 8\,\text{kpc}$, which correspond to the distances of the observer lens and the distances of the observer source, respectively. This configuration allows us to investigate how the modifications introduced by the parameter $q$ affect the angular position and radius of the Einstein ring.

In VI, we present a table in which we assign some possible acceptable values for the violation parameter, and observe how the Einstein ring radius and the angular position respond to this variation.

## IX. QUASINORMAL MODES

The response of the Palatini Kalb Ramond geometry to a small disturbance provides a complementary view of the Lorentz symmetry violation parameters. Whereas the critical photon radius determines the location of null trapping, the oscillation frequency and the attenuation time describe how a wave samples its neighborhood. We examine this response through massless scalar, electromagnetic, gravitational, and Dirac perturbations, retaining the normalization of the Killing time and the areal radius adopted in the preceding sections. The frequency convention is $\exp(-i\omega t)$, so that $\mathrm{Re}\,\omega$ determines the oscillation rate and $-\,\mathrm{Im}\,\omega > 0$ determines the attenuation rate.

The scalar and Maxwell fields obey their usual equations on the physical metric $g_{\mu\nu}$. For the Dirac field we employ the spin connection determined by an orthonormal tetrad of this metric. These choices specify the test field model and preserve the usual electromagnetic gauge symmetry. The gravitational calculation follows a homogeneous axial sector of the Palatini equations, with its auxiliary metric and antisymmetric field restriction stated explicitly below. It is useful to introduce the dimensionless quantities

$$\begin{aligned} x &= \frac{r}{\bar{M}}, \qquad x_* = \frac{r_*}{\bar{M}} = \sigma\left[x + 2\ln\left(\frac{x}{2} - 1\right)\right], \\ \widehat{\omega} &= \bar{M}\omega, \qquad U = \bar{M}^2 V, \qquad \mathcal{D} = \frac{\mathrm{d}}{\mathrm{d}x_*} = \frac{F(x)}{\sigma}\frac{d}{dx}. \end{aligned} \tag{129}$$

An additive constant in $x_*$ changes the phase of a radial solution and leaves its frequency unchanged. Since $F \to 1$ at large radius, the coordinate $t$ is normalized to the proper time of a static observer at infinity, even though the spatial geometry retains its conical radial deformation. After separating the angular and temporal dependence, every decoupled sector considered here assumes the form

$$\begin{aligned} &\frac{\mathrm{d}^2\psi}{\mathrm{d}x_*^2} + \left[\widehat{\omega}^2 - U(x)\right]\psi = 0, \\ &\psi \sim e^{-i\widehat{\omega}x_*} \quad (x_* \to -\infty), \\ &\psi \sim e^{+i\widehat{\omega}x_*} \quad (x_* \to +\infty). \end{aligned} \tag{130}$$

The first condition selects propagation into the future horizon, while the second selects outward propagation in the distant region. Both are defined using the physical time in Eq. (**??**). The resulting complex frequencies are associated with an open wave problem, and their imaginary parts measure the loss of amplitude through its two boundaries.

We evaluate the frequencies with the approximation of Wentzel, Kramers and Brillouin through third order, using derivatives at the exterior maximum of the potential [58, 59, 83]. To present the prescription in a compact form, let $x_0$ denote that maximum and define

$$\begin{aligned} U_k &= \mathcal{D}^k U\Big|_{x=x_0}, \qquad U_1 = 0, \qquad U_2 < 0, \\ \alpha &= n + \frac{1}{2}, \qquad R_k = \frac{U_k}{U_2}, \qquad n = 0, 1, 2, \ldots. \end{aligned} \tag{131}$$

Here $n$ is the overtone number, and every derivative is taken with respect to $x_*$. In terms of these quantities the approximation reads

$$\widehat{\omega}^2_{(3)} = U_0 + A_2 - i\alpha\sqrt{-2U_2}\left(1 + \frac{A_3}{-2U_2}\right), \tag{132}$$

with

$$A_2 = \frac{1}{8}R_4\left(\frac{1}{4}+\alpha^2\right) - \frac{1}{288}R_3^2\left(7+60\alpha^2\right),$$
$$A_3 = \frac{5}{6912}R_3^4\left(77+188\alpha^2\right) - \frac{1}{384}R_3^2R_4\left(51+100\alpha^2\right) + \frac{1}{2304}R_4^2\left(67+68\alpha^2\right) + \frac{1}{288}R_3R_5\left(19+28\alpha^2\right) - \frac{1}{288}R_6\left(5+4\alpha^2\right). \tag{133}$$

The root is chosen with positive real part and negative imaginary part. The first order estimate retains $U_0 - i\alpha\sqrt{-2U_2}$, and the second order estimate adds $A_2$. We use the relative variation

$$\delta_{23} = \frac{\left|\widehat{\omega}_{(3)} - \widehat{\omega}_{(2)}\right|}{\left|\widehat{\omega}_{(3)}\right|} \tag{134}$$

as an internal indication of sensitivity to the retained terms. This quantity measures the variation between two approximations and is reported as such. The tables emphasize small overtone numbers and angular eigenvalues for which the potential maximum governs the response. The calculation uses 70 decimal digits in its working arithmetic and reports six decimal places in the frequency tables. This numerical precision makes the evaluation of the WKB expression reproducible; the approximation sensitivity is assessed separately. We use $q = -0.6, -0.3, 0, 0.3, 0.6$ for the tables and $-0.8 \le q \le 0.8$ for the continuous spectral figures. These values illustrate the exact geometry and do not represent an observational fit. The tabulated bosonic modes have $\ell = 2, 3$ and $n = 0, 1$, while the spinor modes have $\kappa = 2, 3$ and $n = 0, 1$, so each example obeys $n$ smaller than its angular index.

### A. Scalar perturbations

Let us consider $\Phi$ be a neutral massless scalar with negligible influence on the background. Its covariant equation and angular decomposition are

$$\frac{1}{\sqrt{-g}}\partial_\mu\left(\sqrt{-g}\,g^{\mu\nu}\partial_\nu\Phi\right) = 0, \qquad \sqrt{-g} = \sigma r^2\sin\theta,$$
$$\Phi = \frac{\psi_{\rm S}(r)}{r}Y_{\ell m}(\theta,\phi)e^{-i\omega t}, \qquad \ell = 0, 1, 2, \dots. \tag{135}$$

The angular eigenvalue is $\ell(\ell+1)$, and the constant factor $\sigma$ in the metric determinant cancels from the radial divergence. Combining the remaining terms with the definition of $r_*$ gives Eq. (130) with

$$V_{\rm S}(r) = F(r)\left[\frac{\ell(\ell+1)}{r^2} + \frac{F'(r)}{\sigma^2 r}\right],$$
$$U_{\rm S}(x) = F(x)\left[\frac{\ell(\ell+1)}{x^2} + \frac{2}{\sigma^2 x^3}\right]. \tag{136}$$

The first term records the angular momentum barrier, whereas the second describes the radial variation of the physical geometry. Their separate dependence on $\sigma$ shows that changing $q$ at fixed $\bar{M}$ affects both the height of the scalar potential and its width in the propagation coordinate. At $q = 0$, Eq. (136) gives the Schwarzschild potential with mass $\bar{M}$. Its positivity throughout $x > 2$ also supplies a useful check on the radial reduction.

The scalar panel of Fig. 28 displays the potential in the coordinate $x_*$, making its width and decay toward the two boundaries explicit. Table VII gives the associated complex frequencies, while the scalar panels of Figs. 29 and 30 display the oscillation and attenuation rates as functions of $q$. These representations separate the two parts of the frequency and allow the change in oscillation period to be compared directly with the change in decay time. The scalar results use the same mass normalization as all subsequent sectors. For $\ell = 2$ and $n = 0$, increasing $q$ from $-0.6$ to $0.6$ changes $\widehat{\omega}$ from $0.480580 - 0.084785i$ to $0.487965 - 0.116007i$. The oscillation rate grows over this interval, while the attenuation rate increases by about 36.8%.

Table VII. Scalar frequencies $\bar{M}\omega$ from WKB at order three. The columns are labelled by $(\ell, n)$. Each comparison holds $\bar{M}$ fixed.

| $q$ | $(2,0)$ | $(2,1)$ | $(3,0)$ | $(3,1)$ |
|---|---|---|---|---|
| $-0.6$ | $0.480580 - 0.084785\,i$ | $0.464863 - 0.258063\,i$ | $0.673271 - 0.084588\,i$ | $0.661779 - 0.255675\,i$ |
| $-0.3$ | $0.481731 - 0.090199\,i$ | $0.464120 - 0.275021\,i$ | $0.674115 - 0.089963\,i$ | $0.661180 - 0.272176\,i$ |
| $0.0$ | $0.483211 - 0.096805\,i$ | $0.463192 - 0.295810\,i$ | $0.675206 - 0.096512\,i$ | $0.660414 - 0.292344\,i$ |
| $0.3$ | $0.485188 - 0.105113\,i$ | $0.462005 - 0.322120\,i$ | $0.676672 - 0.104738\,i$ | $0.659403 - 0.317772\,i$ |
| $0.6$ | $0.487965 - 0.116007\,i$ | $0.460446 - 0.356898\,i$ | $0.678746 - 0.115503\,i$ | $0.658006 - 0.351225\,i$ |

## B. Vector perturbations

For electromagnetic disturbances we introduce a test potential $A_\mu$ and its field strength $f_{\mu\nu} = \partial_\mu A_\nu - \partial_\nu A_\mu$. The equations

$$\nabla_\mu f^{\mu\nu} = 0, \qquad \nabla_{[\alpha} f_{\mu\nu]} = 0 \tag{137}$$

are evaluated with the physical metric. A decomposition into vector spherical harmonics separates the two electromagnetic parities, which share the same radial potential. For example, an axial harmonic can be written as $A_A = \psi_{\rm V}(r)\epsilon_A{}^B D_B Y_{\ell m}\, e^{-i\omega t}$, where $A, B$ refer to the unit sphere and $D_A$ is its covariant derivative. The radiative multipoles satisfy $\ell \geq 1$, and the radial equation takes the form of Eq. (130), with

$$\begin{aligned} V_{\rm V}(r) &= F(r)\frac{\ell(\ell+1)}{r^2}, \\ U_{\rm V}(x) &= F(x)\frac{\ell(\ell+1)}{x^2}, \qquad x_0 = 3. \end{aligned} \tag{138}$$

The electromagnetic potential has an especially relation to the null orbit. Its maximum remains at $r = 3\bar{M}$, and its height is independent of $q$ at fixed $\bar{M}$. Its derivatives in Eq. (131) nevertheless depend on $\sigma$ because the propagation coordinate carries the radial deformation. The electromagnetic spectrum distinguishes geometries that have the same photon radius and the same critical impact parameter.

The vector panel of Fig. 28 exhibits this change in width directly. Its consequences for the complex frequencies are given in Table VIII and in the vector panels of Figs. 29 and 30. In particular, the geometric comparison is given by the normalized time $t$ and by $\bar{M}$, so that a change in the computed attenuation rate has an immediate interpretation as a change in the lifetime of the electromagnetic response. The common Maxwell potential applies to both parities and shows the numerical calculation to be performed once for each $(\ell, n, q)$. For the fundamental mode with $\ell = 2$, the real part decreases from 0.460505 to 0.450737 between $q = -0.6$ and $q = 0.6$, while the attenuation rate increases from 0.083604 to 0.113073. In addition, the fixed potential height accompanies a measurable change in the complex resonance through its radial width.

Table VIII. Maxwell frequencies $\bar{M}\omega$ from WKB at order three. The columns are labelled by $(\ell, n)$. Each comparison holds $\bar{M}$ fixed. The displayed digits describe the approximation.

| $q$ | $(2,0)$ | $(2,1)$ | $(3,0)$ | $(3,1)$ |
|---|---|---|---|---|
| $-0.6$ | $0.460505 - 0.083604\,i$ | $0.444015 - 0.254719\,i$ | $0.659054 - 0.083992\,i$ | $0.647281 - 0.253936\,i$ |
| $-0.3$ | $0.459044 - 0.088784\,i$ | $0.440453 - 0.271043\,i$ | $0.658047 - 0.089247\,i$ | $0.644755 - 0.270096\,i$ |
| $0.0$ | $0.457131 - 0.095065\,i$ | $0.435828 - 0.290972\,i$ | $0.656733 - 0.095631\,i$ | $0.641471 - 0.289796\,i$ |
| $0.3$ | $0.454518 - 0.102904\,i$ | $0.429581 - 0.316062\,i$ | $0.654947 - 0.103616\,i$ | $0.637031 - 0.314549\,i$ |
| $0.6$ | $0.450737 - 0.113073\,i$ | $0.420682 - 0.349012\,i$ | $0.652379 - 0.114007\,i$ | $0.630692 - 0.346968\,i$ |

### C. Tensor perturbations

The gravitational response requires the field equations of the Palatini theory in addition to its physical metric. We obtain an axial tensor sector by keeping the covariant antisymmetric field fixed, $\delta B_{\mu\nu} = 0$, and selecting the odd metric perturbations with $\ell \geq 2$. Let $a, b \in \{t, r\}$ and $A, B \in \{\theta, \phi\}$. These perturbations satisfy $\delta g_{ab} = 0$ and have vanishing angular trace. The background antisymmetric field has only a radial electric component $b_{tr}$, so the norm constraint $X = B_{\mu\nu}B^{\mu\nu} + b^2 = 0$ has $\delta X = 0$. Moreover, $\delta H = \mathrm{d}\delta B = 0$, and the perturbation of the scalar multiplier vanishes in the axial sector, $\delta\lambda = 0$. These conditions select a homogeneous gravitational response with the antisymmetric background retained in the coefficients of its equation [84].

The independent connection is the metric connection of an auxiliary geometry $h_{\mu\nu}$. Its defining relation follows directly from the variation of the nonminimal curvature couplings,

$$\begin{gathered} \sqrt{-h}\, h^{\mu\nu} = \sqrt{-g}\left[\left(1 + \xi_1 B^{\alpha\beta}B_{\alpha\beta}\right) g^{\mu\nu} + \xi_2 B^{\mu\rho}B^{\nu}{}_{\rho}\right], \\ \ell_1 = \xi_1 b^2, \qquad \ell_2 = \xi_2 b^2. \end{gathered} \tag{139}$$

For the radial electric background this relation gives a particularly simple map. Let us define

$$\begin{gathered} a = 1 - \ell_1 > 0, \qquad c = 1 - \ell_1 - \frac{\ell_2}{2} = a\sigma^2 > 0, \\ h_{ab} = a\, g_{ab}, \qquad h_{AB} = c\, g_{AB}. \end{gathered} \tag{140}$$

The auxiliary background becomes exactly Schwarzschild under the changes

$$\begin{gathered} T = \sqrt{a}\, t, \qquad R = \sqrt{c}\, r, \qquad m_h = \sqrt{c}\, \bar{M}, \\ \mathrm{d}s_h^2 = -\left(1 - \frac{2m_h}{R}\right)\mathrm{d}T^2 + \frac{\mathrm{d}R^2}{1 - 2m_h/R} + R^2 \mathrm{d}\Omega_2^2. \end{gathered} \tag{141}$$

The restrictions above also preserve the axial gauge freedom. An odd gauge vector has angular components only, and $\mathcal{L}_\xi b = i_\xi \mathrm{d}b + \mathrm{d}(i_\xi b) = 0$. Also, the condition $\delta B_{\mu\nu} = 0$ remains valid under the allowed gauge transformations. The auxiliary perturbations obey $\delta h_{aA} = c\,\delta g_{aA}$ and $\delta h^{\mathrm{odd}}_{AB} = c\,\delta g^{\mathrm{odd}}_{AB}$, with vanishing determinant variations at first order.

Table IX. Homogeneous axial gravitational frequencies $\bar{M}\omega$ from WKB at order three, with $\delta B_{\mu\nu} = 0$. The columns are labelled by $(\ell, n)$, with $\ell \geq 2$. Each comparison holds $\bar{M}$ fixed. The displayed digits describe the approximation.

| $q$ | $(2,0)$ | $(2,1)$ | $(3,0)$ | $(3,1)$ |
|---|---|---|---|---|
| $-0.6$ | $0.327285 - 0.078249\,i$ | $0.303477 - 0.241117\,i$ | $0.525590 - 0.081328\,i$ | $0.510759 - 0.246809\,i$ |
| $-0.3$ | $0.347975 - 0.083196\,i$ | $0.322663 - 0.256360\,i$ | $0.558818 - 0.086470\,i$ | $0.543049 - 0.262413\,i$ |
| $0.0$ | $0.373162 - 0.089217\,i$ | $0.346017 - 0.274915\,i$ | $0.599265 - 0.092728\,i$ | $0.582355 - 0.281406\,i$ |
| $0.3$ | $0.404751 - 0.096770\,i$ | $0.375308 - 0.298188\,i$ | $0.649994 - 0.100578\,i$ | $0.631652 - 0.305228\,i$ |
| $0.6$ | $0.446014 - 0.106635\,i$ | $0.413570 - 0.328587\,i$ | $0.716259 - 0.110832\,i$ | $0.696047 - 0.336345\,i$ |

On this background $H = 0$, $\lambda = 0$, and $R_{\mu\nu}(h) = 0$. Under this consideration, variations of the algebraic coefficients that multiply the background Ricci tensor drop out of the linear equations. Their nontrivial axial components reduce to

$$c\,\delta R_{aA}(h) = 0, \qquad a\,\delta R^{\rm odd}_{AB}(h) = 0. \tag{142}$$

The remaining odd Ricci components vanish by their harmonic parity, giving $\delta R_{\mu\nu}(h) = 0$. The linear antisymmetric field equation is satisfied as well, since its curvature source, $\delta H$, and $\delta\lambda$ vanish in this sector. Equation (142) establishes an axial gravitational tower of the Palatini system directly, with its coupling dependence carried by the map between the two metrics.

The Schwarzschild axial master equation for $h_{\mu\nu}$ can now be expressed in the physical coordinates. Its auxiliary tortoise coordinate satisfies $R_* = \sqrt{a}\,r_*$, and a mode with frequency $\omega$ in $t$ has frequency $\omega/\sqrt{a}$ in $T$. The resulting physical radial equation is Eq. (130), with

$$\begin{aligned} V_{\rm T}(r) &= \frac{F(r)}{\sigma^2}\left[\frac{\ell(\ell+1)}{r^2} - \frac{6\bar{M}}{r^3}\right], \\ U_{\rm T}(x) &= \frac{F(x)}{\sigma^2}\left[\frac{\ell(\ell+1)}{x^2} - \frac{6}{x^3}\right], \qquad \ell \geq 2. \end{aligned} \tag{143}$$

For $\sigma = 1$ this is the Schwarzschild axial potential. More generally, writing $y_* = x_*/\sigma$ removes the overall deformation from the differential equation and yields the exact scaling relation

$$\widehat{\omega}^{\rm T}_{\ell n}(q) = \frac{\widehat{\omega}^{\rm Sch}_{\ell n}}{\sigma} = \frac{\widehat{\omega}^{\rm Sch}_{\ell n}}{\sqrt{1 - q/2}}. \tag{144}$$

Here $\widehat{\omega}^{\rm Sch}_{\ell n}$ is the dimensionless Schwarzschild axial frequency. Equation (144) holds for the exact eigenvalue problem and also for the WKB prescription applied at a fixed order. It gives an additional algebraic check on the numerical implementation. At fixed conserved mass $M$ it also yields $M\omega^{\rm T}_{\ell n} = (1-\ell_1)\widehat{\omega}^{\rm Sch}_{\ell n}$, so the dependence on $\ell_2$ cancels in that normalization.

The tensor panel of Fig. 28 displays the simultaneous change in the potential height and width. Table IX lists the frequencies computed from Eq. (143), while the tensor panels of Figs. 29 and 30 display their common $1/\sigma$ dependence. Both the oscillation rate and the attenuation rate increase with $q$ at fixed $\bar{M}$, and their ratio is independent of $q$ for every fixed $(\ell, n)$.

### D. Spinor perturbations

The fermionic response is described by a massless Dirac field obeying

$$
\begin{aligned}
i\gamma^a e_a{}^\mu \left(\partial_\mu + \Gamma_\mu\right)\Psi &= 0, \\
\Gamma_\mu = \frac{1}{4}\omega_{\mu ab}\gamma^a\gamma^b, \qquad \{\gamma^a, \gamma^b\} &= 2\eta^{ab}.
\end{aligned}
\tag{145}
$$

where $e_a{}^\mu$ is a tetrad of Eq. (**??**) and $\omega_{\mu ab}$ is its metric spin connection. Separation into spinor spherical harmonics introduces the positive angular eigenvalue $\kappa = j + \frac{1}{2} = 1, 2, \ldots$, with $j = \frac{1}{2}, \frac{3}{2}, \ldots$. This notation keeps the fermionic angular label distinct from the integer multipole used for the bosonic fields. A suitable rescaling of the radial components produces two combinations $Z_+$ and $Z_-$ satisfying Eq. (130) with the partner potentials [85]

$$
\begin{aligned}
U_\pm = W_\kappa^2 \pm \mathcal{D}W_\kappa, \qquad W_\kappa(x) &= \frac{\kappa\sqrt{F(x)}}{x}, \\
U_\pm(x) = \frac{\kappa^2 F(x)}{x^2} \pm \frac{\kappa\sqrt{F(x)}(3-x)}{\sigma x^3}&.
\end{aligned}
\tag{146}
$$

The two potentials share the nonzero quasinormal frequencies. Their relation follows from the first order radial system, whose differential maps become multiplication by finite nonzero constants in the asymptotic wave regions because $W_\kappa \to 0$ at both boundaries. Either potential can be used to determine the spectrum. We evaluate the frequencies from $U_+$ and retain the comparison with $U_-$ as a useful indication of the sensitivity of a finite WKB approximation.

The factorized form also clarifies the role of potential tails. With $\mathcal{A} = \mathcal{D} + W_\kappa$ and its formal adjoint $\mathcal{A}^\dagger = -\mathcal{D} + W_\kappa$, the radial operators satisfy

$$
\begin{aligned}
-\mathcal{D}^2 + U_- &= \mathcal{A}^\dagger\mathcal{A}, \\
-\mathcal{D}^2 + U_+ &= \mathcal{A}\mathcal{A}^\dagger.
\end{aligned}
\tag{147}
$$

For a hypothetical exponentially growing mode, the quasinormal boundary conditions give a square integrable radial function. Equation (147), which makes the radial quadratic form nonnegative and excludes the negative eigenvalue required for such growth. The same energy argument applies directly to the positive scalar and Maxwell potentials and to the axial tensor potential for $\ell \geq 2$. These statements apply to the three specified test fields and to the homogeneous axial gravitational sector on the static exterior, complementing the attenuation rates obtained from the WKB calculation.

The spinor panel of Fig. 28 displays $U_+$ using the same axes and color assignment as the bosonic sectors. The frequencies are listed in Table X, and their real and imaginary parts appear in the spinor panels of Figs. 29 and 30. For $\kappa = 2$ and $n = 0$, the same interval in $q$ changes the frequency from $0.380310 - 0.084578i$ to $0.375244 - 0.115733i$, with a decrease in oscillation rate and a larger increase in attenuation. The variation between the two partner calculations is examined together with the variation between WKB orders in Fig. 31 and Table XI.

Table X. Massless Dirac frequencies $\bar{M}\omega$ from $V_+$ at WKB order three. The columns are labelled by $(\kappa, n)$, where $\kappa = j + 1/2$. Each comparison holds $\bar{M}$ fixed.

| $q$ | $(2,0)$ | $(2,1)$ | $(3,0)$ | $(3,1)$ |
|---|---|---|---|---|
| −0.6 | 0.380310 − 0.084578 *i* | 0.360587 − 0.259903 *i* | 0.574609 − 0.084459 *i* | 0.561081 − 0.256059 *i* |
| −0.3 | 0.379593 − 0.089965 *i* | 0.357515 − 0.277312 *i* | 0.574213 − 0.089808 *i* | 0.558952 − 0.272662 *i* |
| 0.0 | 0.378627 − 0.096542 *i* | 0.353604 − 0.298746 *i* | 0.573685 − 0.096324 *i* | 0.556185 − 0.292981 *i* |
| 0.3 | 0.377269 − 0.104828 *i* | 0.348484 − 0.326014 *i* | 0.572948 − 0.104503 *i* | 0.552452 − 0.318649 *i* |
| 0.6 | 0.375244 − 0.115733 *i* | 0.341548 − 0.362276 *i* | 0.571849 − 0.115199 *i* | 0.547152 − 0.352515 *i* |

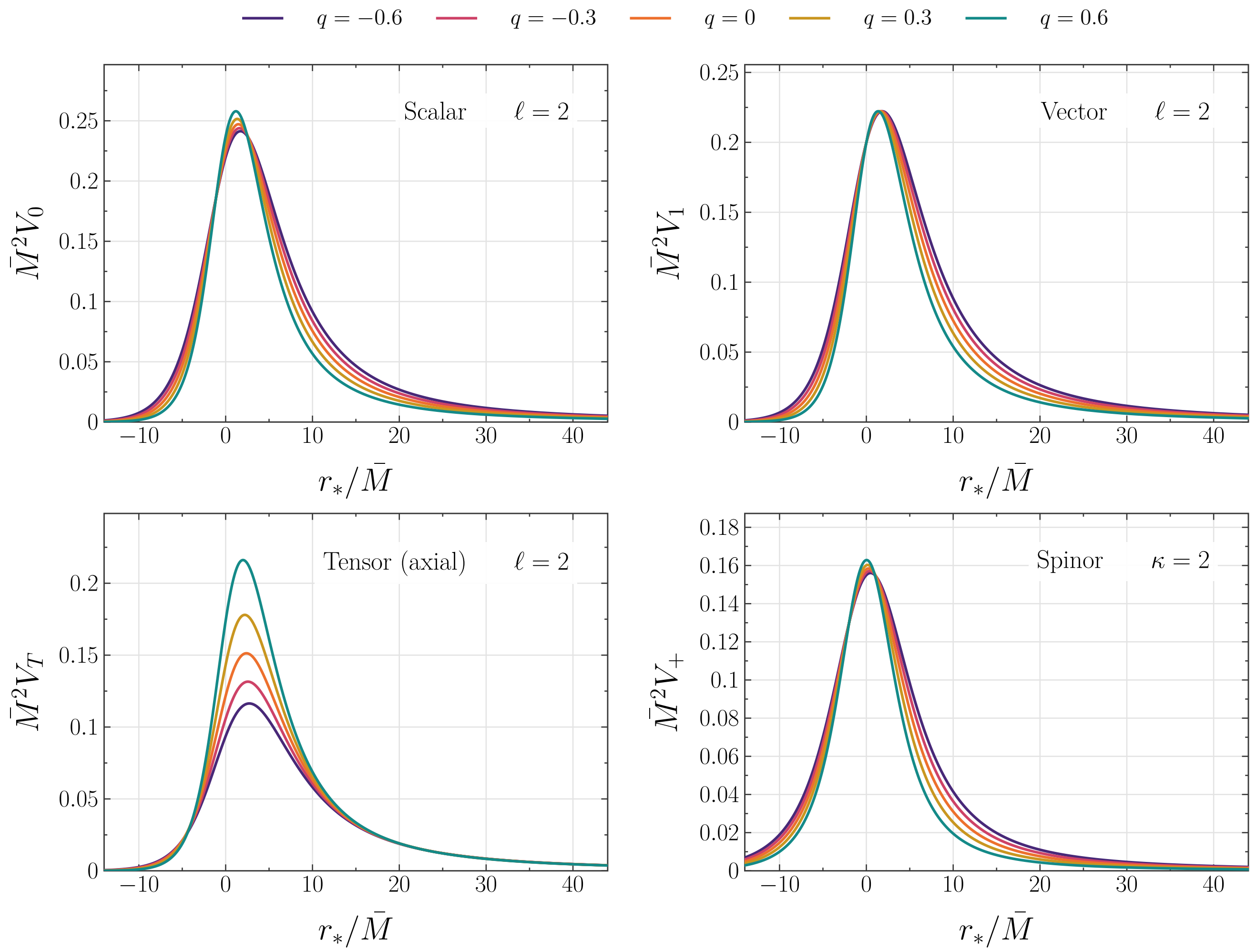

Figure 28. Dimensionless effective potentials for scalar, vector, tensor, and spinor perturbations as functions of $x_* = r_*/\bar{M}$, at fixed $\bar{M}$. Each panel identifies the angular eigenvalue and the values of $q$ used in the calculation. The spinor panel displays $U_+$. The tensor interpretation is specified in Sec. IX C.

### E. Geometric interpretation and numerical analysis

The connection with the unstable null orbit gives an independent way to interpret the leading behavior at large angular momentum [86]. In the normalization used here the circular orbit and its radial instability

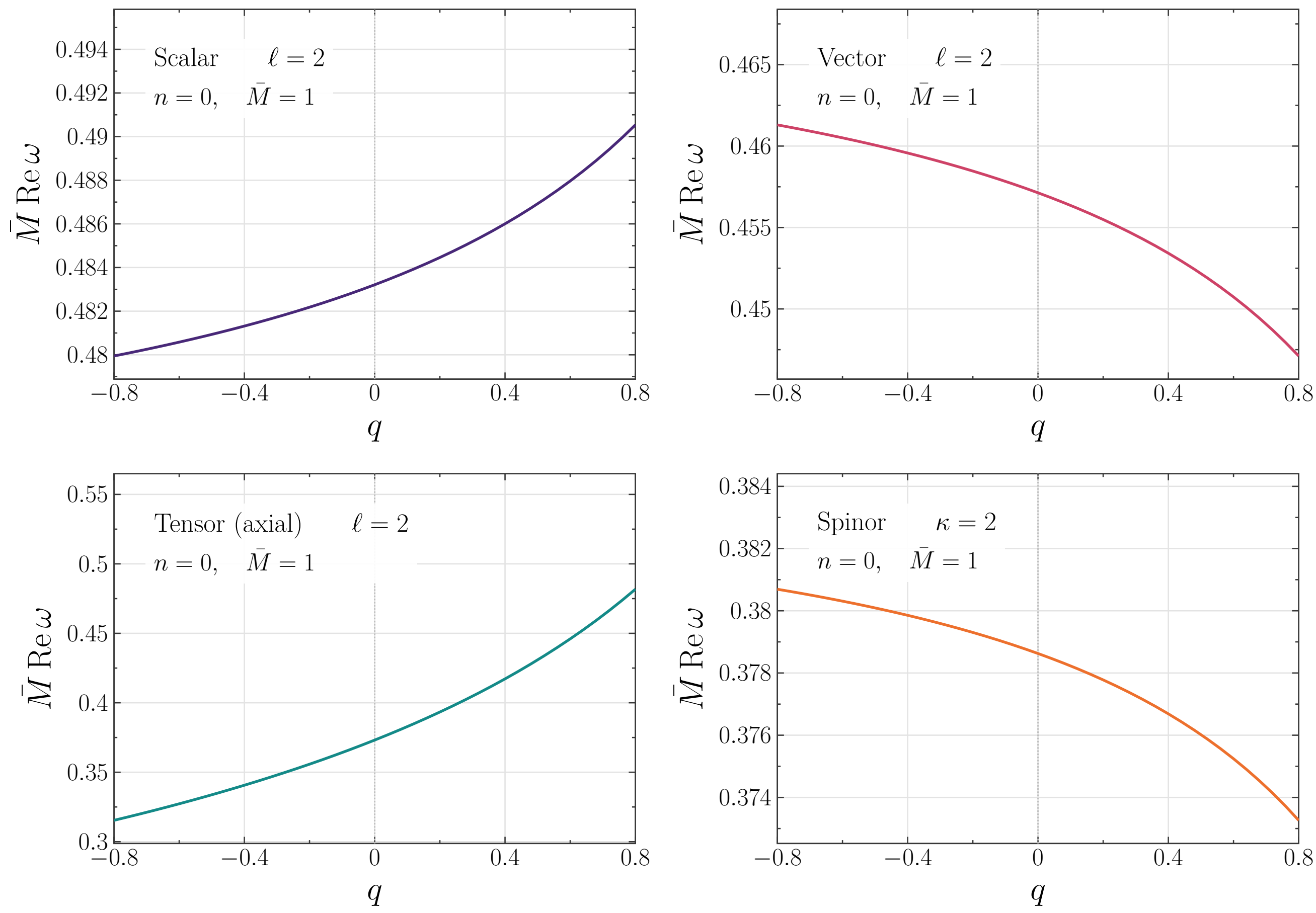

Figure 29. Real parts $\mathrm{Re}(\bar{M}\omega)$ of the quasinormal frequencies as functions of $q$, with the mass $\bar{M}$ held fixed. The panels correspond to scalar, vector, tensor, and spinor perturbations. The angular label and overtone number are indicated inside each panel.

exponent satisfy

$$r_{\rm ph} = 3\bar{M}, \qquad \bar{M}\Omega_{\rm ph} = \frac{1}{3\sqrt{3}},$$
$$\bar{M}\lambda_t = \frac{1}{3\sqrt{3}\,\sigma}. \tag{148}$$

For the scalar and electromagnetic fields the leading eikonal frequencies are

$$\widehat{\omega}_{\ell n} = \frac{\ell+\frac{1}{2}}{3\sqrt{3}} - i\frac{n+\frac{1}{2}}{3\sqrt{3}\,\sigma} + O(\ell^{-1}),$$
$$\ell\to\infty, \qquad n \text{ fixed}, \tag{149}$$

while the Dirac field has the corresponding form

$$\widehat{\omega}_{\kappa n} = \frac{\kappa}{3\sqrt{3}} - i\frac{n+\frac{1}{2}}{3\sqrt{3}\,\sigma} + O(\kappa^{-1}),$$
$$\kappa\to\infty, \qquad n \text{ fixed}. \tag{150}$$

For the axial tensor sector the exact map to the auxiliary geometry instead gives

$$\widehat{\omega}^{\rm T}_{\ell n} = \frac{\ell+\frac{1}{2} - i(n+\frac{1}{2})}{3\sqrt{3}\,\sigma} + O(\ell^{-1}),$$
$$\ell\to\infty, \qquad n \text{ fixed}. \tag{151}$$

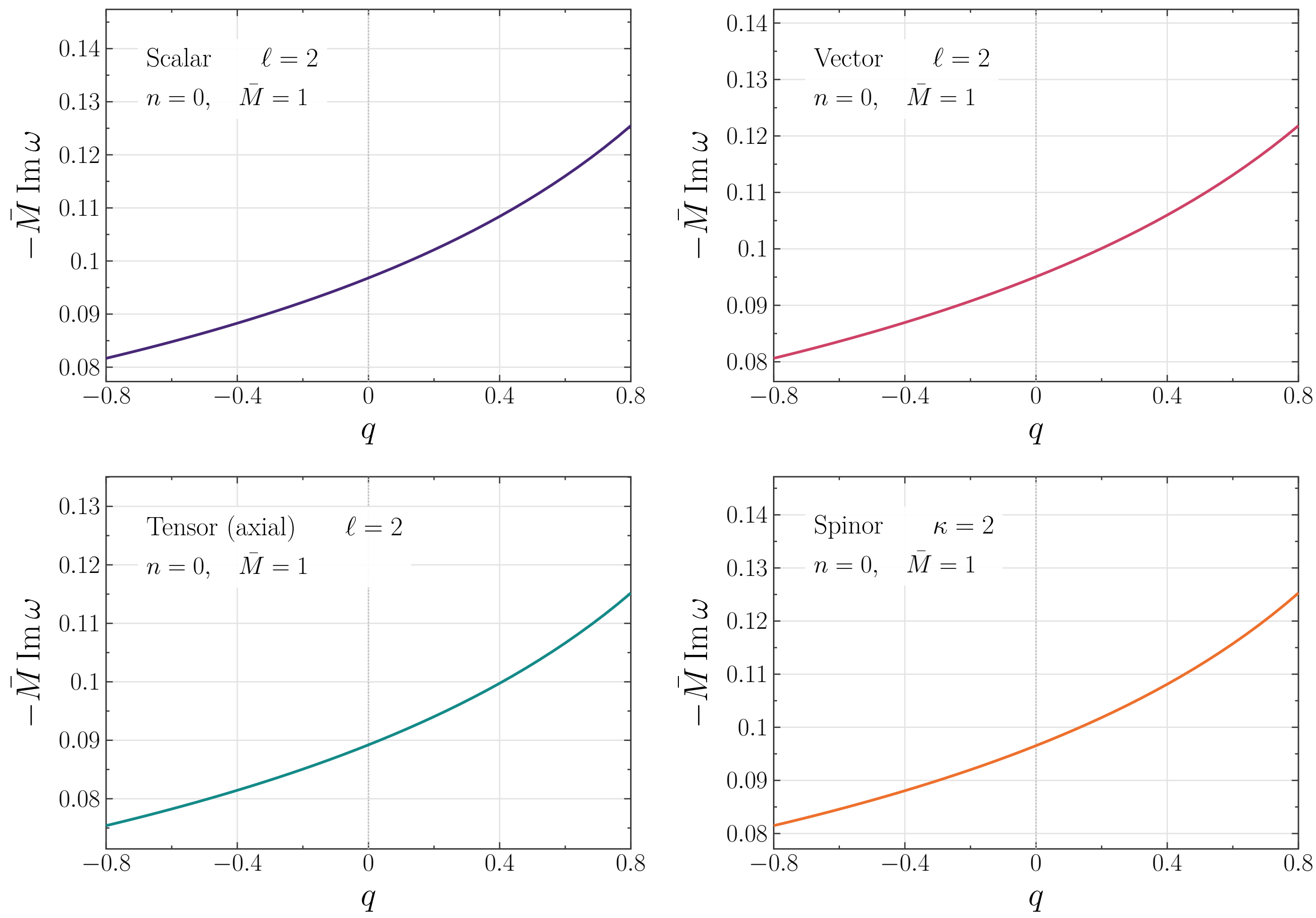

Figure 30. Attenuation rates $-\mathrm{Im}(\bar{M}\omega)$ as functions of $q$ for the same configurations as Fig. 29. Larger values correspond to shorter lifetimes in the normalized Killing time.

The angular term of the tensor potential in Eq. (143) carries the same $1/\sigma^2$ factor as its curvature term. Its leading oscillation rate follows the auxiliary gravitational geometry, whereas the test field oscillation rates are tied to the angular motion of the physical null orbit. At fixed $\bar{M}$, the test field oscillation rates remain unchanged at this order and their attenuation rates vary as $(1-q/2)^{-1/2}$. The same shadow radius can accompany different relaxation times for the test fields. Also, finite angular eigenvalues remain the additional information encoded in the terms of Eqs. (136), (138), and (146).

The radial equations also possess an exact relation with their Schwarzschild counterparts, extending the geometric comparison beyond the leading eikonal limit. Set $y_* = x_*/\sigma$ and multiply Eq. (130) by $\sigma^2$. The resulting frequency is $\sigma\widehat{\omega}$, and the scalar and vector angular terms become those of a Schwarzschild equation with an effective radial index $\nu$, defined by

$$\nu = \frac{\sqrt{1+4\sigma^2\ell(\ell+1)}-1}{2}, \qquad \nu(\nu+1) = \sigma^2\ell(\ell+1). \tag{152}$$

Writing $\mathcal{W}_s(\nu, n)$ for the dimensionless Schwarzschild radial frequency of spin $s$ evaluated at this index, and $\mathcal{W}_{\mathrm{D}}(K, n)$ for its Dirac counterpart at parameter $K$, gives

$$\begin{aligned} \widehat{\omega}^{\mathrm{S}}_{\ell n} &= \sigma^{-1}\mathcal{W}_0(\nu, n), \qquad \widehat{\omega}^{\mathrm{V}}_{\ell n} = \sigma^{-1}\mathcal{W}_1(\nu, n), \\ \widehat{\omega}^{\mathrm{D}}_{\kappa n} &= \sigma^{-1}\mathcal{W}_{\mathrm{D}}(\sigma\kappa, n). \end{aligned} \tag{153}$$

The physical angular harmonics show their integer or half integer labels; $\nu$ and $K$ are parameters of the equivalent radial equations. For the scalar equation the curvature term becomes exactly $2F/x^3$ after multiplication by $\sigma^2$, while the Dirac map follows by replacing $W_\kappa$ with $\sigma W_\kappa$. The ingoing and outgoing conditions transform with the same frequency factor, so Eq. (153) holds for the eigenvalue problem itself.

Table XI. Internal diagnostics for the fundamental modes with angular index 2. Here $\epsilon_{23} = 100|\omega^{(3)} - \omega^{(2)}|/|\omega^{(3)}|$, and $\epsilon_{\pm} = 100|\omega_+^{(3)} - \omega_-^{(3)}|/|\omega_+^{(3)}|$.

| $q$ | $\epsilon_{23}^{\rm S}$ | $\epsilon_{23}^{\rm V}$ | $\epsilon_{23}^{\rm T}$ | $\epsilon_{23}^{\rm D}$ | $\epsilon_{\pm}^{\rm D}$ |
|---|---|---|---|---|---|
| −0.6 | 0.536 | 0.586 | 1.300 | 1.271 | 0.162 |
| −0.3 | 0.639 | 0.705 | 1.300 | 1.528 | 0.209 |
| 0.0 | 0.778 | 0.872 | 1.300 | 1.881 | 0.279 |
| 0.3 | 0.976 | 1.116 | 1.300 | 2.387 | 0.384 |
| 0.6 | 1.274 | 1.500 | 1.300 | 3.152 | 0.556 |

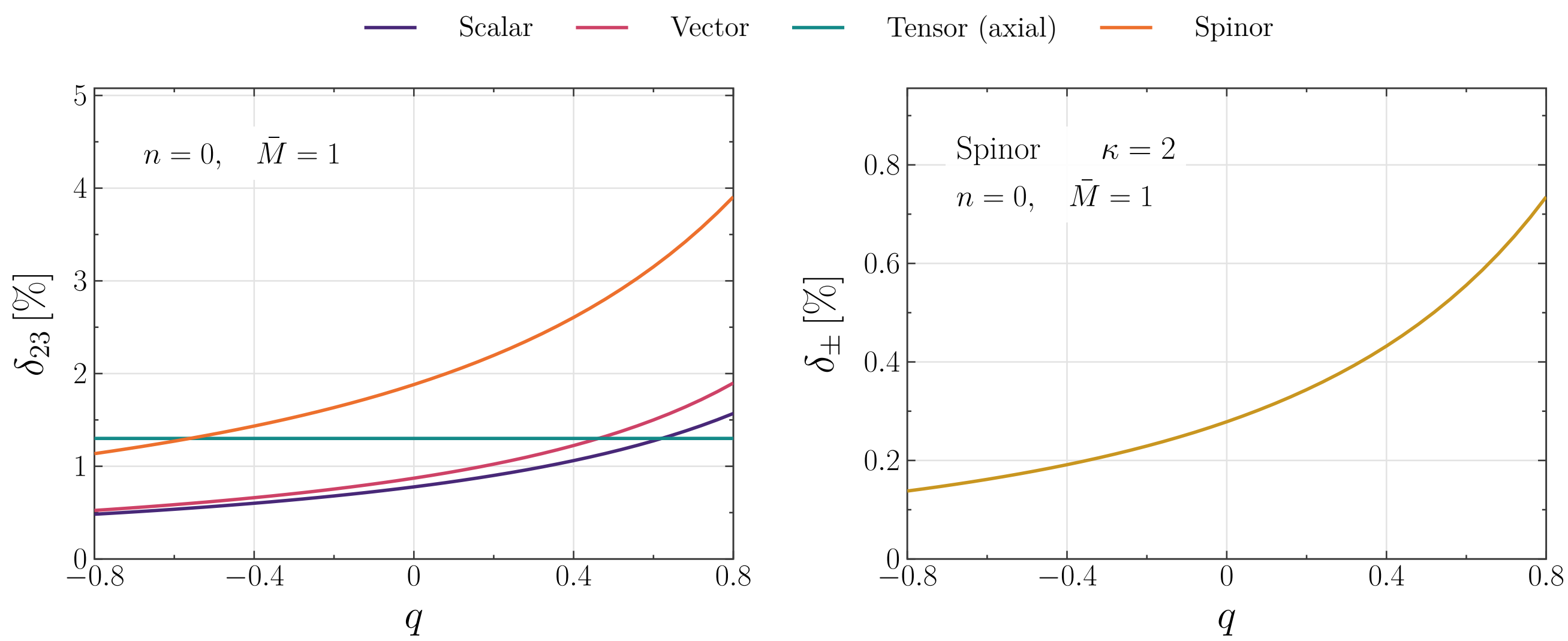

Figure 31. Internal diagnostics of the WKB calculation. The left panel displays the percentage order variation $100\delta_{23}$ defined from Eq. (134). The right panel displays the percentage frequency difference obtained by applying the same finite WKB prescription to the two Dirac partner potentials. Both quantities measure sensitivity of the approximation; their definitions and numerical values are provided with Table XI.

The numerical procedure determines the exterior maximum directly and evaluates the derivatives through repeated application of $\mathcal{D}$, avoiding numerical differentiation of an interpolated potential. It also verifies the vanishing of $U_1$, the sign of $U_2$, and the sign of the computed imaginary part. Figure 31 and Table XI report the WKB order variation and the partner comparison. These checks serve distinct purposes: the first examines sensitivity to the retained expansion terms, and the second tests the approximation against an exact relation between two radial equations. For angular index 2 and $n = 0$, the percentage variation between orders two and three ranges from 0.536% to 1.274% for the scalar field, from 0.586% to 1.500% for the Maxwell field, and from 1.271% to 3.152% for the Dirac field over the tabulated interval. The tensor variation is 1.300% throughout, as required by its exact frequency scaling. The Dirac partner difference ranges from 0.162% to 0.556%.

## X. TIME DOMAIN EVOLUTION

The temporal response supplies a direct view of the propagation, oscillation, and attenuation described by the spectra of Sec. IX. A localized disturbance first excites the exterior potential and subsequently produces a decaying signal at a fixed observation point. We follow this evolution for the four perturbations considered in the previous sections, using the same mass normalization and field equations. The metric, propagation coordinate, angular labels, and effective potentials retain their previous definitions. Only the dimensionless time $\tau = t/\bar{M}$ is introduced here, so the radial amplitudes obey

$$\left[\partial_\tau^2 - \partial_{x_*}^2 + U_s(x)\right] \Psi_s(\tau, x_*) = 0, \qquad s \in \{\mathrm{S}, \mathrm{V}, \mathrm{T}, +\}. \tag{154}$$

The label $+$ selects the Dirac amplitude associated with the potential $U_+$ in Eq. (146). The scalar, electromagnetic, and gravitational entries use Eqs. (136), (138), and (143), respectively. This formulation permits the evolution and the frequency calculation to address the same radial problem.

We integrate Eq. (154) on a characteristic mesh, following the prescription of Gundlach, Price and Pullin [60]. With $u = \tau - x_*$ and $v = \tau + x_*$, the equation becomes $4\partial_u\partial_v\Psi_s + U_s\Psi_s = 0$. Let us consider that $S = (u, v)$, $E = (u, v+h)$, $W = (u+h, v)$, and $N = (u+h, v+h)$ denote the corners of an elementary null cell. The update is

$$\Psi_s(N) = \Psi_s(E) + \Psi_s(W) - \Psi_s(S) - \frac{h^2}{8} U_s(S)\left[\Psi_s(E) + \Psi_s(W)\right] + O(h^4). \tag{155}$$

Here, $U_s(S)$ is evaluated at $x_* = (v-u)/2$ and the corresponding exterior radius. Equation (155) has global error of order $h^2$ for smooth characteristic data. The mesh covers the complete causal past of every reported sample, so the recorded signal is obtained directly from the specified null data.

The same unit pulse is prescribed in all comparisons. A difference of two Gaussian profiles gives an exactly vanishing value at the intersection of the two initial rays while retaining a localized disturbance centered at $v = 20$. In a explicitly manner, we choose

$$\Psi_s(0, v) = G(v) = \exp\left[-\frac{(v-20)^2}{18}\right] - \exp\left[-\frac{(v+20)^2}{18}\right], \qquad \Psi_s(u, 0) = 0. \tag{156}$$

The Gaussian width is 3, and the observation point is $x_*^{\rm obs} = 30$. We hold $\bar{M}$ fixed, set $\ell = 2$ for the bosonic fields and $\kappa = 2$ for the spinor, and compare $q = -0.6, 0, 0.6$. The curves show the raw radial amplitudes produced by Eq. (156), with the same initial normalization. The full evolutions extend from the first observer sample at $\tau = 30$ to $\tau = 350$.

The complex frequency is extracted from the oscillatory portion of each signal with a matrix pencil calculation [61]. Samples separated by $\Delta\tau = 0.5$ are represented by a finite sum

$$\Psi_s(\tau_j) \simeq \sum_{a=1}^{N_{\rm exp}} C_a z_a^j, \qquad z_a = \exp(-i\widehat{\omega}_a \Delta\tau), \qquad \tau_j = \tau_0 + j\Delta\tau. \tag{157}$$

For the real evolutions the retained complex contributions occur in conjugate pairs. Singular value decomposition selects a finite rank for the two shifted sample matrices, whose generalized eigenvalues determine $z_a$. The fundamental frequency is identified through the pair that remains stable under changes in the fitted interval, matrix rank, and mesh spacing. The other retained exponentials improve the reconstruction of the finite signal and carry no individual physical interpretation in the present calculation.

Table XII. Fundamental scalar frequencies for $\ell = 2$ at fixed $\bar{M}$. The WKB values are those of Table VII. The last two columns give the percentage spectral difference and the absolute extraction sensitivity defined in Eq. (159). Six decimal places identify the computed frequency estimate.

| $q$ | $\widehat{\omega}_{\rm TD}$ | $\widehat{\omega}_{(3)}$ | $\Delta_{\rm WKB}$ [%] | $\epsilon_{\rm TD}$ |
|---|---|---|---|---|
| −0.6 | $0.480842 - 0.084760\,i$ | $0.480580 - 0.084785\,i$ | 0.054 | $1.17 \times 10^{-6}$ |
| 0.0 | $0.483644 - 0.096759\,i$ | $0.483211 - 0.096805\,i$ | 0.088 | $4.15 \times 10^{-7}$ |
| 0.6 | $0.488812 - 0.115901\,i$ | $0.487965 - 0.116007\,i$ | 0.170 | $2.74 \times 10^{-6}$ |

The central extraction uses rank eight on $90 \le \tau \le 130$. We also examine the intervals $80 \le \tau \le 120$ and $100 \le \tau \le 140$, and repeat these extractions at rank six. Evolutions with $h = 0.1$ and $0.05$ cover the complete displayed interval; the additional mesh $h = 0.025$ extends to $\tau = 160$ for the frequency calculation. The quoted time domain frequency removes the leading mesh contribution through

$$\widehat{\omega}_{\rm TD} = \frac{4\widehat{\omega}_{0.025} - \widehat{\omega}_{0.05}}{3}. \tag{158}$$

The numerical indicator accompanying the tables records the largest absolute change associated with the fitted intervals, retained rank, and remaining mesh variation,

$$\epsilon_{\rm TD} = \max\{\delta_{\rm window}, \delta_{\rm rank}, \delta_{\rm mesh}\}. \tag{159}$$

Here, $\delta_{\rm window}$ is the largest change from the central rank eight estimate among the tested intervals, and $\delta_{\rm rank}$ includes every rank six estimate within those intervals. The mesh contribution $\delta_{\rm mesh}$ compares Eq. (158) with the analogous extrapolation constructed from $h = 0.1$ and $0.05$. The comparison with the WKB value uses $\Delta_{\rm WKB} = 100|\widehat{\omega}_{\rm TD} - \widehat{\omega}_{(3)}|/|\widehat{\omega}_{\rm TD}|$. The two quantities address separate parts of the calculation: one examines the evolution and extraction, and the other compares the temporal result with the approximation of Sec. IX.

### A. Scalar perturbations

The scalar evolution follows the potential already derived in Eq. (136). Figure 32 displays the signed signal together with its absolute amplitude on a logarithmic scale. The first panel resolves the phase of the oscillations after the pulse reaches the scattering region, and the second follows their attenuation over the full recorded interval. Their common initial normalization permits the dependence on $q$ to be examined through both the excitation amplitude and the subsequent decay. The complex frequencies extracted from this response appear in Table XII.

The extracted scalar frequency changes from $0.480842 - 0.084760i$ at $q = -0.6$ to $0.488812 - 0.115901i$ at $q = 0.6$. The oscillation rate grows, while the attenuation becomes more pronounced. At $q = 0$ the temporal calculation gives $0.483644 - 0.096759i$. Across the three geometries, the differences from the WKB values range from 0.054% to 0.170%, giving a direct temporal confirmation of the scalar ringing identified in the preceding section.

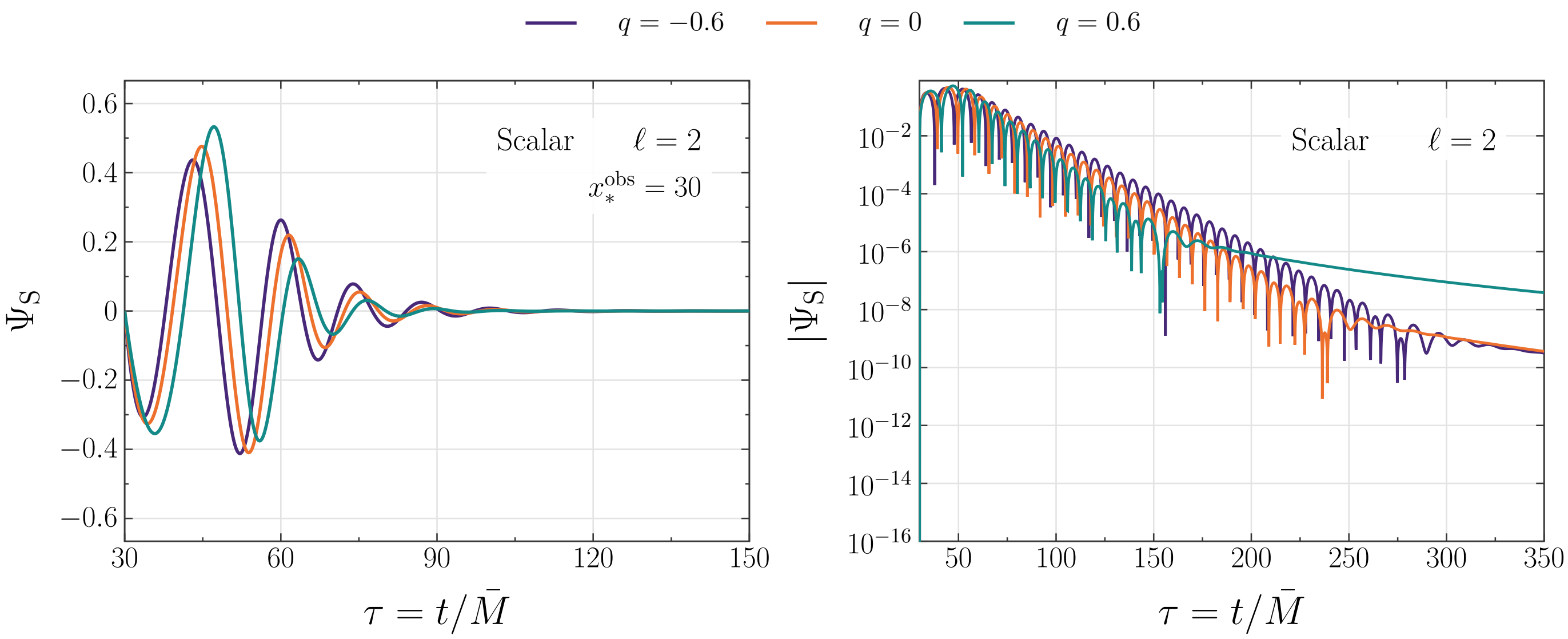

Figure 32. Scalar time domain response for $\ell = 2$ at $x_*^{\rm obs} = 30$. The left panel resolves the signed amplitude, and the right panel follows its absolute value on a logarithmic scale. The same pulse is used for every value of $q$, with $\bar{M}$ held fixed.

Table XIII. Fundamental Maxwell frequencies for $\ell = 2$ at fixed $\bar{M}$. The columns follow Table XII, and the WKB values are those of Table VIII.

| $q$ | $\widehat{\omega}_{\rm TD}$ | $\widehat{\omega}_{(3)}$ | $\Delta_{\rm WKB}$ [%] | $\epsilon_{\rm TD}$ |
|---|---|---|---|---|
| $-0.6$ | $0.460781 - 0.083573\,i$ | $0.460505 - 0.083604\,i$ | 0.059 | $8.72 \times 10^{-7}$ |
| 0.0 | $0.457596 - 0.095004\,i$ | $0.457131 - 0.095065\,i$ | 0.100 | $5.39 \times 10^{-7}$ |
| 0.6 | $0.451681 - 0.112920\,i$ | $0.450737 - 0.113073\,i$ | 0.206 | $1.09 \times 10^{-6}$ |

### B. Vector perturbations

For the Maxwell field both angular parities propagate with Eq. (138). The signal in Fig. 33 describes the radial evolution of either parity for the corresponding initial master amplitude. The fixed height of the potential at fixed $\bar{M}$ is accompanied by a changing width in $x_*$, as shown in Fig. 28. The temporal response gives us a direct way to follow the effect of that width on the duration and phase of the electromagnetic oscillations. Table XIII gives the measured fundamental frequencies and their comparison with the earlier WKB calculation.

The temporal frequencies evolve from $0.460781 - 0.083573i$ to $0.451681 - 0.112920i$ as $q$ increases across the displayed interval. Their real part decreases and their attenuation rate increases. The relative WKB differences lie between 0.059% and 0.206%. In this sense, the temporal calculation resolves the effect of radial propagation even when the electromagnetic potential retains the same height as a function of areal radius.

### C. Tensor perturbations

The gravitational evolution concerns the homogeneous axial sector established in Sec. IX C. We retain the restrictions on the antisymmetric field used in that derivation and evolve its master amplitude with Eq. (143). Figure 34 displays the resulting waveform, while Table XIV records the extracted frequency. This calculation supplies a temporal realization of the axial gravitational tower whose auxiliary description led to Eq. (144).

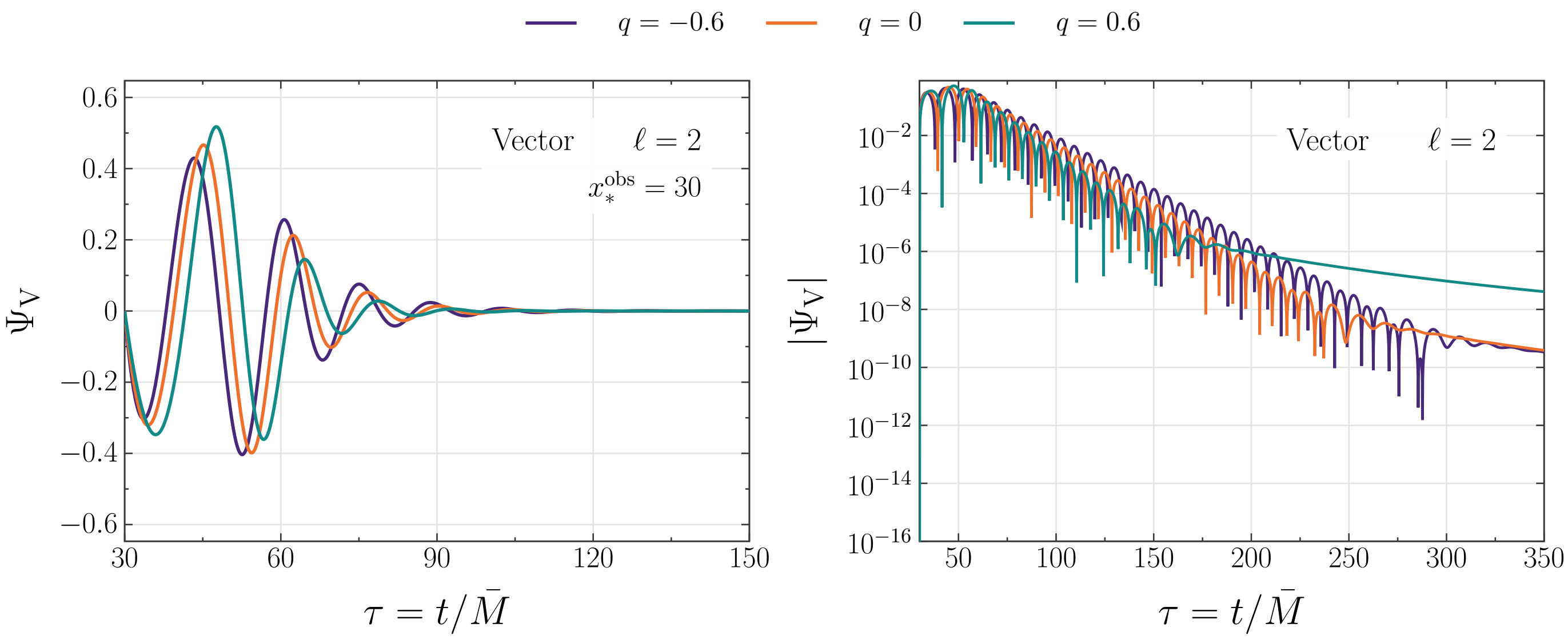

Figure 33. Electromagnetic time domain response for $\ell = 2$ at $x_*^{\rm obs} = 30$. The signed waveform and its absolute value are displayed using the same initial pulse and parameter assignment as Fig. 32. Both Maxwell parities obey the radial equation used here.

Table XIV. Fundamental homogeneous axial gravitational frequencies for $\ell = 2$ at fixed $\bar{M}$. The columns follow Table XII, and the WKB values are those of Table IX. The gravitational sector is specified in Sec. IX C.

| $q$ | $\widehat{\omega}_{\rm TD}$ | $\widehat{\omega}_{(3)}$ | $\Delta_{\rm WKB}$ [%] | $\epsilon_{\rm TD}$ |
|---|---|---|---|---|
| $-0.6$ | $0.327732 - 0.078025\,i$ | $0.327285 - 0.078249\,i$ | 0.148 | $1.06 \times 10^{-6}$ |
| 0.0 | $0.373672 - 0.088962\,i$ | $0.373162 - 0.089217\,i$ | 0.148 | $3.08 \times 10^{-7}$ |
| 0.6 | $0.446623 - 0.106330\,i$ | $0.446014 - 0.106635\,i$ | 0.148 | $7.54 \times 10^{-7}$ |

The common factor $1/\sigma$ in the tensor frequencies controls both the oscillation rate and the attenuation rate. The pulse, its width, and the observation point are fixed in the physical propagation coordinate throughout the figures, so the histories also retain their excitation dependence. The spectral relation nevertheless remains an exact check on the extracted tensor poles for the three geometries.

The measured frequencies are $0.327732 - 0.078025i$, $0.373672 - 0.088962i$, and $0.446623 - 0.106330i$ for the three values of $q$. Their agreement with the common spectral scaling is also reflected in an approximately constant WKB difference of 0.148%. A separate evolution transforms the initial pulse, its width and location, the observer, and the mesh together according to the auxiliary description. The resulting histories agree to a maximum absolute difference of $2.7 \times 10^{-14}$. This additional test checks the temporal map while preserving the fixed physical data used in Fig. 34.

### D. Spinor perturbations

The spinor evolution follows the amplitude associated with $U_+$ in Eq. (146). Figure 35 gives its signed signal and attenuation, and Table XV gives the fundamental frequency for $\kappa = 2$. The spectrum can be compared directly with the WKB entries computed from the same partner potential. The equality of the exact nonzero partner spectra follows from the first order Dirac system discussed in Sec. IX D. The present Gaussian specifies the amplitude evolved with $U_+$.

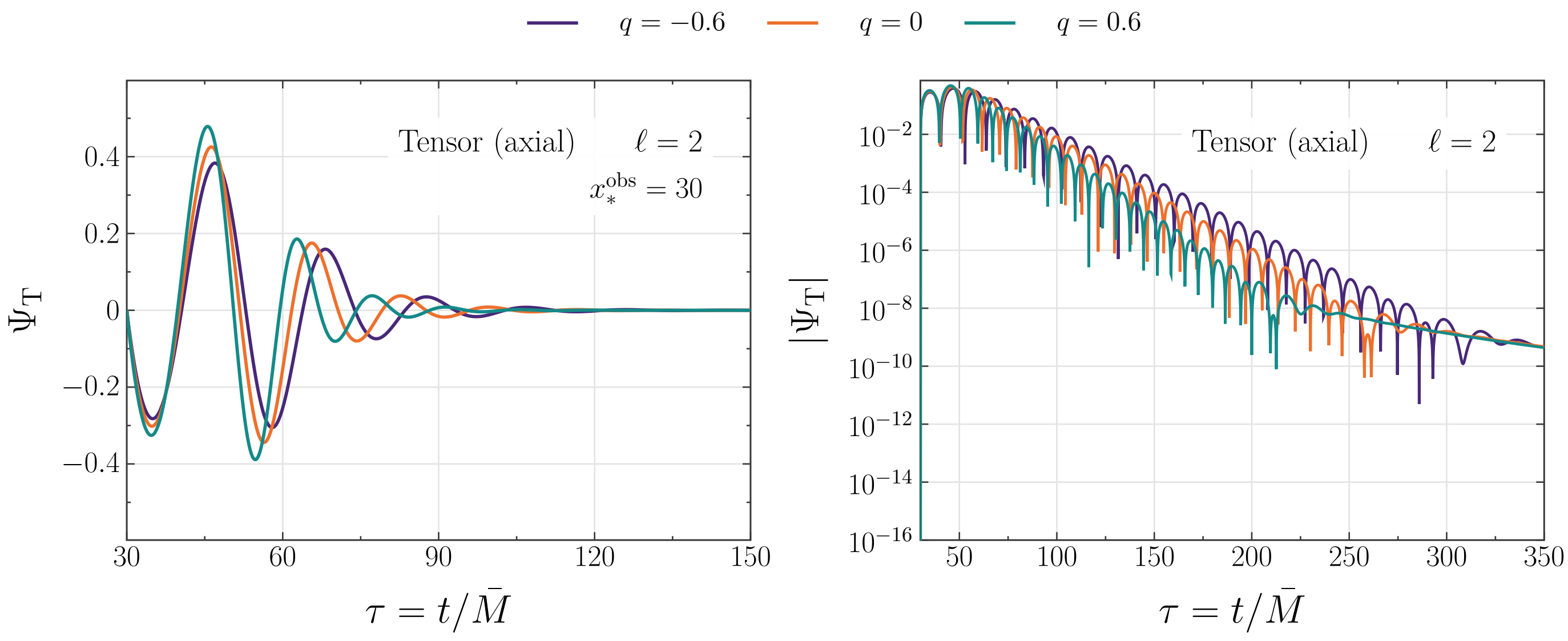

Figure 34. Time domain response of the homogeneous axial tensor sector for $\ell = 2$ at $x_*^{\rm obs} = 30$. The two panels show the signed amplitude and its absolute value.

Table XV. Fundamental Dirac frequencies for $\kappa = 2$ at fixed $\bar{M}$, extracted from the amplitude evolved with $U_+$. The columns follow Table XII, and the WKB values are those of Table X.

| $q$ | $\widehat{\omega}_{\rm TD}$ | $\widehat{\omega}_{(3)}$ | $\Delta_{\rm WKB}$ [%] | $\epsilon_{\rm TD}$ |
|---|---|---|---|---|
| $-0.6$ | $0.381152 - 0.084516\,i$ | $0.380310 - 0.084578\,i$ | 0.216 | $2.07 \times 10^{-6}$ |
| 0.0 | $0.380037 - 0.096405\,i$ | $0.378627 - 0.096542\,i$ | 0.361 | $3.16 \times 10^{-6}$ |
| 0.6 | $0.377976 - 0.115320\,i$ | $0.375244 - 0.115733\,i$ | 0.699 | $5.13 \times 10^{-6}$ |

The spinor frequency changes from $0.381152 - 0.084516i$ to $0.377976 - 0.115320i$ over the same parameter interval. The temporal oscillation rate decreases moderately, and the attenuation grows more visibly. At $q = 0$ the extracted value is $0.380037 - 0.096405i$. The relative WKB difference increases from 0.216% to 0.699% as $q$ grows. In other words, the evolution supplies an independent determination of the fermionic frequency, including the dependence that is most sensitive to the retained WKB order.

### E. Numerical convergence and spectral comparison

The evolution is checked by refining the null mesh while preserving the pulse and the observation point. At $q = 0$ we compare $h = 0.2$, 0.1, and 0.05 on their common observer samples. For the expected global error of order $h^2$, the leading differences satisfy

$$|\Psi_{0.2} - \Psi_{0.1}| \simeq 4|\Psi_{0.1} - \Psi_{0.05}|. \tag{160}$$

Figure 36 places these two quantities on the same axes for each perturbation. Their comparison tests the evolution of the waveform directly, complementing the variation of the extracted frequency under the additional mesh refinement used in Eq. (158).

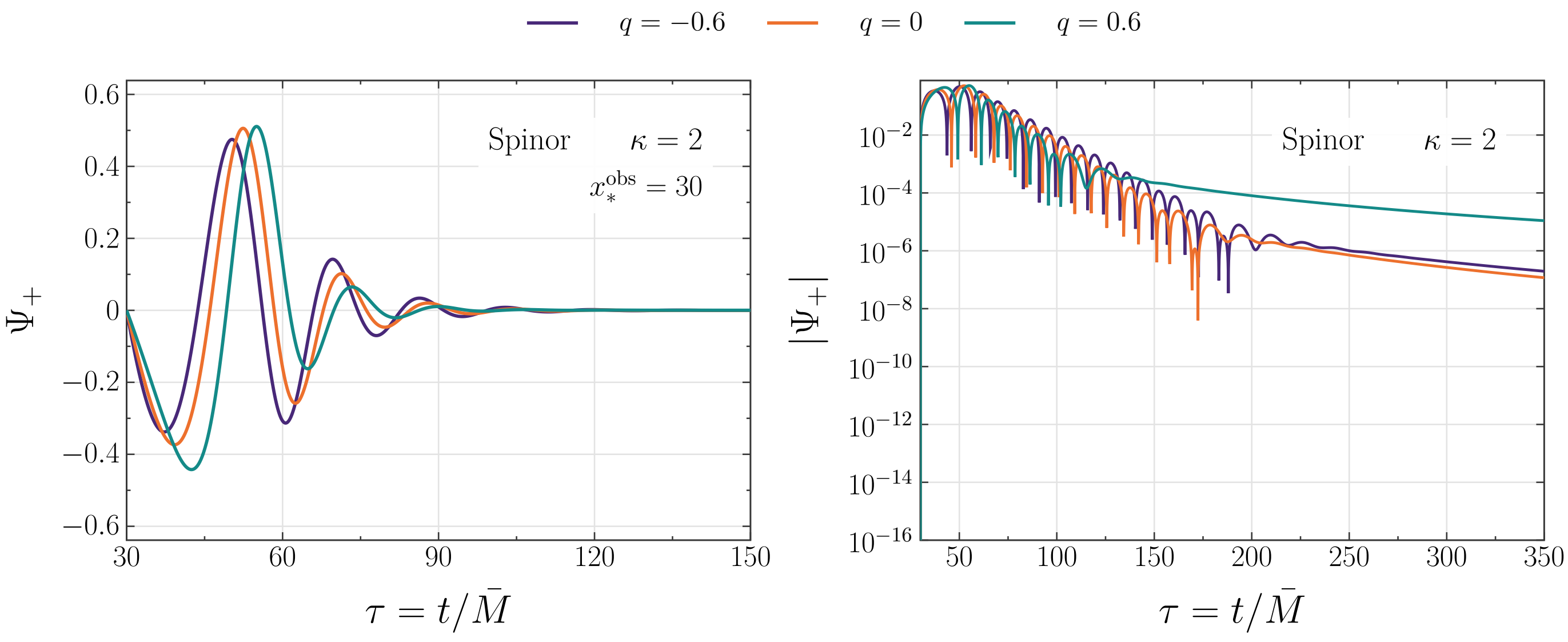

Figure 35. Spinor time domain response for $\kappa = 2$ at $x_*^{\rm obs} = 30$, using $U_+$. The signed amplitude and its absolute value follow from the same Gaussian prescribed in the bosonic evolutions.

Table XVI. Temporal convergence at $q = 0$, evaluated on common samples in $40 \leq \tau \leq 150$. The ratio $\mathcal{R}$ and the relative estimate $E_{0.05}$ are defined in Eq. (161). The observed order is $p = \log_2 \mathcal{R}$.

| Perturbation | $\mathcal{R}$ | $p$ | $E_{0.05}$ |
|---|---|---|---|
| Scalar | 4.001548 | 2.000558 | $1.64 \times 10^{-5}$ |
| Vector | 4.001451 | 2.000523 | $1.60 \times 10^{-5}$ |
| Tensor | 4.001174 | 2.000424 | $1.47 \times 10^{-5}$ |
| Spinor | 4.000987 | 2.000356 | $9.52 \times 10^{-6}$ |

Figure 37 compares each computed signal at $q = 0$ with the rank eight reconstruction on $90 \leq \tau \leq 150$. The fit uses the samples up to $\tau = 130$, and the remaining samples test its subsequent prediction. This comparison examines the temporal information used to identify the stable frequency pair. The table indicators additionally include the interval and rank variations, so the visual agreement of a reconstruction is accompanied by an explicit extraction diagnostic.

For a quantitative comparison, let $\|\cdot\|_2$ denote the Euclidean norm on the common samples in $40 \leq \tau \leq 150$, and define

$$\mathcal{R} = \frac{\|\Psi_{0.2} - \Psi_{0.1}\|_2}{\|\Psi_{0.1} - \Psi_{0.05}\|_2}, \qquad E_{0.05} = \frac{\|\Psi_{0.1} - \Psi_{0.05}\|_2}{3\|\Psi_{0.05}\|_2}. \tag{161}$$

The observed order $p = \log_2 \mathcal{R}$ lies between 2.000356 and 2.000558, as reported in Table XVI. The relative field estimates $E_{0.05}$ are between $9.52 \times 10^{-6}$ and $1.64 \times 10^{-5}$ on this interval. A free propagation calculation also reproduces the characteristic solution with maximum absolute difference $1.5 \times 10^{-14}$, which gives a separate check on the update and observer extraction, as we should experct.

The reconstructions have relative residuals between $1.0 \times 10^{-12}$ and $1.5 \times 10^{-10}$ on the fitted samples. Over the subsequent interval $130.5 \leq \tau \leq 150$, which is excluded from the extraction, the relative differences remain below $1.8 \times 10^{-4}$. This extension is included in Fig. 37, with the end of the fitted interval marked by a vertical dotted line. The additional retained exponentials represent the finite temporal background as well as the oscillatory contribution; the physical interpretation is assigned to the frequency pair stable across the extraction tests.

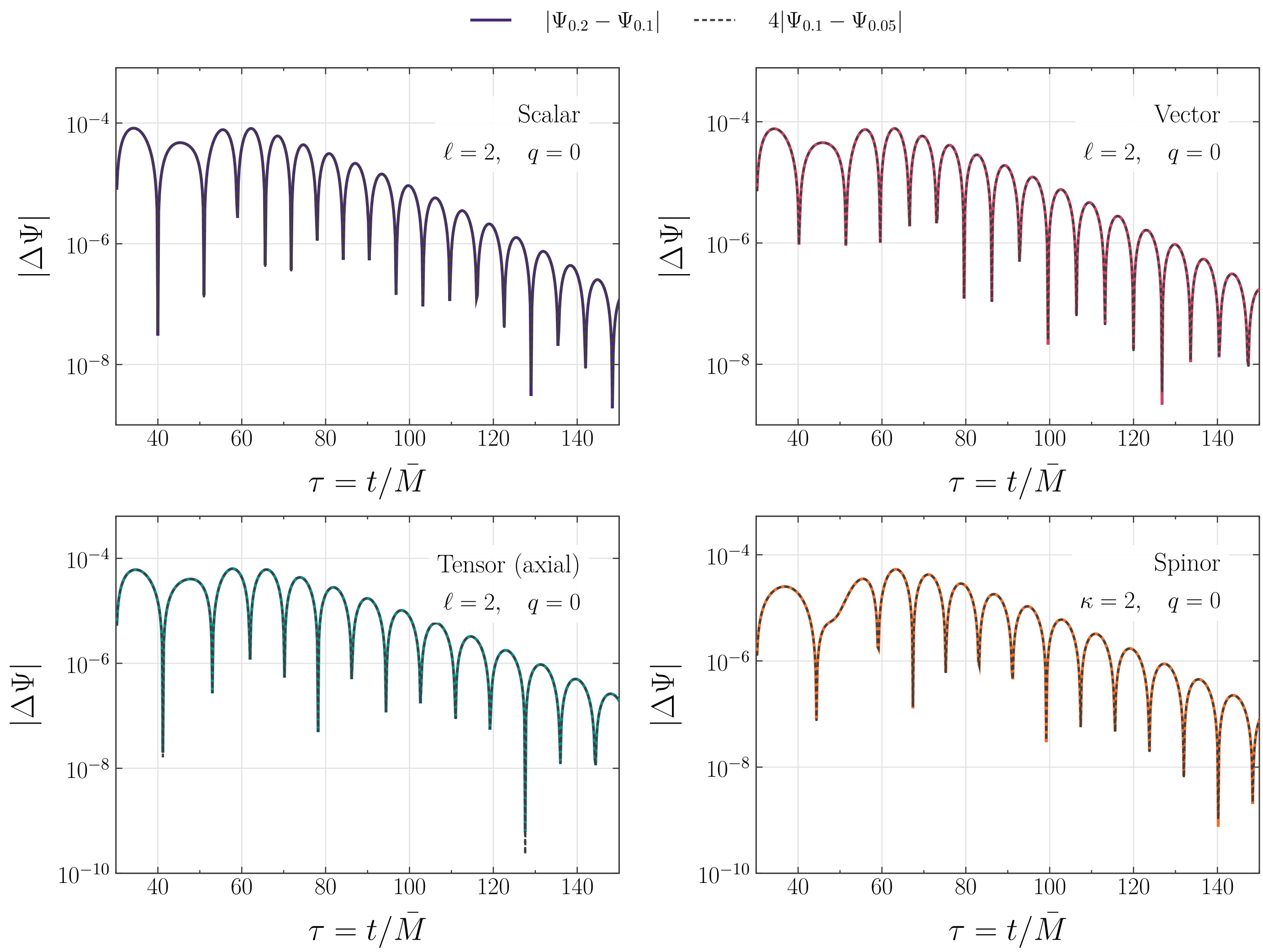

Figure 36. Convergence of the temporal waveforms at $q = 0$ for the four perturbations. Each panel compares $|\Psi_{0.2} - \Psi_{0.1}|$ with $4|\Psi_{0.1} - \Psi_{0.05}|$ on common observer samples. The subscripts denote the null mesh spacing.

The logarithmic panels also display a later response that falls more slowly than the preceding exponential oscillations. Its magnitude and shape depend on the asymptotic potential and on the initial disturbance. Such behavior is expected from curvature scattering, whose decay properties depend on the asymptotic structure of the radial operator [87]. We compare the $h = 0.1$ and $0.05$ signals in intervals of length 25 beginning at $\tau = 150, 200, 250, 300, 325$. The largest relative norm difference among all fields and parameter values is $1.84 \times 10^{-3}$, supporting the resolved late profiles shown in the figures. These finite evolutions characterize the transition in the measured signal without assigning a common asymptotic power to all four potentials. The curves show the actual absolute field values, including their minima at sign changes; the lower plotting limit introduces no artificial amplitude plateau. The temporal waveforms and extracted frequencies are computed in Mathematica, with the WKB entries supplied by the preceding spectral calculation.

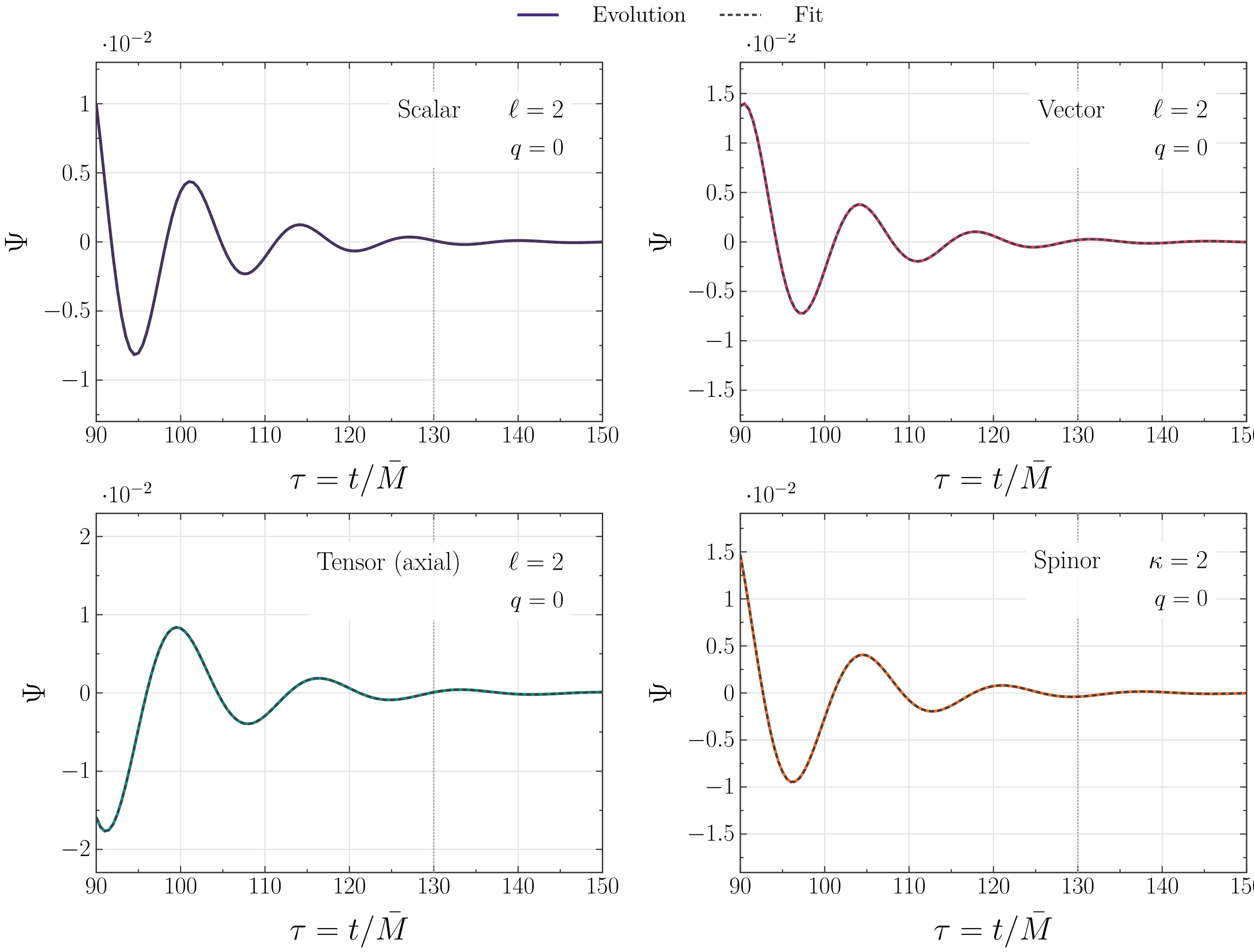

Figure 37. Temporal signals at $q = 0$ and their rank eight matrix pencil reconstructions. The fit uses $90 \le \tau \le 130$; the vertical dotted line marks the start of the subsequent prediction, displayed through $\tau = 150$. The four panels use the same angular labels and observation point as the preceding figures.

## XI. CONCLUSION

We investigated the gravitational and optical properties of a static black hole in Palatini Kalb Ramond gravity with a vanishing cosmological constant. The analysis separated the conserved Iyer Wald mass $M$ from the orbital mass $\bar{M} = M/\rho$ and used the latter to compare configurations with the same gravitational potential. This normalization isolated the radial deformation carried by $\sigma$ and highlilthed which effects survived after the orbital mass had been specified.

At fixed $\bar{M}$, the horizon, photon sphere, critical impact parameter, and innermost stable circular orbit retained their Schwarzschild values. The accumulated azimuth of null and bound timelike trajectories nevertheless changed with $\sigma$. Radial epicyclic motion and the instability of the circular photon orbit also retained sensitivity to this parameter. In particular, the photon orbital frequency remained fixed, while its radial instability exponent varied as $\sigma^{-1}$.

The radial frequency exchange led to a similar distinction. At fixed orbital mass and the same areal release and particle radii, the measured Doppler ratios coincided with their Schwarzschild profiles. Proper radial distances, infall durations, and radial signal propagation times instead acquired a factor $\sigma$. The tidal calculation gave radial stretching and transverse compression with a common factor $\sigma^{-2}$, while their relative magnitude and the vanishing trace of the radial tidal tensor were preserved. The accumulated deformation at a fixed fraction of the release radius remained unchanged, although the proper time required to reach that configuration depended on the couplings.

The optical analysis identified the importance of the conical asymptotics. The angular sweep of a scattered ray factorized into $\sigma$ times the Schwarzschild sweep at the same orbital mass and impact parameter. Subtraction of the corresponding zero mass conical reference geometry gave the same factor for the angular excess, including the logarithmic coefficient near the capture. The shadow angular radius, evaluated in the static observer's frame, depended on $\bar{M}$ and the observer's areal radius. In this manner, the diameter comparison with Sgr A* and M87* left the couplings degenerate when the adopted dynamical fixed $\bar{M}/D_A$.

Moreover, the perturbative response provided further information about radial propagation. Scalar, electromagnetic, and massless Dirac fields evolved on the physical metric, whereas the homogeneous axial gravitational sector followed from the Palatini equations under $\delta B_{\mu\nu} = 0$. Its auxiliary geometry yielded an exact spectral relation to Schwarzschild, with all axial gravitational frequencies multiplied by $\sigma^{-1}$ at fixed $\bar{M}$. The reported fundamental modes displayed increasing attenuation as $q$ increased. Their oscillation rates depended on the sector: the scalar rate increased gently, the electromagnetic and Dirac rates decreased, and the axial gravitational rate followed the exact inverse scaling with $\sigma$. At leading order in large angular momentum, the test field oscillation frequencies followed the physical photon orbit, while the axial gravitational frequencies reflected the auxiliary geometry.

The characteristic evolution checked the third order WKB estimates, with relative differences below 0.7% for the reported fundamental modes. Mesh refinement supported convergence of order two, while changes in the extraction interval and retained rank tested the temporal estimates. The positive radial potentials and Dirac factorization excluded exponential growth in the sectors considered. This feature applied to the specified test fields and homogeneous axial gravitational tower; polar perturbations and independent Kalb Ramond excitations remained outside the analysis.

Moreover, further work could examine matter accretion, neutrino oscillations, and the statistical mechanics of particles in this spacetime, following Refs. [88–92].

## ACKNOWLEDGEMENTS

A. A. Araújo Filho is supported by Conselho Nacional de Desenvolvimento Científico e Tecnológico (CNPq) – [150223/2025-0]. M. S. is supported by Conselho Nacional de Desenvolvimento Científico e Tecnológico CNPq/PDE 200218/2025-5.

---